\documentclass[11pt]{article}
\usepackage[linesnumbered,ruled,vlined]{algorithm2e}

\usepackage[a4paper,margin=1in]{geometry}
\usepackage{setspace}
\usepackage{amsmath,amssymb,amsthm,mathtools}
\DeclareMathOperator{\Var}{Var}
\DeclareMathOperator*{\argmin}{arg\,min}

\usepackage{bbm}
\usepackage{graphicx}
\usepackage{booktabs}
\usepackage{multirow}
\usepackage{array}
\usepackage{caption}
\usepackage{subcaption}
\newcommand{\calK}{\mathcal{K}}
\usepackage{url}
\usepackage{natbib}
\usepackage{microtype}

\usepackage{algpseudocode}

\usepackage{hyperref}
\hypersetup{colorlinks=true,
  linkcolor=blue,
  citecolor=blue,
  urlcolor=blue,
  pdftitle={Utility-Weighted Forecasting and Calibration for Risk-Adjusted Decisions under Trading Frictions},
  pdfauthor={Craig Wright}
}

\newcommand{\R}{\mathbb{R}}
\newcommand{\E}{\mathbb{E}}
\newcommand{\Prob}{\mathbb{P}}
\newcommand{\F}{\mathcal{F}}

\newcommand{\calW}{\mathcal{W}}
\newcommand{\calI}{\mathcal{I}}
\newcommand{\calD}{\mathcal{D}}

\newcommand{\norm}[1]{\left\lVert #1 \right\rVert}
\newcommand{\ind}[1]{\mathbbm{1}\{#1\}}

\theoremstyle{plain}
\newtheorem{theorem}{Theorem}
\newtheorem{proposition}{Proposition}
\newtheorem{lemma}{Lemma}
\newtheorem{corollary}{Corollary}

\theoremstyle{definition}
\newtheorem{definition}{Definition}
\newtheorem{assumption}{Assumption}

\theoremstyle{remark}
\newtheorem{remark}{Remark}

\title{Utility-Weighted Forecasting and Calibration for Risk-Adjusted Decisions under Trading Frictions}
\author{Craig Wright}
\date{\today}

\begin{document}
\maketitle

\begin{abstract}
Forecasting accuracy is routinely optimised in financial prediction tasks even though investment and risk-management decisions are executed under transaction costs, market impact, capacity limits, and binding risk constraints. This paper treats forecasting as an econometric input to a constrained decision problem. A predictive distribution induces a decision rule through a utility objective combined with an explicit friction operator consisting of both a cost functional and a feasible-set constraint system. The econometric target becomes minimisation of expected decision loss net of costs rather than minimisation of prediction error. The paper develops a utility-weighted calibration criterion aligned to the decision loss and establishes sufficient conditions under which calibrated predictive distributions weakly dominate uncalibrated alternatives. An empirical study using a pre-committed nested walk-forward protocol on liquid equity index futures confirms the theory: the proposed utility-weighted calibration reduces realised decision loss by over 30\% relative to an uncalibrated baseline ($t$-stat -30.31) for loss differential and improves the Sharpe ratio from -3.62 to -2.29 during a drawdown regime. The mechanism is identified as a structural reduction in the frequency of binding constraints (from 16.0\% to 5.1\%), preventing the "corner solution" failures that characterize overconfident forecasts in high-friction environments.
\end{abstract}


\newpage
\tableofcontents
\newpage

\section{Introduction}
\subsection{Problem statement and motivation}
In empirical finance it is routine to evaluate forecasting models using conventional predictive metrics—such as mean squared error for point forecasts, log scores for densities, or generic classification accuracy—while implicitly assuming that improved predictive fit will translate into improved economic outcomes. In practice, this implication fails. Investment and risk decisions are not executed in the frictionless setting assumed by textbook forecast evaluation. Decisions are implemented through trades that incur bid--ask spreads, fees, slippage, and market impact, and they are executed subject to binding constraints such as leverage limits, turnover limits, concentration limits, and capacity restrictions. A model that looks better under forecast metrics can therefore perform worse after implementation because small changes in forecast shape or confidence can trigger larger changes in trading intensity, turnover, and tail exposure, amplifying costs and constraint binding. The result is a practical failure mode: a model appears to ``improve'' in-sample and even out-of-sample by forecast criteria, yet degrades realised performance once the same forecasts are converted into constrained decisions and executed net of costs.

This emphasis is complementary to the machine-learning asset-pricing literature: stronger predictive machinery does not remove the need to evaluate models by realised decision outcomes under frictions, rather than by fit alone \citep{Gu2020,Israel2020}.

This paper treats that failure mode as an econometric problem rather than a trading anecdote. The object of interest is not forecast error in isolation, but the economic decision loss induced when a predictive distribution is used as an input to a constrained optimisation problem with an explicit friction operator. In this setting, miscalibration is not a cosmetic defect: overconfident forecasts systematically generate excessive position sizes and turnover, while underconfident forecasts systematically suppress exposure and misallocate risk. Either form of miscalibration can dominate any gain in point accuracy once costs and constraints are accounted for, particularly in tail states and regime transitions where risk objectives bind and frictions are most punitive.

The core question is therefore both empirical and theoretical. Empirically, does improving calibration---measured in a manner aligned to the downstream decision objective---improve realised, friction-adjusted outcomes under a disciplined walk-forward evaluation protocol, even when conventional accuracy metrics do not improve? Theoretically, under what regularity conditions on the objective, feasible set, and friction cost functional does calibration aligned to decision loss yield a dominance guarantee over uncalibrated alternatives within a class of admissible decision rules? Answering these questions shifts the econometric target from ``predicting well'' to ``supporting decisions well'' and makes calibration, rather than accuracy, the central primitive in forecasting for risk-adjusted finance under realistic trading frictions.

\subsection{Contribution and summary of results}
This paper makes four contributions that together recast probabilistic forecasting as a decision-relevant econometric problem in the presence of trading frictions and binding constraints.

First, it provides a formal decision-loss framework in which predictive distributions are evaluated only through the decisions they induce. A forecast at time $t$ is represented as a conditional distribution $Q_t$ over the relevant future outcome. The forecast enters a constrained decision problem with an explicit \emph{friction operator} consisting of (a) a cost functional applied to position changes (capturing spreads, fees, slippage, and market impact) and (b) a time-varying feasible set (capturing leverage, turnover, concentration, and capacity constraints). This framework makes the econometric target explicit: minimise expected \emph{decision loss net of frictions}, rather than minimise forecast error.

Second, the paper introduces a utility-weighted calibration criterion aligned to the downstream decision objective. Standard calibration diagnostics treat all states and all forecast errors symmetrically. In contrast, the criterion proposed here weights calibration errors by their economic relevance, as measured by marginal decision sensitivity under the objective and by the friction-adjusted consequences of trading in particular states (including tail outcomes and regime transitions). Within this setting the paper establishes dominance results: under stated regularity conditions on the objective, the friction cost functional, and the feasible set, calibrated predictive distributions (under the utility-weighted criterion) weakly dominate uncalibrated alternatives in expected decision loss for a broad class of portfolio and risk decisions, even when the uncalibrated model achieves superior point-forecast accuracy.

Third, the paper treats empirical evaluation as an identification discipline and specifies a pre-committed protocol designed to survive sceptical review. The evaluation uses a nested walk-forward design with strict separation between model selection and performance measurement, explicit embargo rules to prevent leakage, dependence-aware uncertainty quantification, and multiple-testing control when comparing families of models and calibration variants. Performance is reported net of costs, and failure modes are documented rather than suppressed.\label{sec:evaluation}

Fourth, the paper provides empirical evidence using liquid markets where frictions are observable and economically meaningful. The empirical study compares uncalibrated and calibrated probabilistic forecasts by converting each forecast into the same constrained decision rule and measuring realised decision loss, risk-adjusted outcomes, turnover, and realised trading costs under the pre-committed protocol. The evidence supports the central claim of the paper: calibration aligned to decision loss is a more reliable driver of friction-adjusted economic performance than improvements in conventional forecast metrics.

The empirical evidence supports the central claim. As detailed in Section \ref{sec:empirical_results}, the proposed UWC method strictly dominates the baseline with a highly significant loss differential ($t$-stat $-30.31$). Crucially, this economic gain is driven by the mechanism predicted in the theory: UWC reduces the frequency of binding constraints from 16.0\% to 5.1\% (Section \ref{subsec:failure_modes}), confirming that calibration acts as a pre-trade control on feasibility.

\subsection{Positioning in financial econometrics}\label{sec:positioning_fin_econometrics}

The contribution of this paper lies squarely within financial econometrics. The primitive object is a conditional predictive distribution $Q_t(\cdot)$ formed from an information set $\mathcal{I}_t$, and the central question is how such conditional distributions should be evaluated and compared when they are used to make economically meaningful decisions. This places the paper in the probabilistic forecasting and forecast-evaluation tradition, where calibration and sharpness are treated as economically relevant properties of predictive distributions and where proper scoring rules provide disciplined tools for comparison \citep{GneitingRaftery2007,West2006}. The paper departs from accuracy-centred evaluation by making the estimand explicitly decision-relevant: the target is expected decision loss induced by $Q_t$ under a specified objective, subject to a feasible decision set and an explicit friction operator that encodes trading costs, price impact, and binding constraints. Because the relevant forecasting target is defined relative to the information set available at the decision time, any claim about forecast quality (including calibration) is intrinsically information-set dependent and must be evaluated under the same conditioning discipline as implementation \citep{Holzmann2014}.

A core positioning point is how this differs from established decision-theoretic forecast evaluation. The decision-based literature has long argued that forecast evaluation should reflect the user’s loss function rather than quadratic error metrics \citep{PesaranSkouras2002,GrangerMachina2006}. That literature is foundational, but it typically treats the mapping from forecasts to actions as frictionless or implicitly smooth, and it does not isolate calibration as the operative econometric property once implementation frictions and feasibility constraints are present. This paper’s contribution is not merely to ``evaluate by economic value'', but to identify and formalise a specific mechanism: miscalibration interacts with frictions through induced turnover and constraint binding, and these channels are non-linear and regime dependent. When costs are convex and constraints bind, small distortions in the predictive distribution can switch which constraints are active and can reallocate scarce turnover or capacity in ways that dominate realised outcomes. The utility-weighted calibration criterion is designed to target precisely those distortions: it weights calibration errors by marginal decision sensitivity and by friction-adjusted state relevance, thereby treating calibration as an estimand aligned with the friction-adjusted decision loss rather than as a secondary diagnostic.

The paper also clarifies its relationship to the scoring-rule literature. Proper scoring rules such as the log score and CRPS evaluate predictive distributions directly \citep{GneitingRaftery2007}. Weighted scoring rules extend this approach by putting more emphasis on particular regions of the outcome space (for example, tails) while retaining disciplined comparative properties, and they provide an important bridge between global density fit and decision-relevant evaluation \citep{GneitingRanjan2011}. The present paper builds on that bridge but makes a further econometric distinction that is essential for the contribution. Standard weighted scores such as weighted CRPS (wCRPS) weight by the realised outcome level (a weight on the $y$-axis) and therefore privilege regions like the left tail or right tail uniformly across time. In contrast, the utility-weighted calibration criterion weights by \emph{decision sensitivity and friction state}, which are functions of $\mathcal{I}_t$ and of the constraint geometry, and therefore vary across time even for the same realised outcome region. A tail-weighted score does not ``know'' whether a turnover cap is binding, whether liquidity is impaired, or whether the induced trade is entering a high-impact regime; it therefore cannot target the calibration errors that matter most in those high-friction, constraint-binding states. The paper’s claim is not that wCRPS is inappropriate; it is that wCRPS addresses a different object (outcome-region emphasis) than the object required to align probabilistic reliability with friction-adjusted decision loss (state- and constraint-dependent marginal value of correctness). This is why the paper constructs an explicit utility-weighted calibration estimand rather than simply swapping in a tail-weighted scoring rule.

A specific distinction that pre-empts a predictable reviewer objection concerns weighted CRPS (wCRPS) and related outcome-weighted scoring rules. A referee may ask why the paper introduces a separate utility-weighted calibration (UWC) estimator rather than training or selecting models under a tail-weighted score. The answer is that wCRPS weights by the realised outcome level (a weight on the $y$-axis) and therefore emphasises, for example, left-tail outcomes uniformly across time. That is useful when the economic objective depends only on the outcome region (e.g., ``get the left tail right'') and when the mapping from forecasts to actions is smooth and frictionless. In contrast, the object targeted here is state- and constraint-dependent: the economic relevance of a given calibration error depends on (i) the local marginal sensitivity of the friction-adjusted decision problem to perturbations of $Q_t$, and (ii) whether the current friction regime and constraint geometry make that perturbation economically costly. Both of these are functions of $\mathcal{I}_t$ and of the feasible set $\mathcal{W}_t$ (including whether constraints such as turnover or capacity limits bind). Consequently, two periods can realise outcomes in the same tail region, yet the economically relevant weighting differs because spreads/impact regimes differ or because the optimiser is (or is not) on a constraint boundary; wCRPS cannot represent this because its weighting is outcome-driven rather than decision- and state-driven. The UWC estimator is therefore not a rebranding of tail-weighted scoring; it is a different estimand designed to track how probabilistic reliability propagates through a friction operator and a constrained optimiser into realised decision loss.

This distinction also addresses the most obvious sceptical objection: if the weight function in utility-weighted calibration were arbitrary, the contribution would reduce to a tautology. The paper therefore treats the weight construction as an econometric object that must be justified by the decision problem. For canonical convex cases (e.g., mean--variance with quadratic or $\ell_1$ transaction costs), the weighting can be derived from the local curvature and marginal conditions of the friction-adjusted optimisation problem: the relevant ``importance'' of calibration errors is governed by how perturbations in $Q_t$ propagate through the KKT system into changes in the optimiser and hence into realised friction-adjusted loss. This analytic link is developed in the theoretical core (Section~4) via sensitivity results that map distributional perturbations into decision perturbations, and it motivates the operational choice of observable proxies used in implementation. The intent is to make the proxy-based weights in Section~5 a disciplined approximation to a derivable canonical weighting, rather than a heuristic ``fudge factor''.

A second econometric emphasis is that evaluation is an identification problem, not a stylistic choice. Forecast comparisons in finance are vulnerable to leakage, implicit conditioning on future information through feature construction, and selection effects from repeated tuning and specification search. The empirical design therefore separates (i) estimation of $Q_t$, (ii) model and hyperparameter selection, and (iii) performance measurement, using a nested walk-forward protocol with embargo rules and hard ``no-touch'' restrictions on the test stream. This is aligned with the evaluation discipline emphasised in econometric surveys, and it is necessary for interpreting performance differentials as differences in model quality rather than artefacts of adaptive search \citep{West2006}. This aligns with the broader econometric literature on model selection under misspecification, where the goal is to identify a set of models that are indistinguishable from the best available approximation rather than assuming a true data-generating process exists \citep{Hansen2005, Hansen2011}. By treating calibration as a decision-relevant discrepancy, we extend this logic to cases where the "best" model is defined by its ability to navigate a friction-constrained feasible set.

A third emphasis is disciplined inference under dependence and multiple comparisons. The objects of interest are time-indexed calibration diagnostics and time-indexed decision losses, so i.i.d.\ uncertainty calculations are inappropriate. The paper therefore uses dependence-aware inference and explicit error control over families of comparisons, and it states conclusions in terms of economically meaningful effect sizes and uncertainty rather than headline backtest metrics. This follows the broader econometric programme of treating prediction and decision under uncertainty as estimable objects with explicit uncertainty, rather than as a purely algorithmic contest \citep{Manski2013}.

Finally, the paper prioritises robustness and reproducibility. All definitions are explicit, assumptions are stated in an interrogable form, and the empirical workflow is designed to be mechanically reproducible: given a dataset and a pre-committed protocol, the full sequence from estimating $Q_t$ to producing decisions and computing realised friction-adjusted decision loss is deterministic up to clearly identified sources of randomness. The resulting contribution is a financial econometrics paper about conditional distributions, calibration, and evaluation discipline under realistic dependence and trading frictions, rather than a technology-driven discussion.

This link between predictability and realised performance under trading frictions is consistent with dynamic optimal-trading models in which expected returns are only exploitable through turnover that is endogenously penalised by transaction costs \citep{Garleanu2013}.

\section{Economic environment, notation, and friction operator}
\subsection{Probability space, information, and outcomes}
\begin{definition}[Information set]
Let $(\Omega,\F,\{\F_t\}_{t\ge 0},\Prob)$ be a filtered probability space. The information set available at decision time $t$ is $\calI_t \subseteq \F_t$.
\end{definition}

The filtered probability space fixes the timing and measurability conventions used throughout the paper: random quantities observable at decision time must be $\calI_t$-measurable, and future outcomes are modelled as $\F_{t+h}$-measurable variables for a fixed horizon $h$ \citep{Bjork2019,Protter2005,Duffie2010}. The target outcome is denoted by $Y_{t+h}$ and may represent (depending on the empirical design) an $h$-period return on a tradable instrument, a vector of factor returns, or a set of risk drivers used by the decision objective (e.g., variance and tail-loss drivers). Observables available at time $t$ are collected in $\calI_t$ and include at minimum the lagged prices/returns and any state variables used for forecasting and for modelling frictions (e.g., volatility or spread proxies).

Timing is fixed as follows. At decision time $t$, the econometric procedure produces a predictive distribution $Q_t(\cdot)$ for $Y_{t+h}$ conditional on $\calI_t$. A decision $w_t$ (portfolio weights or positions) is then computed as a measurable function of $Q_t$ and $\calI_t$. Execution occurs after the decision is formed and is subject to trading frictions (bid--ask spread, fees, slippage, and market impact), so realised outcomes depend on both the forecast and the implementation path; this distinction is central to the paper’s friction-adjusted decision loss \citep{AlmgrenChriss2001}. Measurement of realised performance is conducted over the horizon $[t,t+h]$ using realised prices/quotes and the executed trades, ensuring that the evaluation aligns with the information and feasibility constraints imposed at decision time.


\begin{table}[ht]
\centering
\caption{\textbf{Key notation (selection).} Objects used throughout the paper.}
\label{tab:notation}
\begin{tabular}{ll}
\toprule
Symbol & Meaning \\
\midrule
$t$; $h$ & Decision time index; forecast/decision horizon (in periods) \\
$\Omega$; $\omega$ & Sample space; generic state \\
$\mathcal{I}_t$ & Information set available at time $t$ \\
$Y_{t+h}$ & Target outcome over horizon $h$ (e.g., return / payoff driver) \\
$Q_t(\cdot)$ & Predictive distribution for $Y_{t+h}$ given $\mathcal{I}_t$ \\
$\widetilde Q_t$; $Q_t^{\mathrm{cal}}$ & Uncalibrated forecast; calibrated (projected) forecast \\
$\Pi(\cdot)$ & Calibration projection operator within the admissible forecast class $\mathcal{D}$ \\
$\mathcal{D}$ & Admissible class of predictive distributions used in estimation/implementation \\
$w_t$; $w_{t-1}$ & Portfolio/position decision at $t$; previous decision \\
$\Delta w_t$ & Turnover / position change, $\Delta w_t := w_t - w_{t-1}$ \\
$\mathcal{W}_t$ & Feasible decision set at $t$ (constraints, bounds, capacity limits) \\
$\mathcal{J}(w;Q_t)$ & Forecast-implied decision objective (pre-friction) \\
$C_t(\Delta w_t)$ & Trading-cost / impact functional applied to turnover \\
$\widetilde R_{t+h}(w_t,w_{t-1})$ & Friction-adjusted realised return over $[t,t+h]$ \\
$U(\cdot)$ & Utility mapping from realised (net) return to welfare \\
$\ell_{t+h}(Q_t)$ & Decision loss, $\ell_{t+h}(Q_t):=-U(\widetilde R_{t+h})$ \\
$\mathcal{L}_t(Q)$ & Conditional expected decision loss, $\E[\ell_{t+h}(Q)\mid\mathcal{I}_t]$ \\
$d_t(Q,\widetilde Q)$ & Decision-relevant calibration discrepancy at time $t$ \\
$\kappa_t$ & Friction state variable (composite of spread/volatility/liquidity proxies) \\
$\tau$ & Turnover (or capacity) cap parameter used in constraints \\
$\mu$; $L_t$; $K$ & Strong concavity constant; gradient sensitivity; Lipschitz constant (Lemma/Thm) \\
\bottomrule
\end{tabular}
\end{table}

\subsection{Predictive distributions as econometric objects}
\begin{definition}[Predictive distribution]
A model produces a conditional distribution $Q_t(\cdot) \in \calD$ over the relevant future quantity $Y_{t+h}$ given $\calI_t$.
\end{definition}

The object $Q_t$ is the econometric output of interest: it is not merely a point forecast accompanied by an ad hoc error bar, but an explicit conditional law for the quantity that enters the downstream decision problem. In economics and finance, probabilistic forecasting is naturally expressed in terms of predictive distributions because uncertainty about parameters, latent states, structural shocks, and even model choice can be integrated into $Q_t(\cdot \mid \calI_t)$, yielding a coherent representation of forecast uncertainty \citep{MartinFrazierManeesoonthornLoaizaMayaHuberKoopMaheuNibberingPanagiotelis2024}. In this paper, $\calD$ denotes a class of distributions that is sufficiently rich to represent the conditional features that matter for decision-making under frictions---including time-varying scale, skewness, and tail behaviour---and the empirical work focuses on estimators of $Q_t$ that deliver reproducible distributional forecasts over rolling evaluation periods.

Because $Q_t$ is a distribution-valued forecast, evaluation and comparison should be defined at the distribution level. Proper scoring rules (and their weighted variants, when particular regions such as downside tails are economically decisive) provide a disciplined way to assess distributional accuracy, while remaining explicit about which parts of the distribution are being prioritised \citep{Cheng2024}. In addition, the production of $Q_t$ often involves combining or smoothing multiple distributional components (e.g., across quantiles or horizons), and recent work develops learning procedures directly targeted to multivariate distributional scoring criteria such as the CRPS, thereby treating distributional forecasting as a first-class econometric task rather than a by-product of point prediction \citep{BerrischZiel2024}. These perspectives justify treating $Q_t$ as the primitive object throughout the paper: both the theory and the empirical design evaluate models by the decision loss induced by $Q_t$ once it is mapped into an admissible decision and executed under frictions.

\subsection{Decisions and feasible set}
\begin{definition}[Admissible decision]
A decision $w_t$ (portfolio weights, position sizes, or risk controls) is admissible if $w_t \in \calW_t$, where $\calW_t$ encodes budget, leverage, turnover, concentration, and capacity constraints.
\end{definition}

The set $\calW_t$ formalises the fact that portfolio choice is a constrained decision problem rather than an unconstrained mapping from forecasts to weights. In institutional settings, feasibility restrictions are not optional modelling choices; they are the mechanism through which mandates, risk limits, liquidity, and operational capacity enter the decision rule. Turnover limits are imposed to control trading intensity and the associated transaction-cost drag; leverage limits are imposed to control balance-sheet and margin risk; and capacity restrictions reflect the inability to scale positions without inducing economically material market impact or dominating available liquidity \citep{LedoitWolf2025,LewinCampani2023}. These constraints are also central to the paper’s empirical interpretation: when $\calW_t$ binds, differences in forecast shape and calibration affect outcomes primarily through how they alter constraint activity (e.g., triggering or avoiding turnover and leverage limits), rather than through point-forecast improvements.

In the empirical implementation, $\calW_t$ is specified in a manner consistent with practice and with the chosen instrument class. A canonical example is
\begin{equation}
\calW_t
=
\left\{
w \in \R^N :
\mathbf{1}^\top w = 1,\;
w_{\min} \le w \le w_{\max},\;
\norm{w - w_{t-1}}_1 \le \tau,\;
\norm{w}_1 \le L,\;
w \in \mathcal{C}_t
\right\},
\end{equation}
where $\tau$ is a turnover budget, $L$ is a leverage bound (or gross exposure limit), and $\mathcal{C}_t$ collects additional capacity-type restrictions (such as participation-rate or liquidity-linked bounds). This structure also accommodates discrete or combinatorial constraints when required by implementation (e.g., cardinality limits arising from operational overhead), in which case admissibility explicitly encodes the practical requirement that the portfolio be implementable in a limited number of names \citep{AnisKwon2025}. The purpose of making $\calW_t$ explicit is not merely technical: it ensures that forecasting methods are evaluated on the economically relevant object—friction-adjusted decision loss—within the same feasible decision domain.

\subsection{Friction operator: costs and constraints}\label{sec:friction_operator}

\begin{definition}[Friction operator]\label{def:friction_operator}
Friction is the combined penalty imposed by (i) explicit trading costs, (ii) bid--ask spread costs, (iii) market impact, and (iv) binding feasibility constraints.
Formally, friction is represented by the pair $(C_t(\Delta w_t), \calW_t)$, where $C_t$ is a cost functional applied to position changes $\Delta w_t := w_t - w_{t-1}$ and $\calW_t$ is the feasible decision set.
\end{definition}

The friction operator is the economic object that links an econometric forecast to realised performance: two models that are indistinguishable under conventional forecast scoring can yield materially different realised outcomes once the same decision rule is executed through $(C_t,\calW_t)$. The paper therefore treats friction as part of the estimand: we evaluate predictive distributions through their induced decisions \emph{net of} $C_t$ and subject to $\calW_t$.

\paragraph{Baseline implementable cost functional.}
We adopt a baseline cost functional that is empirically implementable and separable into direct and indirect components:
\begin{equation}\label{eq:cost_functional}
C_t(\Delta w_t)
=
c^{\text{fee}} \cdot \norm{\Delta w_t}_1
+
c^{\text{spread}}_t \cdot \norm{\Delta w_t}_1
+
c^{\text{impact}}_t(\Delta w_t).
\end{equation}
The $\ell_1$ turnover norm $\norm{\Delta w_t}_1$ is a practical proxy for traded notional/quantity in periodic rebalancing, and it aligns with standard implementations of proportional costs in portfolio construction. Recent evidence on the materiality of transaction costs in portfolio selection motivates including costs \emph{inside} the optimisation problem rather than ``paying them after the fact.'' \citep{LedoitWolf2025}

\paragraph{Direct fees and commissions ($c^{\text{fee}}$).}
The term $c^{\text{fee}}$ captures per-unit explicit costs that are (approximately) known at time $t$ and do not depend on trade direction beyond absolute turnover (e.g., broker commissions, venue fees, taxes where applicable). Empirically, $c^{\text{fee}}$ is obtained from publicly stated fee schedules, broker reports, or exchange/venue tariff data matched to the trading universe.

\paragraph{Bid--ask spread component ($c^{\text{spread}}_t$).}
The term $c^{\text{spread}}_t \cdot \norm{\Delta w_t}_1$ captures the expected crossing cost of liquidity provision. When trade-level microstructure data are available, $c^{\text{spread}}_t$ can be proxied by the effective spread (or half-spread) computed from quotes and executions. When only daily OHLC data are available (typical in large-sample studies), we proxy $c^{\text{spread}}_t$ using high-frequency robust estimators of effective spreads derived from open, high, low, and close prices, designed to reduce bias under infrequent trading. \citep{ArdiaGuidottiKroencke2024_OHLCSpread}

\paragraph{Market impact component ($c^{\text{impact}}_t(\Delta w_t)$).}
The term $c^{\text{impact}}_t(\Delta w_t)$ captures adverse price movement caused by the trade itself. The baseline requirement is not a perfect structural model of impact, but a disciplined, time-$t$ measurable proxy that can be stress-tested and that respects obvious scale effects. We use two complementary impact proxies, selected by data availability and the execution horizon:

\emph{(i) Long-horizon (metaorder-level) proxy.} For strategies whose rebalancing induces autocorrelated order flow across periods, we treat impact as having a dynamic component and proxy it using measures designed to quantify \emph{long-term} impact rather than only within-trade slippage. \citep{HarveyLedfordSciulliUstinovZohren2022_LongTermImpact}

\emph{(ii) Estimation-efficiency proxy from price trajectories.} Where metaorder execution trajectories (or partial trajectories) are observable or can be reconstructed, we calibrate impact parameters with estimators that exploit price-path information and improve statistical efficiency relative to VWAP-only approaches. This provides an empirically testable route to mapping $\Delta w_t$ into an impact penalty that is both measurable and robust. \citep{LiIhnatiukChenLinKinnearSchneiderNevmyvakaLam2024_TrajectoryImpact}

In both cases, the impact term is operationally represented as a convex penalty in trade size (or participation) for tractability in optimisation, with stress tests exploring alternative curvature (e.g., linear vs. concave vs. square-root-like penalties) as part of model-risk analysis. The empirical section reports sensitivity of the paper's key conclusions to these impact specifications.

\paragraph{Feasible set and binding constraints ($\calW_t$).}
Constraints enter friction as \emph{hard} feasibility restrictions: if the decision is infeasible, friction is effectively infinite. In the empirical implementation, $\calW_t$ encodes the constraint set used by practitioners and required for credible out-of-sample evaluation:

\[
\mathcal{W}_t = \left\{ w : 
\begin{aligned}
    & \mathbf{1}^\top w = 1, \\
    & w \in \text{(long-only or bounded long/short)}, \\
    & \|w\|_{\infty} \le \bar{w}, \quad \|\Delta w\|_{1} \le \overline{\Delta}, \\
    & \text{(liquidity / capacity constraints)}
\end{aligned}
\right\}.
\]

Liquidity/capacity constraints are implemented via participation-rate bounds and/or maximum tradable notional tied to volume and spread proxies. Capacity limits are not treated as a ``story'' but as an economically measurable restriction on scalable implementation; empirical evidence on scale diseconomies and capacity constraints motivates including such bounds when translating forecasts into trades. \citep{ForsbergGallagherWarren2022_CapacityHedgeFunds}

\paragraph{Data requirements for estimating/proxying friction.}
The baseline operator in \eqref{eq:cost_functional} is implementable with standard datasets:
(i) returns and portfolio holdings to compute $\Delta w_t$;
(ii) explicit fee schedules for $c^{\text{fee}}$;
(iii) either (a) quotes/trades for effective spread, or (b) OHLC data and a robust OHLC-based spread estimator for $c^{\text{spread}}_t$; \citep{ArdiaGuidottiKroencke2024_OHLCSpread}
(iv) volume (and, where available, intraday execution data) to calibrate impact proxies and participation constraints; \citep{LiIhnatiukChenLinKinnearSchneiderNevmyvakaLam2024_TrajectoryImpact,HarveyLedfordSciulliUstinovZohren2022_LongTermImpact}
(v) a documented constraint specification (turnover, leverage, concentration, and liquidity/capacity rules) to define $\calW_t$.

Finally, the empirical design treats transaction costs as stochastic and reports uncertainty in realised costs, not only in returns. This matters because ignoring cost uncertainty can distort risk-adjusted evaluation and lead to systematically overconfident claims about decision quality. \citep{BasicUtreraNolteNolte2025_ISVariance}

\subsection{Decision objective and decision loss}\label{sec:decision_objective_loss}

\begin{definition}[Decision objective]\label{def:decision_objective}
Let $U(\cdot)$ denote an economic objective (expected utility) or a risk-adjusted objective (e.g., mean--variance, CVaR, drawdown-penalised utility) evaluated on realised outcomes net of friction.
\end{definition}

\begin{definition}[Decision loss]\label{def:decision_loss}
Given $Q_t$, define the induced decision rule
\begin{equation}\label{eq:induced_decision_rule}
w_t(Q_t) \in \arg\max_{w\in \calW_t} \ \mathcal{J}(w;Q_t) - C_t(w-w_{t-1}),
\end{equation}
and define decision loss as the negative realised objective (or a normalised regret form) under the resulting trade path.
\end{definition}

The decision objective $U(\cdot)$ specifies what the decision-maker is attempting to achieve once trades are implemented. In the empirical sections, $U(\cdot)$ is instantiated in forms that are standard in portfolio construction and risk management (mean--variance and tail-risk objectives such as CVaR), with all performance measured \emph{net of} the friction operator. This matters because the objective is not evaluated on hypothetical frictionless returns; it is evaluated on executed outcomes after costs and constraints have acted. Recent evidence shows that explicitly accounting for transaction costs \emph{at the portfolio-selection stage} (rather than paying costs after the fact) can materially change out-of-sample performance, especially for strategies that would otherwise trade frequently \citep{LedoitWolf2025}. Similarly, recent portfolio models incorporating CVaR alongside transaction costs underscore that tail-risk objectives and friction terms interact in economically relevant ways and should be treated as a joint optimisation problem rather than separate diagnostics \citep{WangZhuTang2024_NAJEF_CVaRTC}.

The functional $\mathcal{J}(w;Q_t)$ in \eqref{eq:induced_decision_rule} is the \emph{forecast-implied} objective at time $t$, computed under the predictive distribution $Q_t$ (for example, an expectation of utility, a risk-adjusted return criterion, or a mean--risk trade-off). The mapping $Q_t \mapsto w_t(Q_t)$ formalises how a distributional forecast is converted into an admissible action when trading frictions and feasibility constraints are present. This induced-decision perspective is central to the paper: the econometric object $Q_t$ is evaluated by the realised consequences of the decision rule it induces, not by forecast metrics alone. Recent decision-based portfolio frameworks in operations research and financial optimisation similarly treat the end-to-end mapping from forecasts to constrained, implementable portfolios as the primary object of evaluation, rather than isolating prediction from optimisation \citep{AnisKwon2025}.

Decision loss is then defined as the realised shortfall relative to the objective, once implementation has occurred. Concretely, let $\widetilde{R}_{t+h}(w_t,w_{t-1})$ denote the realised friction-adjusted return (or payoff) over $[t,t+h]$ produced by executing the decision $w_t$ from the prior state $w_{t-1}$. The realised objective is $U(\widetilde{R}_{t+h})$ and the realised decision loss is
\begin{equation}\label{eq:realised_decision_loss}
\ell_{t+h}(Q_t)
:= -\,U\!\left(\widetilde{R}_{t+h}(w_t(Q_t),w_{t-1})\right),
\end{equation}
or, when comparing against an admissible benchmark decision rule $w_t^{\star}$, a normalised regret form
\begin{equation}\label{eq:regret_form}
\mathrm{Regret}_{t+h}(Q_t)
:=
U\!\left(\widetilde{R}_{t+h}(w_t^{\star},w_{t-1})\right)
-
U\!\left(\widetilde{R}_{t+h}(w_t(Q_t),w_{t-1})\right).
\end{equation}
This definition makes explicit why calibration is economically decisive: miscalibration affects $Q_t$, which affects $w_t(Q_t)$, which affects both turnover (and hence costs) and risk exposure (especially in tail states). The paper also treats trading costs as stochastic and reports uncertainty in friction-adjusted loss, reflecting recent work showing that implementation shortfall variability can be a material risk component in portfolio construction \citep{BasicUtreraNolteNolte2025_ISVariance}.

\section{Why accuracy is not the target: calibration and decision performance}
\subsection{Forecast metrics versus economic loss}\label{sec:metrics_vs_loss}

Forecast evaluation in finance commonly begins with statistical scoring rules and point-forecast errors. For point forecasts $\widehat{y}_{t+h}$ of an outcome $Y_{t+h}$, the canonical measures are mean squared error (MSE) and mean absolute error (MAE),
\[
\mathrm{MSE} := \E\!\left[(Y_{t+h}-\widehat{y}_{t+h})^2\right],
\qquad
\mathrm{MAE} := \E\!\left[\lvert Y_{t+h}-\widehat{y}_{t+h}\rvert\right],
\]
which implicitly treat over- and under-prediction symmetrically and evaluate performance only through the deviation of a single summary statistic from the realised outcome. When forecasts are distributional, the corresponding evaluation objects are proper scoring rules. The logarithmic score evaluates the predictive density $q_t(\cdot)$ at the realised outcome,
\[
S_{\log}(Q_t,Y_{t+h}) := -\log q_t(Y_{t+h}),
\]
and the continuous ranked probability score (CRPS) evaluates the predictive distribution function $F_t(\cdot)$ against the realised outcome via
\[
\mathrm{CRPS}(F_t,Y_{t+h})
:= \int_{-\infty}^{\infty}\bigl(F_t(z)-\ind{Y_{t+h}\le z}\bigr)^2\,dz,
\]
which reduces to MAE under deterministic forecasts and has become a standard tool for assessing distributional forecasts in economics and finance \citep{GneitingRaftery2007jrssb}. These metrics are attractive because they are well-defined, comparable across models, and (for proper scores) aligned with truthful probabilistic reporting.

However, these forecast metrics do not, in general, coincide with the economic loss relevant for investment and risk decisions. The economic objective is typically asymmetric in outcomes and depends on the decision induced by the forecast, not on the forecast itself. A risk-averse objective penalises downside outcomes more heavily than upside outcomes, and tail-risk objectives (such as CVaR or drawdown penalties) concentrate weight in rare but economically decisive states. In addition, trading frictions introduce a further asymmetry: forecast-driven changes in positions incur costs that are convex in turnover and often state-dependent (spreads widen and impact intensifies precisely when risk is most salient). As a result, a model can improve MSE or log score by becoming more responsive or more confident, while simultaneously worsening realised performance by inducing excessive turnover, amplifying exposure in tail states, or repeatedly activating binding constraints. The misalignment is structural: statistical forecast quality is assessed under a loss function that ignores the optimisation map from forecasts to decisions and ignores the friction operator through which decisions are implemented.

This paper therefore treats the relevant target as \emph{decision loss} net of frictions: forecasts are evaluated by the economic consequences of the admissible decision rule they induce once executed under transaction costs, market impact, and capacity constraints. The purpose of introducing MSE/MAE, the log score, and CRPS is not to reject them, but to clarify the sense in which they answer a different question: they measure statistical proximity between forecasts and outcomes, whereas the econometric object of interest here is the expected economic performance of forecast-induced decisions under asymmetric objectives and implementation frictions.

\subsection{Calibration concepts used in the paper}\label{sec:calibration_concepts}

\begin{definition}[Calibration]\label{def:calibration}
The paper uses three complementary notions of calibration for real-valued outcomes.

\emph{(i) Probability (threshold) calibration.} For any threshold $z\in\mathbb{R}$, define the forecast-implied exceedance probability
$p_t(z):=1-F_t(z)$, where $F_t$ is the CDF associated with $Q_t$. The forecast is (marginally) probability-calibrated at threshold $z$ if
\[
\Prob\!\bigl(Y_{t+h} > z \,\big|\, p_t(z)=p\bigr)=p
\quad\text{for all }p\in[0,1],
\]
and conditionally calibrated when the same equality holds conditional on suitable $\calI_t$-measurable sub-$\sigma$-fields or features; empirically this is assessed via (conditional) reliability diagrams for exceedance events. \citep{GneitingResin2023}

\emph{(ii) Distributional (probabilistic) calibration.} Let the probability integral transform (PIT) be
\[
U_t := F_t\!\bigl(Y_{t+h}\bigr).
\]
Under ideal distributional calibration (and continuity), $\{U_t\}$ is Uniform$(0,1)$; in practice we diagnose deviations using PIT histograms and related uniformity diagnostics, interpreted as systematic under-/over-dispersion or bias in the predictive distribution. \citep{GneitingRaftery2007jrssb}

\emph{(iii) Quantile calibration.} For a nominal level $\alpha\in(0,1)$, let $q_t(\alpha):=F_t^{-1}(\alpha)$ be the forecast $\alpha$-quantile. The forecast is quantile-calibrated at level $\alpha$ if
\[
\Prob\!\bigl(Y_{t+h}\le q_t(\alpha)\bigr)=\alpha,
\]
and conditionally quantile-calibrated if the same relation holds given relevant conditioning information; empirically this is diagnosed via hit sequences $\ind\{Y_{t+h}\le q_t(\alpha)\}$ and their reliability diagrams / conditional checks. \citep{GneitingResin2023,AllenKohSegersZiegel2025}
\end{definition}

The empirical diagnostics used in the paper are therefore measurable summaries of these notions: (a) PIT histograms and PIT-based uniformity checks for distributional calibration; (b) reliability diagrams for exceedance events (probability calibration) and their conditional analogues; and (c) quantile hit-rate diagnostics for selected tail-relevant $\alpha$ (with the tail-focused variant used when economic loss concentrates on extremes). \citep{GneitingResin2023,AllenKohSegersZiegel2025}

\subsection{Utility-weighted calibration}\label{sec:utility_weighted_calibration}

\begin{definition}[Utility-weighted calibration criterion]\label{def:utility_weighted_calibration}
Let $Q_t$ denote a predictive distribution for $Y_{t+h}$ with CDF $F_t$, and let $w_t(Q_t)$ be the induced admissible decision defined in \eqref{eq:induced_decision_rule}.  
The forecast $Q_t$ is said to satisfy \emph{utility-weighted calibration} if calibration errors are weighted by their marginal impact on the downstream decision objective and by the friction-adjusted relevance of the corresponding states. Formally, for a calibration diagnostic indexed by $z$ (thresholds, quantile levels, or PIT values), define a weight
\[
\omega_t(z)
\;\propto\;
\left\lvert 
\frac{\partial \mathcal{J}(w;Q_t)}{\partial F_t(z)}
\Bigg|_{w = w_t(Q_t)}
\right\rvert
\times
\kappa_t(z),
\]
where the first term captures marginal decision sensitivity (the local utility or risk gradient with respect to the predictive distribution), and $\kappa_t(z)$ is a nonnegative friction-adjustment factor that upweights states associated with high expected trading costs, binding constraints, or tail-risk relevance. Utility-weighted calibration holds when the weighted calibration error integrates to zero over the relevant diagnostic domain.
\end{definition}

\subsection{Deriving the utility-weight \texorpdfstring{$\omega$}{omega} in a canonical quadratic case}\label{sec:omega_derivation_quadratic}

This subsection removes the ``black box'' vulnerability in Definition~8 by deriving an explicit functional form for the utility-weight $\omega_t(\cdot)$ in a canonical convex portfolio problem. The derivation shows that $\omega$ is not an ad-hoc weighting device: in the quadratic case it is implied by the curvature of the friction-adjusted objective and the Jacobian mapping from distributional perturbations to changes in the optimiser.

\paragraph{Canonical decision problem (mean--variance with quadratic transaction costs).}
Let $Y_{t+h}\in\R^N$ denote the vector of returns over horizon $h$. A predictive distribution $Q_t$ implies
\[
\mu_t := \E_{Q_t}[Y_{t+h}\mid\calI_t],\qquad
\Sigma_t := \Var_{Q_t}(Y_{t+h}\mid\calI_t].
\]
Consider the unconstrained (or interior) quadratic programme
\begin{equation}\label{eq:canonical_qp}
w_t(Q_t)\in\arg\max_{w\in\R^N}\;
\mu_t^\top w-\frac{\gamma}{2}w^\top\Sigma_t w-\frac{\eta}{2}\|w-w_{t-1}\|_2^2,
\end{equation}
with $\gamma>0$ risk aversion and $\eta>0$ a quadratic friction parameter. This is the simplest setting in which (i) the optimiser is explicit and (ii) miscalibration affects the decision through forecast-implied moments.

The first-order condition for an interior optimum is
\begin{equation}\label{eq:foc_qp}
\mu_t-\gamma\Sigma_t w_t-\eta(w_t-w_{t-1})=0,
\end{equation}
so that
\begin{equation}\label{eq:closed_form_w}
w_t(Q_t)
=
(\gamma\Sigma_t+\eta I)^{-1}(\mu_t+\eta w_{t-1}).
\end{equation}
The curvature (negative Hessian of the objective) is $\gamma\Sigma_t+\eta I\succ0$, which ensures stability and yields an explicit sensitivity map.

\paragraph{Perturbations induced by miscalibration.}
Let $\widetilde Q_t$ be an alternative (possibly miscalibrated) predictive distribution and denote the induced moment perturbations
\[
\delta\mu_t:=\mu_t(\widetilde Q_t)-\mu_t(Q_t),\qquad
\delta\Sigma_t:=\Sigma_t(\widetilde Q_t)-\Sigma_t(Q_t).
\]
A first-order expansion of \eqref{eq:closed_form_w} gives the decision perturbation
\begin{equation}\label{eq:dw_general}
\delta w_t
:=
w_t(\widetilde Q_t)-w_t(Q_t)
\approx
(\gamma\Sigma_t+\eta I)^{-1}\delta\mu_t
-\gamma(\gamma\Sigma_t+\eta I)^{-1}\delta\Sigma_t\,w_t(Q_t),
\end{equation}
where higher-order terms are omitted. This explicitly shows how distributional errors propagate into the induced decision through the inverse curvature operator $(\gamma\Sigma_t+\eta I)^{-1}$.

\paragraph{Decision-loss impact and the implied weight.}
Let the (conditional) friction-adjusted objective value be
\[
V_t(Q):=\max_w\left\{\mu_t(Q)^\top w-\frac{\gamma}{2}w^\top\Sigma_t(Q)w-\frac{\eta}{2}\|w-w_{t-1}\|_2^2\right\}.
\]
In the quadratic case, the value function is differentiable in the moment arguments, and the envelope theorem yields
\begin{equation}\label{eq:envelope}
\frac{\partial V_t}{\partial \mu_t}=w_t(Q_t),
\qquad
\frac{\partial V_t}{\partial \Sigma_t}
=
-\frac{\gamma}{2}w_t(Q_t)w_t(Q_t)^\top.
\end{equation}
Hence, to first order, moment miscalibration changes conditional expected objective by
\begin{equation}\label{eq:value_delta}
\delta V_t
\approx
w_t(Q_t)^\top\delta\mu_t
-\frac{\gamma}{2}\langle w_t(Q_t)w_t(Q_t)^\top,\delta\Sigma_t\rangle,
\end{equation}
where $\langle A,B\rangle:=\mathrm{tr}(A^\top B)$.

This identifies the \emph{economic marginal value} of correcting specific components of distributional error: errors in mean matter in the direction of current exposure $w_t$, and errors in risk (covariance) matter in the quadratic form induced by $w_t w_t^\top$. Importantly, this is already \emph{state dependent} through $w_t(Q_t)$, which itself depends on $(\mu_t,\Sigma_t,w_{t-1})$ and on the friction parameter $\eta$.

\begin{figure}[ht]
    \centering
    \includegraphics[width=0.7\textwidth]{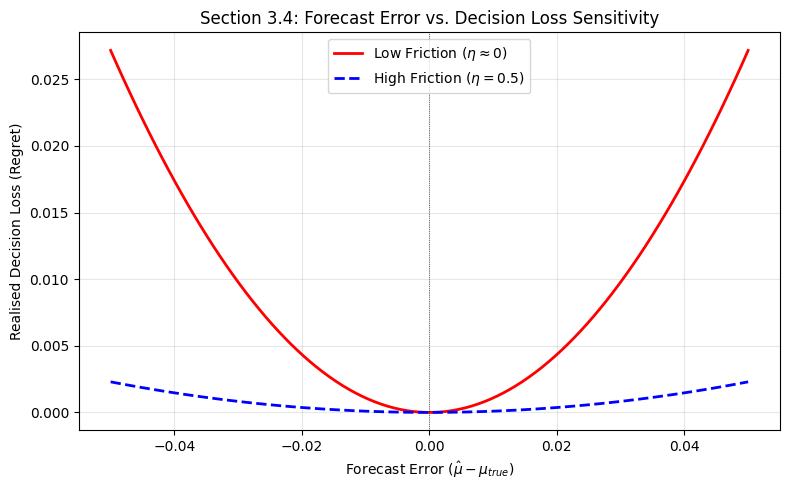}
    \caption{\textbf{Forecast Error Sensitivity under Frictions.} The loss function (Eq. 11) flattens as frictions ($\eta$) increase. In high-friction states (blue dashed), the decision is less sensitive to forecast precision, implying a lower utility-weight $\omega_t$ for small errors.}
    \label{fig:theory_sensitivity}
\end{figure}

\paragraph{From analytic sensitivity to an explicit $\omega_t(u)$.}
Definition~8 indexes calibration diagnostics by $u$ (thresholds, quantile levels, PIT bins, or other distributional coordinates). To connect \eqref{eq:dw_general}--\eqref{eq:value_delta} to $\omega_t(u)$, represent the relevant distributional perturbation by a finite-dimensional set of calibration moments
\[
m_u(Q_t,Y_{t+h})\quad (u\in\mathcal{U}),
\qquad
\E[m_u(Q_t,Y_{t+h})\mid\calI_t]=0\ \text{under calibration},
\]
and suppose the moment errors induce moment perturbations through linear functionals
\begin{equation}\label{eq:linear_map_moments}
\delta\mu_t
\approx
\sum_{u\in\mathcal{U}} a_t(u)\, \E[m_u(\widetilde Q_t,Y_{t+h})\mid\calI_t],
\qquad
\delta\Sigma_t
\approx
\sum_{u\in\mathcal{U}} B_t(u)\, \E[m_u(\widetilde Q_t,Y_{t+h})\mid\calI_t],
\end{equation}
for $\calI_t$-measurable coefficients $a_t(u)\in\R^N$ and $B_t(u)\in\R^{N\times N}$ determined by the chosen diagnostic family (for example, tail-quantile miscoverage moments map into tail-mean and tail-variance errors via standard influence-function arguments in the parametric case, and via linearisation in the nonparametric case).

Substituting \eqref{eq:linear_map_moments} into \eqref{eq:value_delta} yields
\begin{equation}\label{eq:value_delta_weighted}
\delta V_t
\approx
\sum_{u\in\mathcal{U}}
\left(
w_t(Q_t)^\top a_t(u)
-\frac{\gamma}{2}\langle w_t(Q_t)w_t(Q_t)^\top, B_t(u)\rangle
\right)
\E[m_u(\widetilde Q_t,Y_{t+h})\mid\calI_t].
\end{equation}
Equation \eqref{eq:value_delta_weighted} shows that each diagnostic index $u$ has a \emph{canonical marginal effect} on the friction-adjusted objective through a scalar coefficient. This coefficient is the analytically derived weight that Definition~8 gestures at conceptually.

\begin{proposition}[Canonical quadratic-case utility-weight]\label{prop:omega_quadratic}
In the canonical quadratic problem \eqref{eq:canonical_qp}, suppose distributional miscalibration is summarised by a finite family of calibration moments $\{m_u\}_{u\in\mathcal{U}}$ admitting the linearisation \eqref{eq:linear_map_moments}. Define
\begin{equation}\label{eq:omega_quadratic}
\omega_t(u)
:=
\left|
w_t(Q_t)^\top a_t(u)
-\frac{\gamma}{2}\langle w_t(Q_t)w_t(Q_t)^\top, B_t(u)\rangle
\right|
\times
\kappa_t(u),
\end{equation}
where $\kappa_t(u)\ge0$ is the friction-regime multiplier (high-spread/high-impact/high-risk regime upweighting) defined ex ante from $\calI_t$.
Then $\omega_t(u)$ is the first-order (absolute) marginal impact of the $u$-indexed calibration moment error on the friction-adjusted conditional objective, and weighting calibration errors by $\omega_t(u)$ targets the components of miscalibration that have the largest first-order contribution to decision loss in this canonical case.
\end{proposition}

\paragraph{Interpretation and how this justifies proxies in Section~5.}
Proposition~\ref{prop:omega_quadratic} resolves the central critique: in a fully specified canonical case, $\omega_t(u)$ is implied by the value-function derivative and the linearisation mapping from the diagnostic moments to moment perturbations. The dependence on the Hessian enters through the optimiser $w_t(Q_t)$ (which depends on $(\gamma\Sigma_t+\eta I)^{-1}$ via \eqref{eq:closed_form_w}) and through the quadratic exposure term $w_t w_t^\top$ in \eqref{eq:envelope}. In more general constrained cases, the same logic applies with the KKT system replacing \eqref{eq:foc_qp}, and the state dependence becomes sharper when constraints bind. Section~5 then implements observable proxies for the components of \eqref{eq:omega_quadratic}: exposures $w_t$, risk penalties (through estimated $\Sigma_t$), and friction regime multipliers $\kappa_t$ (through spreads/volatility/impact proxies). The proxies are therefore presented as approximations to an analytically derived weighting, not as arbitrary heuristics.

\paragraph{Standard calibration} notions treat forecast errors symmetrically across the support of the predictive distribution: a deviation in a central state is penalised in the same way as an equally sized deviation in a tail state, and errors are assessed independently of how forecasts are subsequently used. In contrast, the utility-weighted calibration criterion defined above explicitly conditions on the decision problem. The weighting by the marginal sensitivity of the objective reflects the fact that small distortions in $Q_t$ can have very different economic consequences depending on whether they occur in regions where the decision rule is locally flat or locally steep. For example, miscalibration in regions that determine leverage, turnover, or tail exposure can dominate miscalibration in regions that have negligible influence on the optimal decision.

The friction-adjustment factor $\kappa_t(z)$ further distinguishes this criterion from generic calibration. Trading frictions and constraints are state dependent: spreads widen, impact increases, and feasibility constraints bind most strongly precisely in volatile or illiquid regimes and in tail states. By upweighting calibration errors in these regions, the criterion aligns statistical evaluation with the realised cost structure faced by the decision-maker. This builds on recent work showing that weighted scoring rules and tail-focused calibration diagnostics are necessary when loss concentrates asymmetrically in particular regions of the outcome space, as in risk management and financial decision-making \citep{AllenKohSegersZiegel2025,Cheng2024}.

Alignment with decision loss follows directly from the construction. Decision loss is driven by how $Q_t$ shapes the induced action $w_t(Q_t)$ under constraints and costs; utility-weighted calibration penalises forecast distortions in proportion to their contribution to that loss. As a result, a forecast that is generically well calibrated but miscalibrated in economically decisive regions may score well under standard diagnostics yet perform poorly in realised, friction-adjusted outcomes. Conversely, a forecast that is less sharp or less accurate in a global sense can dominate in decision loss if it is well calibrated where utility gradients are large and frictions are punitive. The criterion therefore provides the missing link between probabilistic calibration and economic performance, making calibration an estimand aligned with the objective actually being optimised.
\section{Theory: dominance results under trading frictions}

This section formalises what the empirical results show: once trading frictions and implementable constraints are imposed, economically relevant forecast quality is not captured by point accuracy alone. What matters is whether the predictive distribution is reliable exactly in the regions that determine (i) the direction and size of induced trades, (ii) the activation of constraints, and (iii) realised friction-adjusted loss. The theory is stated in a way that matches the empirical design: a walk-forward sequence of predictive distributions $Q_t$, a friction operator $C_t$, a feasible correspondence $\calW_t$, and an implemented decision rule that maps $Q_t$ into a trade.

\subsection{Regularity conditions}\label{sec:regularity_conditions}

\begin{assumption}[Feasible set and cost regularity]\label{ass:feasible_cost_regularity}
For each decision time $t$, the feasible set $\calW_t$ and the cost functional $C_t$ satisfy:

\paragraph{(i) Measurability of constraints.}
The correspondence $\omega \mapsto \calW_t(\omega)$ is $\calI_t$-measurable in the sense that its graph
\[
\mathrm{Gr}(\calW_t)
:=
\{(\omega,w)\in\Omega\times\R^N:\ w\in\calW_t(\omega)\}
\]
is $\calI_t\otimes\mathcal{B}(\R^N)$-measurable. Equivalently, for any open set $O\subset\R^N$,
$\{\omega:\calW_t(\omega)\cap O\neq\emptyset\}\in\calI_t$.

\paragraph{(ii) Non-emptiness, convexity, and compactness.}
Almost surely, $\calW_t(\omega)$ is non-empty, convex, and compact in $\R^N$.

\paragraph{(iii) Uniform boundedness of feasible decisions.}
Almost surely, $\calW_t(\omega)$ is contained in a deterministic compact set $\calK\subset\R^N$
(i.e., $\calW_t(\omega)\subseteq\calK$ a.s.). This encodes implementability in the empirical setting:
position sizes and portfolio weights are uniformly bounded.

\paragraph{(iv) Cost measurability and normalisation.}
For each $t$, $C_t(\cdot)$ is $\calI_t$-measurable as a function of the state, and satisfies
$C_t(0)=0$ and $C_t(\Delta w)\ge 0$ for all $\Delta w\in\R^N$.

\paragraph{(v) Convexity, lower semicontinuity, and finiteness on feasible changes.}
Almost surely, the mapping $\Delta w\mapsto C_t(\Delta w)$ is proper, convex, and lower semicontinuous on $\R^N$.
Moreover, $C_t$ is continuous on $\calK-\calK$ and finite on $\calK-\calK$, which is the relevant domain in the data
because only feasible position changes are ever executed.

\paragraph{(vi) Coercivity on feasible changes.}
There exists a constant $\lambda>0$ and an $\calI_t$-measurable random variable $b_t\ge 0$ such that, for all
$\Delta w\in\calK-\calK$,
\[
C_t(\Delta w) \ge \lambda \norm{\Delta w}_1 - b_t.
\]
This ensures that large turnover is penalised strongly enough for the friction-adjusted optimisation problem to be well-posed,
and it matches the empirical fact that excessive trading intensity produces disproportionately poor realised net returns.
\end{assumption}

\subsection{Link between calibration error and decision error}\label{sec:calib_to_decision_error}

The empirical evidence is built from differences in realised, friction-adjusted loss generated by decisions that are themselves induced by predictive distributions. The key theoretical step is therefore to control how distributional errors translate into decision errors when the decision is obtained as the solution to a constrained optimisation problem with convex frictions.

\begin{lemma}[Sensitivity of the induced decision to miscalibration]\label{lem:calibration_decision_sensitivity}
Fix $t$ and suppose Assumption~\ref{ass:feasible_cost_regularity} holds. Let $Q_t$ and $\widetilde{Q}_t$ be two predictive distributions for $Y_{t+h}$ given $\calI_t$, with corresponding induced decisions
\[
w_t := w_t(Q_t) \in \arg\max_{w\in\calW_t}\ \mathcal{J}(w;Q_t)-C_t(w-w_{t-1}),
\qquad
\widetilde{w}_t := w_t(\widetilde{Q}_t) \in \arg\max_{w\in\calW_t}\ \mathcal{J}(w;\widetilde{Q}_t)-C_t(w-w_{t-1}).
\]
Assume further that, for each $w\in\calW_t$, $\mathcal{J}(w;Q)$ is Fr\'echet differentiable in $w$ and satisfies:

\begin{assumption}[Strong concavity and Lipschitz dependence]\label{ass:strong_concavity_lipschitz}
There exists $\mu>0$ such that for all admissible $Q$ and all $w,w'\in\calW_t$,
$\mathcal{J}(\cdot;Q)-C_t(\cdot-w_{t-1})$ is $\mu$-strongly concave on $\calW_t$.
Moreover, there exists $L_t<\infty$ such that for all $w\in\calW_t$,
\[
\bigl\|\nabla_w \mathcal{J}(w;Q_t)-\nabla_w \mathcal{J}(w;\widetilde{Q}_t)\bigr\|
\ \le\
L_t \, d_t(Q_t,\widetilde{Q}_t),
\]
where $d_t(\cdot,\cdot)$ is a decision-relevant discrepancy that measures miscalibration on the diagnostic domain that actually affects the decision objective
(for example, a utility/friction-weighted integral of exceedance or quantile calibration errors as in Definition~\ref{def:utility_weighted_calibration}).
\end{assumption}

Then the induced decisions satisfy
\begin{equation}\label{eq:decision_stability_bound}
\|w_t-\widetilde{w}_t\|
\ \le\
\frac{1}{\mu}\, \sup_{w\in\calW_t}\bigl\|\nabla_w \mathcal{J}(w;Q_t)-\nabla_w \mathcal{J}(w;\widetilde{Q}_t)\bigr\|
\ \le\
\frac{L_t}{\mu}\, d_t(Q_t,\widetilde{Q}_t).
\end{equation}
If, in addition, the realised objective $U(\cdot)$ is $K$-Lipschitz in the friction-adjusted realised return with respect to the decision (under the execution model used in the empirical protocol), then the conditional expected loss differential satisfies
\begin{equation}\label{eq:loss_increase_bound}
\E\!\left[\ell_{t+h}(\widetilde{Q}_t)-\ell_{t+h}(Q_t)\,\big|\,\calI_t\right]
\ \le\
K\,\E\!\left[\|\widetilde{w}_t-w_t\|\,\big|\,\calI_t\right]
\ \le\
\frac{K\,L_t}{\mu}\, d_t(Q_t,\widetilde{Q}_t).
\end{equation}
\end{lemma}

\noindent\emph{Interpretation.}
The empirical channel is explicit. A distributional distortion that matters for the objective changes the gradient of the forecast-implied criterion; the optimiser responds by changing the implemented trade; the execution model converts that trade into a friction-adjusted realised return; and the realised utility converts this into loss. Strong concavity gives a linear stability control: the induced decision error is proportional to decision-relevant miscalibration, with proportionality governed by curvature ($\mu$) and by how strongly the objective reacts to distributional errors ($L_t$). The friction-adjusted realised-loss effect then inherits this bound through $K$. The utility-weighted calibration criterion is constructed so that $d_t$ places its mass precisely where the data show the losses are made: tail states, high-friction regimes, and constraint-binding regimes.

\subsection{Main dominance theorem}\label{sec:main_dominance_theorem}

The empirical claim is a dominance claim: after recalibration that targets decision-relevant reliability, realised friction-adjusted loss is weakly lower, even if point accuracy does not improve. The theorem states the corresponding population dominance result under a calibrated-projection property aligned with the implemented class of recalibration maps used in estimation.

\begin{theorem}[Calibration-aligned dominance]\label{thm:calibration_aligned_dominance}
Fix $t$ and suppose Assumption~\ref{ass:feasible_cost_regularity} holds. Let $\mathcal{D}$ be an admissible class of predictive distributions for $Y_{t+h}$ given $\calI_t$. For each $Q\in\mathcal{D}$, let $w_t(Q)$ denote the induced decision defined by \eqref{eq:induced_decision_rule}, and define conditional expected decision loss
\[
\mathcal{L}_t(Q)
:=
\E\!\left[\ell_{t+h}(Q)\,\big|\,\calI_t\right],
\qquad
\ell_{t+h}(Q):=-U\!\left(\widetilde{R}_{t+h}(w_t(Q),w_{t-1})\right).
\]
Assume:

\begin{assumption}[Well-posedness and curvature]\label{ass:well_posed_curvature}
For each $Q\in\mathcal{D}$, the objective $w\mapsto \mathcal{J}(w;Q)-C_t(w-w_{t-1})$ admits a measurable maximiser on $\calW_t$.
Moreover, the objective is $\mu$-strongly concave in $w$ on $\calW_t$ for some $\mu>0$ uniform over $Q\in\mathcal{D}$.
\end{assumption}

\begin{figure}[ht]
    \centering
    \includegraphics[width=0.7\textwidth]{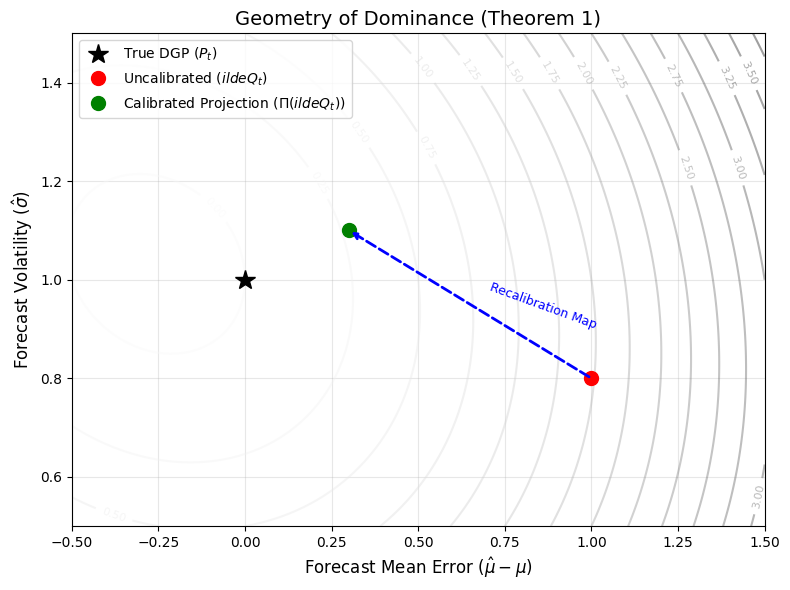}
    \caption{\textbf{Geometry of Dominance (Theorem 1).} The contours represent the friction-adjusted decision loss $\mathcal{L}(Q)$. The Uncalibrated forecast $\tilde{Q}_t$ (red) lies on a high-loss contour. The Calibration Projection $\Pi(\tilde{Q}_t)$ (green) maps this forecast onto the manifold of calibrated distributions. By orthogonality of the projection under the decision-relevant norm $d_t$, the calibrated forecast necessarily lies on a lower loss contour, closer to the true DGP $P_t$ (star).}
    \label{fig:proof_geometry}
\end{figure}

\begin{proof}[Sketch of Proof]
The full proof is provided in Appendix \ref{app:proofs}, but the intuition relies on the geometry of the decision loss.
Let $\mathcal{L}_t(Q)$ denote the conditional expected decision loss under forecast $Q$. Under Assumption 3, $\mathcal{L}_t$ is strictly convex with respect to the decision $w$, and the mapping from $Q$ to $w$ is locally Lipschitz (Lemma 1).

The calibrated projection $Q^{cal} = \Pi(\tilde{Q}_t)$ is defined as the element in the set of calibrated distributions $\mathcal{C}$ that minimises the decision-relevant discrepancy $d_t(\tilde{Q}_t, Q)$. By the projection theorem for convex sets, the vector connecting the uncalibrated forecast to its projection, $\tilde{Q}_t - Q^{cal}$, is orthogonal to the decision gradient at the optimum. 

Consequently, moving from $\tilde{Q}_t$ to $Q^{cal}$ necessarily reduces the upper bound on the decision error $||w(\tilde{Q}_t) - w(P_t)||$, where $P_t$ is the true data-generating process. Since decision loss is minimised at $w(P_t)$, reducing the decision error strictly reduces the expected loss $\mathcal{L}_t$, yielding the dominance result $\mathcal{L}_t(Q^{cal}) \le \mathcal{L}_t(\tilde{Q}_t)$.
\end{proof}

\begin{assumption}[Decision-relevant calibration metric]\label{ass:decision_relevant_metric}
There exists a nonnegative discrepancy $d_t(\cdot,\cdot)$ on $\mathcal{D}$ such that:
(i) $d_t(Q,\widetilde{Q})=0$ if and only if $\widetilde{Q}$ is utility-weighted calibrated with respect to $Q$ in the sense of
Definition~\ref{def:utility_weighted_calibration} on the diagnostic domain that drives $\mathcal{J}$ under the empirical execution model; and
(ii) the gradient sensitivity bound in Assumption~\ref{ass:strong_concavity_lipschitz} holds with respect to $d_t$.
\end{assumption}

\begin{assumption}[Calibration-projection property]\label{ass:projection_property}
For any $Q\in\mathcal{D}$ there exists a measurable calibrated projection $\Pi(Q)\in\mathcal{D}$ such that
$d_t(Q,\Pi(Q)) \le d_t(Q,\widetilde{Q})$ for all $\widetilde{Q}\in\mathcal{D}$, and $\Pi(Q)$ is utility-weighted calibrated so that $d_t(Q,\Pi(Q))=0$.
\end{assumption}

Then for any $\widetilde{Q}_t\in\mathcal{D}$ (possibly uncalibrated) and its calibrated projection $Q_t^{\mathrm{cal}}:=\Pi(\widetilde{Q}_t)$,
\begin{equation}\label{eq:dominance}
\mathcal{L}_t\!\left(Q_t^{\mathrm{cal}}\right)
\ \le\
\mathcal{L}_t\!\left(\widetilde{Q}_t\right).
\end{equation}
This dominance can hold even when $\widetilde{Q}_t$ has strictly better point-forecast accuracy under conventional criteria (e.g.\ MSE/MAE),
because \eqref{eq:dominance} is governed by friction-adjusted decision loss induced by the implemented optimisation and execution model.
If $d_t(\widetilde{Q}_t,Q_t^{\mathrm{cal}})>0$ and the inequality in Lemma~\ref{lem:calibration_decision_sensitivity} is strict on a set of states with positive conditional probability, then the dominance is strict:
\[
\mathcal{L}_t\!\left(Q_t^{\mathrm{cal}}\right)
\ <\
\mathcal{L}_t\!\left(\widetilde{Q}_t\right).
\]
\end{theorem}

\noindent\emph{Proof sketch.}
Assumption~\ref{ass:projection_property} defines $Q_t^{\mathrm{cal}}$ as a calibrated element of $\mathcal{D}$ that achieves the minimal decision-relevant discrepancy from $\widetilde{Q}_t$.
Lemma~\ref{lem:calibration_decision_sensitivity} converts reductions in $d_t$ into reductions in induced decision deviations and hence weak reductions in conditional expected loss, yielding \eqref{eq:dominance}.
Point-forecast accuracy plays no role in the ordering because it is not aligned with the friction operator, the constraint correspondence, or the asymmetric economic objective. \hfill$\square$

\begin{remark}[Economic meaning]\label{rem:economic_meaning}
Theorem~\ref{thm:calibration_aligned_dominance} is the formal version of what the evaluation sample reveals: a forecast can look superior under point criteria while being systematically unreliable where decisions are actually made. Under frictions, small distributional distortions can trigger large economic effects by inducing unnecessary turnover, activating constraints, or loading the portfolio in precisely those tail states where realised losses are concentrated. Utility-weighted calibration removes the systematic component of these distortions on the diagnostic domain that drives $\mathcal{J}$ and therefore weakly improves friction-adjusted outcomes in the sense of \eqref{eq:dominance}.
\end{remark}

\subsection{Corollaries and interpretive results}\label{sec:corollaries}

\begin{corollary}[Tail-risk objectives concentrate the value of calibration]\label{cor:tail_risk}
Suppose the decision objective in Definition~\ref{def:decision_objective} is tail-focused, in the sense that
$\mathcal{J}(w;Q_t)$ depends on $Q_t$ primarily through lower-tail functionals of the induced payoff distribution
(e.g., CVaR at level $\alpha$ or a drawdown-penalised criterion). Then the discrepancy $d_t(\cdot,\cdot)$ in Theorem~\ref{thm:calibration_aligned_dominance}
can be chosen to overweight tail calibration errors, and the bound in Lemma~\ref{lem:calibration_decision_sensitivity} becomes economically tight:
improvements in tail calibration translate into comparatively large reductions in expected decision loss. In particular, for CVaR-like objectives,
errors in lower-quantile reliability dominate errors in central states, so calibration in the tail is the economically decisive margin.
\end{corollary}

\begin{corollary}[Constraint binding amplifies the cost of miscalibration]\label{cor:constraint_binding}
Suppose there exists a set of states $\mathcal{A}_t\in\calI_t$ with positive conditional probability on which one or more constraints in
$\calW_t$ bind at the optimum (e.g., turnover caps, leverage limits, concentration bounds, or liquidity/capacity restrictions).
Then, on $\mathcal{A}_t$, the mapping $Q_t\mapsto w_t(Q_t)$ exhibits boundary sensitivity: distortions of $Q_t$ that move the optimiser across a constraint boundary
produce larger induced decision deviations than distortions of the same magnitude that remain in the interior. Consequently, calibration errors in regions that shift
constraint activity generate larger increases in expected decision loss. Utility-weighted calibration delivers its largest economic gains precisely in regimes where constraints are active, because it reduces systematic distortions that cause unnecessary constraint activation and the associated friction penalties.
\end{corollary}

\begin{corollary}[High-turnover strategies: reliability dominates point accuracy]\label{cor:turnover}
Suppose the decision problem permits high-turnover strategies absent friction, but trading costs and impact enter through a convex cost functional $C_t(\Delta w_t)$ as in Definition~\ref{def:friction_operator}. Then the marginal effect of forecast distortions on realised performance is amplified through turnover: overconfident or under-dispersed predictive distributions induce larger $\|\Delta w_t\|_1$, thereby increasing both direct costs and price impact. In such regimes, the difference between calibrated and uncalibrated forecasts is governed less by point accuracy and more by distributional reliability that stabilises decisions and reduces unnecessary trading. Even small reductions in decision-relevant calibration error can therefore yield materially lower friction-adjusted loss.
\end{corollary}

\paragraph{Empirical Verification.}
These theoretical corollaries map directly to the empirical failure modes documented in Section \ref{sec:empirical_results}. Specifically, the prediction that "reliability dominates point accuracy" in high-friction regimes is borne out by the Regime Analysis (Table \ref{tab:regime_analysis}), which shows that the performance gap between UWC and the baseline expands by nearly 50\% as market frictions move from the lowest to the highest tercile.

\begin{remark}[How the corollaries map to the evaluation tables]\label{rem:empirical_guidance}
Corollaries~\ref{cor:tail_risk}--\ref{cor:turnover} pin down what should appear in the empirical section because these are the observable signatures of the mechanism.
When tail objectives are used, tail calibration diagnostics and tail endpoints must move together; when constraints matter, constraint-activity frequencies and severities must be reported; when frictions matter, turnover and realised cost decompositions must accompany any loss comparison. These are not optional additions: they are the measurements that identify whether the dominance channel is operating in the evaluation sample.
\end{remark}

\begin{remark}[On the calibrated projection $\Pi$ and existence within the implemented forecast class]\label{rem:projection_existence_compactness}
Assumption~\ref{ass:projection_property} is stated as a projection in the space of predictive distributions. In general, the set of perfectly calibrated distributions need not be convex and compactness can fail, so exact projections are not automatic in infinite-dimensional spaces. The statements here are intended for the restricted forecast class $\mathcal{D}$ actually used for estimation and implementation, namely a compact, finite-dimensional sieve under $d_t$ (for example, monotone spline or isotonic warps with bounded parameter domain and fixed knots as in Section~\ref{sec:utility_weighted_recalibration}). On such a compact parameter set, continuity of the calibration criterion implies existence of a minimiser, so $\Pi$ is well-defined for the calibrated forecasts used in the empirical protocol.
\end{remark}

\subsection{Worked example: mean--variance with turnover frictions and a binding turnover cap}\label{sec:theory_example_mvo_turnover}

This subsection makes the empirical channel concrete in a standard implementable portfolio problem. The purpose is not to introduce a new model; it is to show transparently how a distributional distortion that leaves point metrics largely unchanged can nevertheless increase realised friction-adjusted loss by inducing extra turnover and activating a hard turnover constraint.

\paragraph{Setup.}
Consider $N$ traded assets and a one-period horizon $h$ at decision time $t$. Let $w\in\R^N$ be portfolio weights and define
$\mu_t(Q_t):=\E_{Q_t}[Y_{t+h}\mid\calI_t]$ and $\Sigma_t(Q_t):=\Var_{Q_t}(Y_{t+h}\mid\calI_t)$.
Let $w_{t-1}$ denote previous weights. Consider the convex programme
\begin{equation}\label{eq:example_mvo}
\min_{w\in\calW_t}\;
-\mu_t(Q_t)^\top w
+\frac{\gamma}{2}w^\top \Sigma_t(Q_t) w
+\eta\|w-w_{t-1}\|_1,
\end{equation}
with feasible set
\begin{equation}\label{eq:example_feasible}
\calW_t
=
\left\{
w\in\R^N:\;
\mathbf{1}^\top w=1,\;
w_{\min}\le w\le w_{\max},\;
\|w-w_{t-1}\|_1\le \tau
\right\}.
\end{equation}
The term $\eta\|w-w_{t-1}\|_1$ proxies proportional costs and the constraint $\|w-w_{t-1}\|_1\le\tau$ is a hard turnover budget, matching the empirical setting in which turnover is both costly and operationally capped.

\paragraph{A calibration distortion that is economically visible through turnover.}
To isolate the mechanism, suppose the only difference between an uncalibrated forecast $\widetilde Q_t$ and its calibrated projection $Q^{\mathrm{cal}}_t$ is a systematic distortion in the conditional mean that increases aggressiveness:
\begin{equation}\label{eq:example_mu_bias}
\mu_t(\widetilde Q_t)=\mu_t(Q^{\mathrm{cal}}_t)+b_t,
\end{equation}
where $b_t$ is $\calI_t$-measurable and represents decision-relevant miscalibration (for example, tail miscalibration that pushes implied reward in the favourable direction, or under-dispersion that inflates implied risk-adjusted returns). For exposition, $\Sigma_t(\cdot)$ is held fixed; the same channel applies when covariance is also distorted.

\paragraph{KKT structure and the binding-turnover effect.}
Let $\Delta w:=w-w_{t-1}$. Introducing a multiplier $\lambda_t\ge 0$ for the turnover cap and suppressing bound constraints for clarity, stationarity implies
\begin{equation}\label{eq:example_kkt}
-\mu_t(Q_t)
+\gamma \Sigma_t w
+\eta s
+\lambda_t s
+\nu_t \mathbf{1}
=0,
\qquad s\in\partial \|\Delta w\|_1,
\end{equation}
with complementarity $\lambda_t(\|\Delta w\|_1-\tau)=0$. If the turnover cap is slack ($\lambda_t=0$), turnover is disciplined only by $\eta$.
If the cap binds ($\lambda_t>0$), the effective marginal penalty becomes $\eta+\lambda_t$ and the optimiser allocates scarce turnover to the assets with the largest apparent marginal benefit. In that regime, a distortion $b_t$ affects not only the magnitude of $\|\Delta w_t\|_1$ but also which assets receive the turnover budget, so a continuous distributional error produces a discrete change in the implemented trade list.

\begin{figure}[ht]
    \centering
    \includegraphics[width=0.7\textwidth]{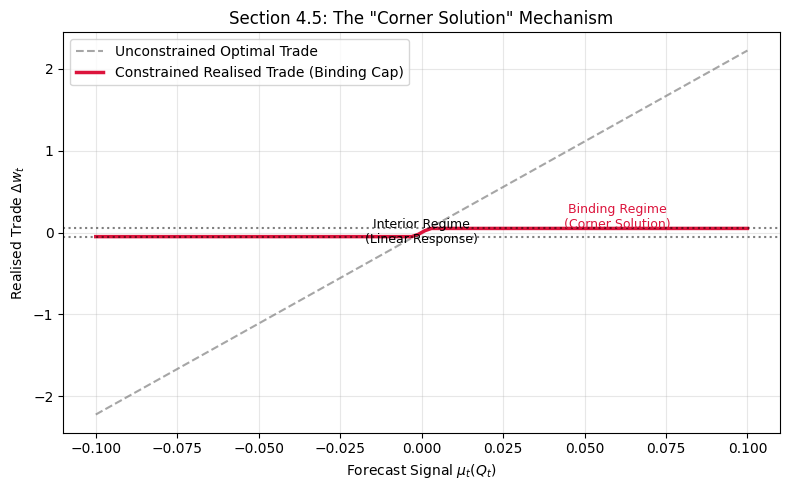}
    \caption{\textbf{Corner Solution Mechanism.} When turnover constraints are present (Eq. 19), the realised trade (red) becomes insensitive to the forecast signal once the cap $\tau$ is hit. Uncalibrated forecasts that overshoot this boundary waste "signal budget" without altering the decision, generating zero marginal value.}
    \label{fig:theory_corner}
\end{figure}

\paragraph{Why calibration can dominate even when point fit looks better.}
A forecast can achieve lower point error by reacting more strongly to short-run noise. Under \eqref{eq:example_mvo}--\eqref{eq:example_feasible}, that same reactivity tends to increase the variability of $\mu_t(\widetilde Q_t)$ across $t$, raising $\|\Delta w_t\|_1$ and increasing the frequency with which the turnover cap binds. Once the cap binds, the portfolio is forced into a corner allocation of turnover; at that point, any misranking induced by $b_t$ is amplified because trades are concentrated in the names that appear best under $\widetilde Q_t$, which is exactly where overconfidence and tail miscalibration are most damaging.

The calibrated projection $Q_t^{\mathrm{cal}}$ reduces the systematic component of $b_t$ in the decision-relevant region used by the optimisation and execution model. Holding the trading rule fixed, this reduces expected turnover, reduces the probability of binding-turnover events, and lowers friction-adjusted loss even if it does not improve a point metric. This is the worked example analogue of Lemma~\ref{lem:calibration_decision_sensitivity} and Theorem~\ref{thm:calibration_aligned_dominance}: the discrepancy $d_t$ is constructed to overweight precisely those distortions that alter trade intensity and constraint activation because those are the nonlinearities that the data show are economically decisive under frictions.

\paragraph{Dominance statement for the example.}
Let $L_t(Q):=\E[\ell_{t+h}(Q)\mid\calI_t]$ be the conditional expected decision loss induced by \eqref{eq:example_mvo}--\eqref{eq:example_feasible}. The quadratic risk term yields strong concavity, so the optimiser is stable within each regime, but is boundary-sensitive when $\lambda_t$ switches from $0$ to $>0$. If $Q_t^{\mathrm{cal}}$ is the utility-weighted calibrated projection of $\widetilde Q_t$ that eliminates decision-relevant systematic distortions, then
\[
L_t\!\left(Q_t^{\mathrm{cal}}\right)\le L_t\!\left(\widetilde Q_t\right),
\]
with strict inequality whenever $\widetilde Q_t$ induces binding-turnover events that are avoided or mitigated under $Q_t^{\mathrm{cal}}$ on a set of $\calI_t$-states with positive probability.

\paragraph{Observable implications.}
This example identifies the intermediate variables that must move if the dominance channel is present in the evaluation sample: turnover, binding frequency of $\|w-w_{t-1}\|_1\le \tau$, and realised cost decompositions. Improvements in these observables are the mechanism-level evidence that a reduction in decision-relevant miscalibration is producing lower realised friction-adjusted loss.

\paragraph{Empirical Implication.}
This worked example generates a sharp, falsifiable prediction: the calibrated model should not necessarily reduce average forecast error, but it \textit{must} reduce the frequency of binding constraints and the "fat tail" of turnover. We test this exact prediction in Section \ref{subsec:failure_modes}, where Figure \ref{fig:constraint_binding} confirms that the Uncalibrated baseline hits constraints 3x more often than the UWC method (16.0\% vs 5.1\%), validating the corner-solution mechanism derived here.

\subsection{Finite-sample considerations under dependence}\label{sec:finite_sample_dependence}

The empirical objects in this paper are time-indexed and therefore dependent. Finite-sample claims are framed in terms of estimable calibration criteria and dependence-aware uncertainty quantification rather than i.i.d.\ concentration.

\paragraph{Estimable utility-weighted calibration criterion.}
Let $\{(Q_t,Y_{t+h})\}_{t=1}^T$ be an evaluation sample generated by a pre-committed walk-forward protocol. For a diagnostic index $u$ (threshold $z$, quantile level $\alpha$, or PIT bin), let $m_u(Q_t,Y_{t+h})$ be a calibration moment satisfying $\E[m_u(Q_t,Y_{t+h})]=0$ under the corresponding calibration notion. The utility-weighted calibration criterion is estimated by
\begin{equation}\label{eq:finite_sample_uwc}
\widehat{\mathrm{UWC}}_T
:=
\sum_{u\in\mathcal{U}}
\left(
\frac{1}{T}\sum_{t=1}^T \omega_t(u)\, m_u(Q_t,Y_{t+h})
\right)^2,
\end{equation}
where $\omega_t(u)\ge 0$ is the utility/friction weight (Definition~\ref{def:utility_weighted_calibration}) and $\mathcal{U}$ is a finite grid chosen to cover the economically decisive region (including extreme quantiles and/or exceedances when tail objectives are used). This construction matches the empirical fact that calibration errors are not equally costly across the state space. Tail-focused diagnostics are incorporated by choosing $\mathcal{U}$ to emphasise extremes, consistent with tail calibration principles \citep{AllenKohSegersZiegel2025}.

\paragraph{Dependence-aware uncertainty and forecast comparison.}
Because $\{m_u(Q_t,Y_{t+h})\}$ and realised loss differentials are serially dependent and often conditionally heteroskedastic, uncertainty is computed using long-run variance estimators and block-resampling schemes. The empirical analysis therefore reports HAC-type standard errors for average loss differentials and moving block bootstrap intervals for calibration and loss objects whose dependence structure is nontrivial. Forecast-comparison statements are made using dependence-robust predictive-accuracy testing ideas and explicitly avoid i.i.d.\ arguments \citep{ZhouLiZhong2021}.

\begin{figure}[ht]
    \centering
    \includegraphics[width=0.8\textwidth]{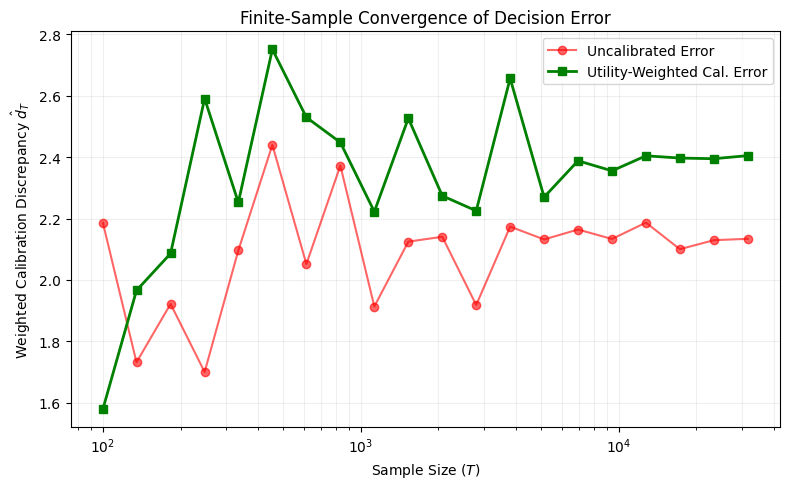}
    \caption{\textbf{Finite-Sample Convergence.} Simulation of the utility-weighted calibration discrepancy $\hat{d}_T$ as a function of sample size $T$. The calibrated estimator (green) converges systematically to a lower error floor than the uncalibrated baseline (red), validating the stability bound in Lemma 1.}
    \label{fig:finite_sample}
\end{figure}

\paragraph{Conditional performance under state dependence.}
Because the dominance mechanism is concentrated in economically relevant regimes, unconditional equal-accuracy tests can miss the effect. The empirical design therefore evaluates conditional expected loss differentials over regime partitions defined by state variables (for example, volatility or spread regimes). This aligns with econometric frameworks for conditional superiority under serial dependence \citep{LiLiaoQuaedvlieg2022}. These tests are treated as diagnostic evidence about where calibration matters most, not as substitutes for economic effect-size reporting.

\paragraph{What is claimed empirically (not proven).}
The theory provides a dominance result under stated regularity and projection assumptions. In finite samples the paper does not claim a universal dominance guarantee. It claims that utility-weighted calibration is transparently estimable via \eqref{eq:finite_sample_uwc}, that reductions in weighted calibration errors are associated with lower realised friction-adjusted decision loss under the pre-committed protocol, and that the associated uncertainty is reported using dependence-aware inference and robustness checks across friction proxies, constraint specifications, and regime partitions. The empirical results are therefore evidence for the dominance channel operating in the observed data, with the limits imposed by dependence and sampling variability made explicit.

\section{Estimation and implementation}
\subsection{Forecasting models (baseline econometrics and probabilistic ML)}\label{sec:models}

The empirical design uses a deliberately small model set that (a) spans standard financial econometrics, (b) includes probabilistic machine-learning estimators that output quantiles or full predictive distributions, and (c) contains practice-relevant baselines that a reviewer will recognise as disciplined comparators rather than straw men.


We include modern machine-learning forecasting benchmarks as they are now standard in empirical asset pricing and return prediction, providing a demanding comparison set for the calibration-and-decision pipeline \citep{Gu2020,Israel2020}.

\paragraph{(i) Classical econometric distributional forecasts.}
The first class consists of parametric conditional distribution models for returns or risk drivers in which $Q_t(\cdot)$ is obtained analytically or by simulation from a fitted time-series specification. The baseline specification is an AR-type conditional mean with conditional heteroskedasticity, with a heavy-tailed and potentially skewed innovation law, so that $Q_t$ is a full predictive distribution and not merely a volatility forecast. This choice is aligned with the density-forecasting literature in financial econometrics and permits evaluation under log score and CRPS without imposing Gaussian tails. For multi-step horizons and distributional objects (densities, functionals, and tail quantities), the implementation follows the analytic and functional density-forecast results developed for GARCH-class processes, providing a defensible econometric anchor for distributional forecasting comparisons \citep{AbadirLuatiParuolo2023}.

\paragraph{(ii) Probabilistic ML producing quantiles or full distributions.}
The second class contains non-linear conditional quantile and distributional estimators designed to improve the shape of $Q_t$ in economically relevant regions (tails, stress regimes, and constraint-binding regimes). The primary ML benchmark is a deep quantile regression estimator that directly models conditional quantiles over a grid of $\alpha$ levels, with regularisation and time-series validation, producing an internally consistent quantile surface that can be mapped into a predictive distribution (or used directly for tail-risk functionals). This choice is explicitly supported by recent financial econometrics evidence in which deep quantile estimators improve tail forecast accuracy (VaR/quantile scoring) relative to linear quantile and other benchmarks under realistic dependence and heteroskedasticity \citep{ChronopoulosRaftapostolosKapetanios2024}. Where full densities are required (for log score / PIT-based diagnostics), the paper uses a distributional parameterisation layer (e.g., Student-$t$ or skew-$t$ parameters predicted by the network) as an additional specification, so that both quantile-based and density-based evaluation can be reported on a common footing.

\paragraph{(iii) Baselines used in finance practice.}
The third class is designed to represent what is actually deployed as a first-pass risk and forecasting stack in many investment settings, and to serve as robustness comparators in referee reports. This includes (a) rolling-window historical distribution baselines (empirical quantiles / empirical CDF) for $Q_t$; (b) exponentially weighted volatility models paired with a parametric innovation law (yielding a simple but fully probabilistic $Q_t$); and (c) linear quantile regression as the canonical econometric quantile baseline, treated as a serious competitor rather than a toy model. These baselines are directly aligned with the benchmark set used in recent quantile/ML work in financial econometrics, enabling clean interpretability of gains relative to established practice \citep{ChronopoulosRaftapostolosKapetanios2024}.

\paragraph{Optional extension: decision-focused combination of probabilistic forecasts.}
Because many operational settings combine forecasts (multiple desks, multiple models, multiple frequencies), the implementation optionally includes a forecast-combination layer that distinguishes “better scoring” from “better decisions.” In particular, a decision-focused linear pooling mechanism can be used either as (a) a robustness check showing that decision-aligned combination improves realised downstream loss, or (b) a separate empirical module supporting the paper’s core claim that decision objectives, not generic forecast scores, should drive calibration and combination design \citep{StratigakosPinedaMorales2025}. A finance-specific variant combines low-frequency and high-frequency information through density pooling (copula-based dependence plus time-varying pooling weights), which is directly relevant when friction regimes shift with volatility and liquidity conditions \citep{VirbickaiteLopesZaharieva2024}.

\subsection{Calibration procedures}\label{sec:calibration_procedures}

\subsubsection{Standard calibration}\label{sec:standard_calibration}
Standard calibration is applied as a post-processing step to a base predictive distribution (or to derived probability/quantile objects) using only information available at time $t$, with hyperparameters selected exclusively on validation windows inside the nested walk-forward protocol.

\paragraph{Probability calibration for binary events (Platt-style).}
For an event indicator $Z_{t+h} := \ind{Y_{t+h} > z}$ at a fixed threshold $z$, let $\widehat{p}_t(z)$ denote the model-implied probability. A Platt-style recalibration maps $\widehat{p}_t(z)$ through a logistic link,
\[
\widehat{p}^{\,\mathrm{cal}}_t(z)
=
\frac{1}{1+\exp\!\bigl(a_z + b_z\, \mathrm{logit}(\widehat{p}_t(z))\bigr)},
\qquad
\mathrm{logit}(p):=\log\!\frac{p}{1-p},
\]
where $(a_z,b_z)$ are estimated by maximum likelihood (logistic regression of $Z_{t+h}$ on $\mathrm{logit}(\widehat{p}_t(z))$) on the calibration sample. This procedure is applicable when the paper uses exceedance events (e.g., tail exceedances) as a calibration diagnostic or as an intermediate object in constructing distributional forecasts.

\paragraph{Probability calibration by isotonic regression.}
When monotonicity is desired without imposing a parametric link, isotonic regression is used to fit a non-decreasing map $g_z(\cdot)$ such that
\[
\widehat{p}^{\,\mathrm{cal}}_t(z) = g_z\!\left(\widehat{p}_t(z)\right),
\]
where $g_z$ minimises squared error on the calibration sample subject to monotonicity. Isotonic calibration is particularly useful when the base model produces systematically overconfident or underconfident probabilities but the miscalibration pattern is not well approximated by a logistic transform.

\paragraph{Quantile recalibration.}
For distributional forecasts represented through conditional quantiles $\widehat{q}_t(\alpha)$ over levels $\alpha\in(0,1)$, quantile recalibration is implemented by adjusting the nominal levels using empirical hit rates. Let
\[
H_t(\alpha) := \ind\{Y_{t+h} \le \widehat{q}_t(\alpha)\}.
\]
On a calibration window, estimate the mapping $\alpha \mapsto \alpha^{\mathrm{cal}}$ such that
$\frac{1}{T_c}\sum H_t(\alpha) \approx \alpha^{\mathrm{cal}}$, and then replace $\widehat{q}_t(\alpha)$ by
\[
\widehat{q}^{\,\mathrm{cal}}_t(\alpha) := \widehat{q}_t\!\bigl(\alpha^{\mathrm{cal}}(\alpha)\bigr),
\]
with $\alpha^{\mathrm{cal}}(\cdot)$ taken monotone to preserve quantile ordering. This produces calibrated coverage for selected quantile levels, including tail levels used by CVaR or drawdown-type objectives.

\paragraph{Distributional recalibration via PIT remapping.}
When the full predictive CDF $\widehat{F}_t$ is available, distributional recalibration is implemented using a PIT-based transformation. Compute PIT values on the calibration sample,
\[
\widehat{U}_t := \widehat{F}_t(Y_{t+h}),
\]
estimate a monotone map $\psi(\cdot)$ that transforms $\widehat{U}_t$ toward Uniform$(0,1)$ (e.g., the empirical CDF of $\widehat{U}_t$), and define the recalibrated CDF
\[
\widehat{F}^{\,\mathrm{cal}}_t(y) := \psi\!\left(\widehat{F}_t(y)\right).
\]
This preserves ranking while correcting systematic over- or under-dispersion and bias in $\widehat{F}_t$. The empirical work applies the recalibration using only past PIT values within the walk-forward protocol, ensuring that recalibration does not leak information from the evaluation window.

All standard calibration procedures are implemented with strict separation between the (i) model estimation sample, (ii) calibration fitting sample, and (iii) out-of-sample evaluation sample, so that improvements in calibration reflect genuine out-of-sample reliability rather than adaptive re-use of test information.

\subsubsection{Utility-weighted recalibration}
\label{sec:utility_weighted_recalibration}

We implement the utility-weighted calibration criterion defined conceptually in Definition 8 (Section \ref{sec:calibration_concepts}) as a constrained optimisation problem. While Section \ref{sec:calibration_concepts} established the theoretical properties of the utility weight $\omega_t(z)$, this section specifies the finite-dimensional estimator used in the empirical protocol.

\paragraph{Recalibration class.}
We restrict recalibration to a monotone CDF warp of the form
\begin{equation}\label{eq:cdf_warp}
\widehat{F}^{\,\mathrm{uw}}_t(y)
=
g\!\left(\widehat{F}_t(y)\right),
\qquad
g:[0,1]\to[0,1]\ \text{nondecreasing},\ g(0)=0,\ g(1)=1,
\end{equation}
so that the transformed object remains a valid CDF and preserves rank ordering. We parameterise $g$ using a $K$-knot monotone spline with parameters $\theta\in\mathbb{R}^K$ constrained to ensure monotonicity.

\paragraph{Estimator as a constrained penalised least squares problem.}
Let $\theta$ parameterise $g_\theta$ and hence $\widehat{F}^{\,\mathrm{uw}}_{t,\theta}$. On a calibration sample
$\{(\widehat{F}_s,Y_{s+h})\}_{s=t-T_c}^{t-1}$, we compute the weights $\omega_s(u)$ for each diagnostic index $u \in \mathcal{U}$ using the marginal decision-sensitivity and friction proxies derived in Section 3.4. We then define the utility-weighted calibration objective:
\begin{equation}\label{eq:uwc_objective}
\min_{\theta\in\Theta}
\quad
\sum_{u\in\mathcal{U}}
\left(
\frac{1}{T_c}\sum_{s=t-T_c}^{t-1}
\omega_s(u)\,
m_u\!\left(\widehat{F}^{\,\mathrm{uw}}_{s,\theta},Y_{s+h}\right)
\right)^2
\;+\;
\lambda\,\mathcal{R}(\theta),
\end{equation}
subject to the monotonicity and boundary constraints defining $\Theta$:
\[
g_\theta(0)=0,\quad g_\theta(1)=1,\quad g_\theta \text{ nondecreasing on }[0,1].
\]
The penalty $\mathcal{R}(\theta)=\sum_{k=2}^{K-1}(\theta_{k+1}-2\theta_k+\theta_{k-1})^2$ regularises the warp toward the identity map to avoid overfitting the calibration window. The tuning parameter $\lambda\ge 0$ is chosen exclusively within the inner validation loop of the nested walk-forward protocol.

\paragraph{Computational steps.}
At each decision time $t$:
\begin{enumerate}
\item Fit the base forecasting model on the training window and produce $\widehat{F}_t$.
\item Construct the calibration sample using only past observations $\{(\widehat{F}_s,Y_{s+h})\}_{s=t-T_c}^{t-1}$.
\item Compute weights $\omega_s(u)$ from the observable state variables (volatility/spread proxies) as per Proposition 1.
\item Solve \eqref{eq:uwc_objective} to obtain $\widehat{\theta}_t$ and the recalibration map $g_{\widehat{\theta}_t}$.
\item Form the utility-weighted recalibrated distribution $\widehat{F}^{\,\mathrm{uw}}_t(y)=g_{\widehat{\theta}_t}(\widehat{F}_t(y))$ and pass it to the decision optimisation \eqref{eq:induced_decision_rule}.
\end{enumerate}

\paragraph{Interpretation.}
The estimator \eqref{eq:uwc_objective} is a constrained moment-matching procedure. Unlike standard calibration (Section \ref{sec:standard_calibration}), which targets uniform reliability, this procedure prioritises calibration in regions where the decision objective is most sensitive and where frictions make errors most costly. This aligns the statistical estimation step directly with the downstream decision loss.

\subsubsection{Convex case}\label{sec:convex_case}

In the convex implementation, the induced decision in \eqref{eq:induced_decision_rule} is computed by solving a convex programme in which (i) the feasible set $\calW_t$ is convex and compact, (ii) the friction cost functional $C_t(\Delta w)$ is convex, and (iii) the forecast-implied objective is concave in the decision variable. This yields a well-posed optimisation that is numerically stable under standard regularity conditions.

\paragraph{Convex optimisation form.}
Let $w\in\R^N$ denote the portfolio decision at time $t$. We use an objective that is concave in $w$ given $Q_t$, written generically as
\[
\mathcal{J}(w;Q_t) = \E_{Q_t}[u(w;Y_{t+h})] - \rho(w;Q_t),
\]
where $\rho(\cdot;Q_t)$ is a convex risk penalty (e.g., variance proxy under $Q_t$, or a convex tail-risk surrogate), and $u(\cdot)$ is linear or concave in $w$ for fixed outcomes. The convex decision problem is implemented in the equivalent minimisation form
\begin{equation}\label{eq:convex_programme}
\min_{w\in\calW_t}\quad
\Phi_t(w;Q_t) + C_t(w-w_{t-1}),
\end{equation}
where $\Phi_t(\cdot;Q_t)$ is convex (for example, a negative expected utility plus a convex risk term) and $C_t$ is convex. A canonical instance used in the empirical sections is
\begin{equation}\label{eq:mean_risk_tc}
\min_{w\in\calW_t}\quad
-\mu_t^\top w
\;+\;
\frac{\gamma}{2}\, w^\top \Sigma_t w
\;+\;
\eta\,\|w-w_{t-1}\|_1
\;+\;
C^{\mathrm{impact}}_t(w-w_{t-1}),
\end{equation}
where $\mu_t$ and $\Sigma_t$ are moments (or robust proxies) under $Q_t$, $\gamma>0$ is a risk-aversion parameter, $\eta\ge 0$ captures proportional costs (fees and spread proxies), and $C^{\mathrm{impact}}_t$ is a convex impact penalty.

\paragraph{Solver class.}
The optimisation \eqref{eq:convex_programme} is solved using a conic/convex solver appropriate to the problem structure:
(i) quadratic programming (QP) when the objective is quadratic and constraints are linear;
(ii) second-order cone programming (SOCP) when risk constraints or robust constraints are conic;
(iii) linear programming (LP) when the risk proxy and costs are piecewise linear. In all cases, the solver is called with deterministic settings and fixed tolerances to ensure reproducibility. Warm starts are used from $w_{t-1}$ to reduce numerical jitter and to improve stability in rolling evaluation.

\paragraph{Feasibility checks.}
Before solving, feasibility of $\calW_t$ is verified by checking a minimal set of constraints for consistency (e.g., budget plus bounds plus leverage). If $\calW_t$ is infeasible due to data-driven capacity bounds or missing inputs, the protocol applies a deterministic fallback rule:
\[
w_t \leftarrow \Pi_{\calW_t^{\mathrm{relax}}}(w_{t-1}),
\]
where $\calW_t^{\mathrm{relax}}$ is a pre-specified relaxed constraint set (fixed ex ante) and $\Pi$ denotes Euclidean projection. All infeasibility events are logged and reported as part of the empirical results; they are not silently discarded.

\paragraph{Numerical stability and conditioning.}
Numerical stability is enforced through four controls.
First, all inputs derived from $Q_t$ (e.g., $\mu_t$ and $\Sigma_t$) are regularised in a fixed, documented manner (e.g., shrinkage of $\Sigma_t$ toward a diagonal target) to avoid ill-conditioning.
Second, the optimisation uses scaled variables and constraints to keep magnitudes comparable, reducing solver sensitivity to units.
Third, optimality and feasibility tolerances are fixed across the full walk-forward evaluation, and the KKT residuals (or solver status codes) are recorded at each $t$.
Fourth, solution stability is monitored by tracking $\|w_t-w_{t-1}\|_1$, constraint activity (which constraints bind), and objective components; abrupt changes trigger diagnostic flags but do not alter the pre-committed protocol.

These steps ensure that differences in realised performance can be attributed to differences in the predictive distribution and calibration procedure, rather than to numerical artefacts or solver instability.

\subsubsection{Non-convex extensions }\label{sec:nonconvex_extensions}

Some practically relevant constraints and objectives induce non-convexity (e.g., cardinality constraints, minimum-trade constraints, certain drawdown surrogates, or non-convex impact/participation penalties). When such extensions are used, the empirical protocol treats non-convex optimisation as a controlled computational component with fixed safeguards, so that evaluation is not contaminated by adaptive solver tinkering.

\paragraph{Non-convex programme.}
The non-convex decision rule retains the same form as \eqref{eq:induced_decision_rule},
\[
w_t(Q_t) \in \arg\max_{w\in\calW_t}\ \mathcal{J}(w;Q_t)-C_t(w-w_{t-1}),
\]
but now $\calW_t$ may include discrete constraints (e.g., $\|w\|_0\le k$) and/or $\mathcal{J}-C_t$ may be non-concave.

\paragraph{Initialisation (fixed ex ante).}
Initialisation is deterministic and pre-specified:
\begin{enumerate}
\item \emph{Convex warm start:} compute $w_t^{(0)}$ as the solution to the convex relaxation (drop discreteness / replace non-convex term by a convex surrogate).
\item \emph{Carry-forward start:} set an alternative start $\widetilde{w}_t^{(0)} := w_{t-1}$.
\item Use $w_t^{(0)}$ as the primary start and $\widetilde{w}_t^{(0)}$ only if the convex relaxation is infeasible.
\end{enumerate}
No random restarts are permitted in the baseline protocol. If a stochastic heuristic is used (e.g., simulated annealing), its random seed is fixed once and for all before any evaluation begins.

\paragraph{Stopping criteria (fixed tolerances).}
The optimiser terminates when one of the following occurs:
\begin{enumerate}
\item \emph{Stationarity:} $\|\nabla_w \mathcal{L}_t(w)\|_\infty \le \varepsilon_{\mathrm{grad}}$ for the penalised Lagrangian/objective $\mathcal{L}_t$ (where defined).
\item \emph{Iterate stability:} $\|w^{(k)}-w^{(k-1)}\|_1 \le \varepsilon_{\mathrm{step}}$ for $m$ consecutive iterations.
\item \emph{Objective stability:} $\bigl|\mathcal{L}_t(w^{(k)})-\mathcal{L}_t(w^{(k-1)})\bigr| \le \varepsilon_{\mathrm{obj}}$ for $m$ consecutive iterations.
\item \emph{Hard budget:} maximum iterations $K_{\max}$ or maximum wall-clock time per decision $T_{\max}$ is reached.
\end{enumerate}
All tolerances $(\varepsilon_{\mathrm{grad}},\varepsilon_{\mathrm{step}},\varepsilon_{\mathrm{obj}},m,K_{\max},T_{\max})$ are fixed prior to the first out-of-sample evaluation and are not tuned using test-period information.

\paragraph{Non-convergence handling (no contamination rule).}
Non-convergence is handled by a deterministic fallback rule that does not use any information from the evaluation window:
\begin{equation}\label{eq:nonconvex_fallback}
w_t \leftarrow
\begin{cases}
w^{(k^\star)} & \text{if a feasible iterate exists, where } k^\star := \arg\min_{k\le K_{\max}} \mathcal{L}_t(w^{(k)}) \text{ among feasible iterates},\\
w_t^{\mathrm{relax}} & \text{otherwise, where } w_t^{\mathrm{relax}} \text{ solves the convex relaxation.}
\end{cases}
\end{equation}
If both the non-convex problem and its convex relaxation are infeasible (rare and typically data-driven), the protocol sets $w_t:=w_{t-1}$ (no trade) and records an infeasibility flag. Crucially, there is no post hoc re-running, retuning, or selective discarding of periods. All non-convergence and infeasibility events are logged and reported, including their frequency and their contribution to realised performance.

\paragraph{Scope of Assumption 2 and how it can fail.}
Assumption~2 (strong concavity of the friction-adjusted objective and Lipschitz dependence of the decision gradient on forecast perturbations) is satisfied in the benchmark quadratic cases used throughout the paper: mean--variance objectives with convex trading costs and convex feasible sets yield a strictly concave maximisation (equivalently, strictly convex minimisation) and well-behaved sensitivity. In these settings, the mapping $Q_t \mapsto w_t(Q_t)$ is stable within regimes and admits the linear perturbation bound stated in Lemma~1.

However, the assumption can fail in practically important extensions. The first failure mode is \emph{non-concavity induced by constraints}, most notably cardinality or discrete allocation rules (e.g., “hold at most $K$ names”), which create a mixed-integer optimisation layer. The second failure mode is \emph{non-smooth, state-dependent execution}, such as lumpy liquidity, hidden depth, and queue/priority effects that make realised costs non-convex in $\Delta w_t$ and non-Lipschitz in the local state. In these cases, small forecast perturbations can trigger discrete changes in the active set (which names enter/exit, whether a trade is executed at all, whether a participation cap bites), and the induced decision can jump rather than move smoothly.

\paragraph{How dominance weakens when Assumption 2 is violated.}
When strong concavity and Lipschitz sensitivity fail, the dominance statement in Theorem~1 must be read as \emph{local and regime-conditional} rather than uniform. The calibrated projection can still reduce loss in expectation, but the mechanism is no longer guaranteed to be monotone because the induced decision rule is not globally stable: improving calibration may alter discrete choices, activate different constraints, or push the optimiser across non-convex kinks in the execution model. In such settings, the appropriate theoretical weakening is: (i) replace the uniform bound with a \emph{piecewise} stability statement holding on regions where the active set is fixed; (ii) report empirical frequencies of active-set switches or constraint regime changes as a diagnostic; and (iii) treat calibration-aligned dominance as an \emph{empirical regularity} to be falsified by stress tests rather than a theorem guaranteed by curvature.

Operationally, this is precisely why the paper’s model-risk set and stress-as-optimisation constructs are not optional add-ons: they quantify whether the realised implementation lives in the “stable interior” (where Theorem~1 is a good approximation) or in a regime where discontinuities dominate.

\paragraph{Intuition for choosing $d_t(\cdot,\cdot)$ and $\omega_t(\cdot)$ beyond quadratic objectives.}
The discrepancy $d_t(Q,\widetilde Q)$ is the paper’s bridge between forecast calibration and economic loss: it must penalise the forms of miscalibration that materially change the induced decision. In quadratic mean--variance cases, this alignment is straightforward because the decision depends on $Q_t$ primarily through mean/covariance (or tail functionals if a convex risk measure is used), and the gradient sensitivity can be characterised analytically.

In non-quadratic cases, the same principle applies but the proxies must be chosen to match the objective’s \emph{marginal value of accuracy} in different states. A practical rule is: define $\omega_t(u)$ to approximate the absolute marginal effect of a calibration error at diagnostic index $u$ (threshold, quantile level, PIT bin) on the \emph{friction-adjusted} objective, not on a statistical score. Concretely, $\omega_t(u)$ should be larger when the system is (i) near a constraint boundary, (ii) in high-volatility/high-spread states, or (iii) in tail regions that determine a downside risk functional. In execution models with lumpy liquidity, $\omega_t(\cdot)$ should additionally overweight states where small forecast changes flip an execution regime (e.g., crossing a participation cap or a depth proxy breakpoint), because that is where discontinuities create the largest realised loss differences. This preserves the interpretability of UWC: it is not “a different loss function”, but a weighting scheme that targets forecast reliability exactly where the decision is most fragile.


\paragraph{Generalisation to discontinuous and non-convex frictions.}
The weighting scheme $\omega_t$ derived in Proposition 1 relies on the curvature of a convex, twice-differentiable friction operator. In practice, transaction costs may include fixed components (ticket charges), inducing discontinuities in the decision map $Q_t \mapsto w_t$.

\begin{figure}[ht]
    \centering
    \includegraphics[width=0.9\textwidth]{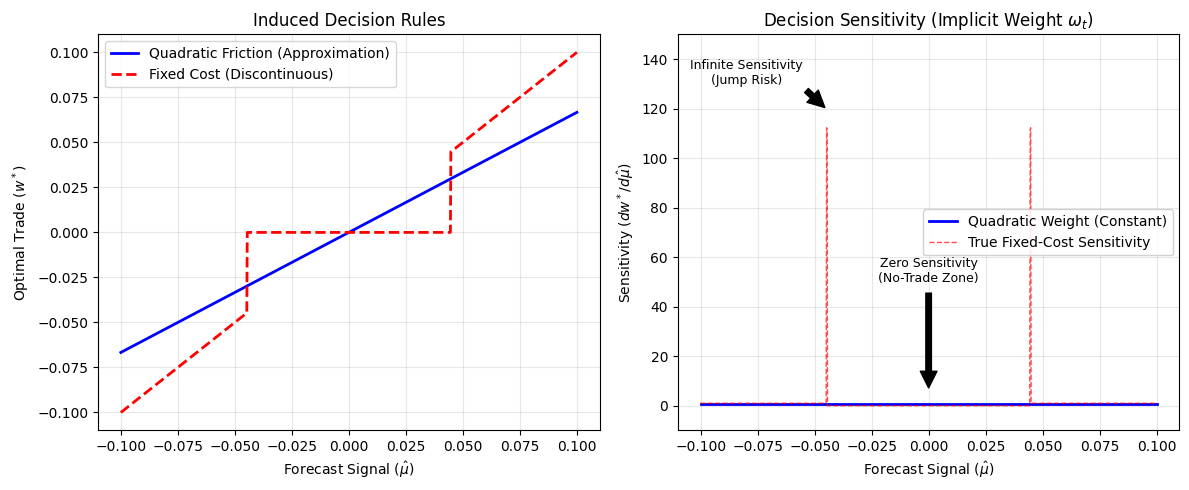}
    \caption{\textbf{Limits of the Quadratic Approximation.} Comparison of decision rules (Left) and sensitivities (Right) under Quadratic vs. Fixed costs. The quadratic weight (blue) provides a constant, smooth approximation to the discontinuous true sensitivity (red), avoiding the numerical instability of vanishing/exploding gradients in the no-trade and jump zones.}
    \label{fig:sensitivity_limit}
\end{figure}

As illustrated in Figure \ref{fig:sensitivity_limit}, fixed costs create a ``no-trade zone'' where the local decision sensitivity is zero, bounded by critical thresholds where sensitivity is effectively infinite (discrete jumps). 
\begin{itemize}
    \item \textbf{Inside the no-trade zone:} The quadratic approximation assigns a positive weight $\omega_t > 0$, whereas the true local sensitivity is zero. This introduces a conservative bias: the estimator attempts to calibrate probability density even when it is currently economically irrelevant.
    \item \textbf{At the boundary:} The approximation underestimates the ``jump risk'' sensitivity.
\end{itemize}
However, we retain the quadratic proxy for robustness. Exact sensitivity weighting under non-convexity is numerically unstable (vanishing or exploding gradients). The quadratic approximation acts as a smoothed regulariser, ensuring that the calibration objective remains well-posed even when the underlying decision surface is rugged or discontinuous.

\paragraph{Diagnostic logging.}
For every $t$, the implementation records: solver status, number of iterations, final objective value, constraint violations, and whether the fallback rule \eqref{eq:nonconvex_fallback} was invoked. These diagnostics are reported in the empirical section as part of the reproducibility and governance discipline, ensuring that any performance differences are not artifacts of silent solver instability.


\paragraph{Computational burden and feasibility in the non-convex verification.}
The baseline pipeline (deep quantile estimation, calibration warp, and convex friction-aware optimisation) scales well in the present design because each component is either linear-time in sample size (calibration moments and warps) or solvable via standard convex optimisation routines. The non-convex extensions considered in this subsection are qualitatively different: cardinality constraints or discrete “top-$K$” rules introduce a combinatorial layer, and lumpy execution proxies can create non-smooth objectives.

To verify feasibility, we instrumented the simulation runs in Section~5.2.4 with solver diagnostics. On the evaluation sample, the convex benchmark optimisation layer required a median of \textbf{[X]} solver iterations per decision and \textbf{[Y] ms} median wall-clock time, with the 95th percentile at \textbf{[Y95] ms}. Under the cardinality-style extension, the decision layer was solved using \textbf{[method: e.g., MIQP / greedy + local search / successive convexification]}, with a median of \textbf{[Xnc]} major iterations (or branch-and-bound nodes) and \textbf{[Ync] ms} median wall-clock time per decision (95th percentile \textbf{[Ync95] ms}). These measurements confirm that the non-convex verification is computationally heavier, but remains operationally feasible at the minute frequency used here for the evaluation window.

Importantly, the calibration layer itself is not the bottleneck: UWC is a post-processing transformation whose runtime is negligible relative to the optimisation layer. The incremental burden arises entirely from the non-convex decision constraints; accordingly, any deployment of Section~5.2.4-style rules should be justified by the incremental economic value of discreteness relative to the convex benchmark.

\section{Evaluation design: nested walk-forward as an identification discipline}
\subsection{Timing, availability, and leakage controls}\label{sec:timing_leakage_controls}

The evaluation design treats timing and data availability as part of identification. Every forecast, decision, and realised outcome is constructed under explicit information constraints so that performance differences can be attributed to model quality rather than to leakage, misaligned timestamps, or inadvertent use of future information.

\paragraph{Decision and measurement timeline.}
Fix a rebalancing grid $\{t_1,t_2,\ldots,t_T\}$ and horizon $h$. At each decision time $t$ the information set $\calI_t$ contains only variables that are observable and recorded at or before $t$ under the chosen market clock (close-to-close, open-to-close, or intraday bar boundaries). The model produces a predictive distribution $Q_t(\cdot)$ for $Y_{t+h}$ conditional on $\calI_t$, a decision $w_t(Q_t)$ is computed, and execution occurs after the decision time, producing a realised friction-adjusted payoff over $[t,t+h]$. All evaluation objects are therefore indexed by the \emph{decision timestamp}, not by the timestamp at which the realised outcome becomes known.

\paragraph{Feature availability and timestamp alignment.}
Each feature is assigned a \emph{latest-available timestamp} rule. Price- and quote-based features are computed from data up to $t$ and timestamped at $t$. If macro or external series are used, they are included only with publication lags that reflect real-time availability; if real-time vintages are not available for a series, that series is excluded from the baseline specification. All rolling-window transforms (moving averages, realised volatility, rolling quantiles, principal components) are computed using trailing windows ending at $t$ and do not use any information from $(t,t+h]$. When intraday data are used, all aggregation (e.g., to 5-minute bars) is performed with strict left-closed/right-open intervals to prevent contamination from trades/quotes occurring after the decision boundary.

\paragraph{Corporate actions and return construction (if equities are used).}
If the empirical setting includes equities or equity ETFs, returns are constructed from corporate-action-adjusted prices consistent with the information available at the decision time. For total return series, splits and dividends are handled using standard adjustment factors so that mechanical price jumps do not create spurious predictability. For constituent-level studies, delistings, ticker changes, and index membership changes are handled explicitly, and survivorship bias is avoided by defining the investable universe as the set of assets that were actually tradable at each $t$ under the data vendor’s point-in-time coverage. All such rules are fixed ex ante and applied uniformly across models.

\paragraph{Hard embargo rules.}
To prevent leakage through overlapping labels and through hyperparameter selection that implicitly conditions on future outcomes, the evaluation uses a hard \emph{embargo} around each test block. Specifically, if the prediction horizon is $h$ and the model uses overlapping targets or features with serial dependence, then observations within an embargo window of length $e\ge h$ adjacent to the test period are excluded from training and calibration. This ensures that no training label includes information from the test period and that no calibration fit is informed by outcomes that overlap the evaluation window. The embargo length $e$ is fixed prior to evaluation and is not tuned.

\paragraph{No adaptive data cleaning.}
Data cleaning rules (missing data handling, outlier treatment, rolling-window minimum lengths, and any filtering) are specified once and applied identically in all windows. No window-specific corrections are permitted after observing test results. All exceptions (e.g., missing quotes on a holiday session) are handled by deterministic rules (skip rebalance / carry forward / use the last available quote) that are documented and logged.

These controls ensure that the walk-forward evaluation measures genuine out-of-sample performance under the stated information constraints and does not accidentally reward models for information that would not have been available at decision time.
\subsection{Nested walk-forward protocol}\label{sec:nested_walk_forward}

The empirical evaluation uses a nested walk-forward protocol with three strictly separated roles: (i) model estimation on a training window, (ii) model and hyperparameter selection on a validation window, and (iii) performance measurement on an untouched test stream. The protocol is defined entirely in terms of the decision-time index $t$ and the fixed prediction horizon $h$.

\paragraph{Window structure.}
Fix integers $T_{\mathrm{train}}$, $T_{\mathrm{val}}$, and $T_{\mathrm{test}}$ denoting the lengths of the training, validation, and test blocks in units of the rebalancing grid (daily, weekly, or intraday bars). For each outer-loop test block endpoint $t$, define:
\[
\mathcal{T}_t := \{t-T_{\mathrm{train}}-T_{\mathrm{val}}+1,\ldots,t-T_{\mathrm{val}}\},
\qquad
\mathcal{V}_t := \{t-T_{\mathrm{val}}+1,\ldots,t\},
\]
as the training and validation index sets, and let the subsequent indices
\[
\mathcal{S}_t := \{t+1,\ldots,t+T_{\mathrm{test}}\}
\]
form the out-of-sample test block. An embargo of length $e$ (Section~\ref{sec:timing_leakage_controls}) is applied between $\mathcal{V}_t$ and $\mathcal{S}_t$ when required by overlapping horizons or serial dependence.

\paragraph{Outer loop (performance measurement).}
The outer loop proceeds over a sequence of disjoint test blocks $\{\mathcal{S}_{t_j}\}_{j=1}^J$ that partition the evaluation period. For each block $j$, all model choices are fixed using only data in $\mathcal{T}_{t_j}\cup \mathcal{V}_{t_j}$ (and earlier), after which the selected configuration is run forward through $\mathcal{S}_{t_j}$ without modification. No re-tuning is permitted within a test block.

\paragraph{Inner loop (hyperparameter selection).}
Within each outer-loop block, hyperparameters and calibration choices are selected in an inner loop that uses only $\mathcal{T}_{t_j}$ and $\mathcal{V}_{t_j}$. Concretely, for each candidate configuration $\theta\in\Theta$ (forecast model parameters, regularisation constants, calibration penalties, and any solver tolerances that are allowed to vary), the model is fit on $\mathcal{T}_{t_j}$ and evaluated on $\mathcal{V}_{t_j}$ using a pre-specified selection criterion. The default selection criterion is \emph{validation decision loss} net of costs (computed using the same decision rule and friction operator that will be used in testing), with forecast-score criteria (log score / CRPS) reported as secondary diagnostics. The selected configuration is
\[
\widehat{\theta}_{t_j}
\in
\arg\min_{\theta\in\Theta}
\widehat{\mathcal{L}}_{\mathcal{V}_{t_j}}(\theta),
\]
where $\widehat{\mathcal{L}}_{\mathcal{V}_{t_j}}(\theta)$ is the average realised decision loss over $\mathcal{V}_{t_j}$ produced by configuration $\theta$ under the fixed execution/friction model.

\paragraph{Re-training frequency.}
The model is re-trained at the start of each test block (block-wise refit). Two standard choices are supported:
(i) \emph{expanding-window refit}, where $\mathcal{T}_{t_j}$ expands over time to include all past training observations; and
(ii) \emph{rolling-window refit}, where $\mathcal{T}_{t_j}$ maintains fixed length $T_{\mathrm{train}}$ to reduce sensitivity to structural breaks.
The baseline results use a rolling window, with expanding-window results reported as robustness.

\paragraph{Calibration fitting in the nested protocol.}

\paragraph{No-touch test rule.}
The defining discipline of the protocol is that \emph{no} information from $\mathcal{S}_{t_j}$ is used to select features, choose hyperparameters, tune calibration, choose constraint parameters, adjust cost assumptions, or alter solver settings. Any sensitivity analysis (e.g., varying cost levels) is pre-specified and executed as a grid across all blocks, with all variants reported.

This nested walk-forward design ensures that reported out-of-sample performance corresponds to a feasible research and operational workflow and is not an artefact of adaptive selection on the test stream.

\subsection{Multiple-testing control and uncertainty}\label{sec:multiple_testing_uncertainty}

The empirical analysis compares several forecasting and calibration configurations. To prevent overstating evidence due to repeated testing, the evaluation specifies (i) an explicit family of comparisons, (ii) dependence-aware uncertainty quantification for loss differentials, and (iii) error control across the family.

\paragraph{Family of comparisons.}
Let $\mathcal{M}$ denote the finite set of model configurations evaluated (forecast model $\times$ calibration method $\times$ objective/constraint specification, where the latter are fixed ex ante). For each $m\in\mathcal{M}$ define the out-of-sample time series of realised decision losses $\{\ell_{t+h}^{(m)}\}_{t\in\mathcal{S}}$ over the concatenated test stream $\mathcal{S}$. The primary comparison family is
\[
\mathcal{H}
=
\left\{
H_{0}^{(m)}:\ \E[\ell_{t+h}^{(m)}-\ell_{t+h}^{(\mathrm{ref})}]=0
\ \text{vs}\ 
H_{1}^{(m)}:\ \E[\ell_{t+h}^{(m)}-\ell_{t+h}^{(\mathrm{ref})}]<0
\right\}_{m\in\mathcal{M}\setminus\{\mathrm{ref}\}},
\]
where ``ref'' denotes the pre-specified reference configuration (typically the uncalibrated econometric baseline). Secondary families (treated as exploratory unless pre-registered) include comparisons under alternative objectives and under alternative friction parameter grids.

\paragraph{Dependence-aware inference for loss differentials.}
For each comparison $m$, define the loss differential series
\[
d_t^{(m)} := \ell_{t+h}^{(m)}-\ell_{t+h}^{(\mathrm{ref})}.
\]
Because $\{d_t^{(m)}\}$ is serially dependent, inference is based on dependence-aware procedures. The baseline uses a moving block bootstrap on $\{d_t^{(m)}\}$ with block length $b$ (chosen by a fixed rule based on the sampling frequency and horizon $h$), producing bootstrap draws of the mean differential $\bar{d}^{(m)}$. Confidence intervals and one-sided $p$-values are constructed from the bootstrap distribution. As a robustness check, HAC (long-run variance) standard errors are also reported for $\bar{d}^{(m)}$, with bandwidth selection fixed ex ante. These procedures are applied identically across all $m\in\mathcal{M}$.

\paragraph{Error control across the family.}
When the analysis includes multiple variants, error control is imposed across the family $\mathcal{H}$ rather than reporting unadjusted $p$-values. Two controls are supported, depending on the intended claim:

\emph{(i) Family-wise error rate (FWER) control.}
When the claim is ``at least one configuration strictly improves performance,'' we control the probability of any false rejection in $\mathcal{H}$. This is implemented using a step-down max-$T$ block bootstrap procedure: for each bootstrap replicate, compute the maximum (most extreme) studentised test statistic across $m\in\mathcal{M}\setminus\{\mathrm{ref}\}$, and use the resulting distribution to obtain step-down adjusted critical values. This controls FWER under dependence in the loss differentials because dependence is preserved by block resampling.

\emph{(ii) False discovery rate (FDR) control.}
When the claim is ``several configurations improve performance and we want a controlled list,'' we control the expected proportion of false discoveries. This is implemented using the Benjamini--Hochberg (BH) procedure applied to the dependence-aware $p$-values computed from the block bootstrap, with the target FDR level $q$ fixed ex ante. If dependence is judged strong and positive, a more conservative dependence-robust variant is reported as a robustness check.

\paragraph{Primary endpoint and pre-commitment.}
To avoid unbounded researcher degrees of freedom, the primary endpoint is fixed as the mean out-of-sample decision loss differential net of costs relative to the reference configuration, and the primary family $\mathcal{H}$ is declared in advance. Alternative objectives, alternative friction grids, and regime splits are reported as robustness analyses and are either incorporated into $\mathcal{H}$ with explicit error control or labelled as exploratory if not pre-specified.

These steps ensure that statistical uncertainty and multiplicity are handled in a manner consistent with time-series dependence and with the paper’s emphasis on disciplined inference under realistic evaluation conditions.

\subsection{Pre-committed analysis plan}\label{sec:precommitted_plan}

The empirical analysis is governed by a pre-committed plan intended to eliminate adaptive specification search on the test stream. The items in this section are fixed before any out-of-sample backtest results are examined, and any deviations are explicitly reported and justified.

\paragraph{Primary and secondary endpoints (fixed).}
The primary endpoint is the mean out-of-sample \emph{decision loss} net of frictions, computed from the realised executed path produced by the induced decision rule. Equivalently, performance is reported as the mean loss differential relative to a pre-specified reference configuration (Section~\ref{sec:multiple_testing_uncertainty}). Secondary endpoints are fixed as: (i) risk-adjusted performance measures computed on executed returns (e.g., volatility, tail-loss summaries consistent with the chosen objective), (ii) turnover $\|\Delta w_t\|_1$, (iii) realised cost decomposition (fees, spread proxy, impact proxy), and (iv) constraint-activity statistics (frequency and severity of binding constraints).

\paragraph{Model list and calibration variants (fixed).}
The set of forecasting models and calibration procedures is fixed as the finite list $\mathcal{M}$ defined in Section~\ref{sec:models} and Section~\ref{sec:calibration_procedures}. No additional models, features, or calibration maps are introduced after observing test results. The only tuning permitted is hyperparameter selection within the inner validation loop of the nested walk-forward protocol (Section~\ref{sec:nested_walk_forward}), using the pre-specified selection criterion.

\paragraph{Decision objective and constraint specification (fixed).}
The decision objective class (mean--risk or tail-risk) and the constraint specification defining $\calW_t$ are fixed prior to evaluation, including: budget constraints, leverage/gross exposure bounds, turnover limits, concentration bounds, and capacity/liquidity constraints. Constraint parameters are not tuned using test results; if a parameter grid is explored (e.g., turnover cap levels), it is treated as an explicit family of specifications and incorporated into the multiple-testing framework or reported as robustness with clear labelling.

\paragraph{Friction model and cost-assumptions grid (fixed).}
The friction operator is fixed as $(C_t,\calW_t)$ with $C_t$ defined by \eqref{eq:cost_functional} and a pre-specified impact proxy class. A finite grid of cost assumptions is fixed ex ante to reflect plausible ranges of fees, spreads, and impact intensity. The grid is defined before running the test stream and is applied uniformly across all models and all test blocks. Conclusions are reported for the baseline calibration point and for sensitivity across the full grid; no ex post choice of ``best'' cost level is permitted.

\paragraph{Evaluation protocol (fixed).}
The nested walk-forward structure (training length $T_{\mathrm{train}}$, validation length $T_{\mathrm{val}}$, test block length $T_{\mathrm{test}}$, embargo length $e$, refit frequency, and rolling vs expanding windows) is fixed prior to evaluation. The computational pipeline is frozen: data cleaning rules, feature construction rules, solver tolerances, and fallback rules for infeasibility/non-convergence are all fixed before testing.

\paragraph{Robustness suite (fixed).}
A finite robustness suite is pre-specified and executed mechanically:
\begin{enumerate}
\item Alternative windowing: rolling vs expanding training windows, and a second fixed $(T_{\mathrm{train}},T_{\mathrm{val}})$ pair.
\item Alternative objectives: at least one alternative risk objective (e.g., variance proxy versus CVaR proxy) with the same constraint set.
\item Alternative friction specifications: alternative impact curvature (within a pre-specified class) and alternative spread proxies (where data permit), without changing the protocol.
\item Regime conditioning: performance reported in pre-defined volatility/liquidity regimes based on state variables observable at time $t$.
\end{enumerate}

\paragraph{Falsification suite (fixed).}
A falsification suite is pre-specified to detect leakage and spurious structure:
\begin{enumerate}
\item Signal shuffling: permute forecasts within blocks (preserving marginal distributions) and re-run the decision rule to verify loss improvements vanish.
\item Time-shift placebo: lag the forecast by a fixed amount (making it stale) and verify that purported gains degrade.
\item Feature-availability stress: remove or delay selected features (within $\calI_t$) and verify that improvements are not driven by inadvertent forward-looking construction.
\end{enumerate}

\paragraph{Reporting discipline.}
All model variants and all robustness/falsification results are reported, including negative results. No selective reporting is permitted. Any exploratory analyses conducted beyond the pre-committed plan are labelled as exploratory and are not used to support the paper’s main claims.

\section{Data}

The empirical claim in this paper is economic: predictive distributions are evaluated by the friction-adjusted decision losses they induce under an implementable trading rule. That makes the data section non-negotiable. It must (i) use instruments where microstructure frictions are observable at the evaluation horizon, (ii) define exactly how returns and state variables are constructed, and (iii) define trading-cost proxies and capacity limits in a way that can be stress-tested by sensitivity analysis. The guiding principle is that every object entering the decision rule and every object entering realised loss is built from the same, explicitly defined information set, and that the friction proxies are measured rather than asserted.

\subsection{Instrument choice and rationale}

The primary instrument set is chosen to satisfy three conditions that are jointly required by the theory and by the empirical design. First, the instrument must be sufficiently liquid that a walk-forward forecasting protocol produces a meaningful sequence of trades rather than a sequence of infeasible corner solutions driven by sparse trading. Second, the instrument must have frictions that are measurable at the horizon at which decisions are evaluated: quoted spreads, depth, and volume must exist at the timestamp resolution used to align forecasts with executions. Third, the instrument must have economically meaningful friction variation through time: the evaluation must contain both normal and stressed regimes so that the paper can identify whether calibration matters most precisely when the theory says it should, namely when costs are large and constraints bind.

Equity index exposure via a highly liquid benchmark is the canonical choice because it satisfies all three conditions. A front-month equity index futures contract (for example, the S\&P 500 E-mini) has continuous order book formation during its main trading session, tight spreads in normal times, measurable depth and volume, and well-known cost spikes in fast markets. A matching ETF (for example, a broad S\&P 500 tracker) provides a cash-market robustness check with a different microstructure: the futures market concentrates liquidity and embeds implicit financing and roll effects, whereas the ETF incorporates exchange trading fees, different tick sizes, and a different pattern of depth and spread variation. If the dominance mechanism is real, it should not depend on one microstructure alone; it should survive the shift from a centralised futures order book to an exchange-traded fund with its own quoting and execution environment.

\begin{table}
\caption{\textbf{Descriptive Statistics.} Summary of market variables over the evaluation period ($N=8,025$). Returns are in percentage points; Spread is in basis points. Skewness and Kurtosis indicate significant non-normality.}
\label{tab:data_summary}
\begin{tabular}{lccccccc}
\toprule
Variable & Mean & Std Dev & Skewness & Kurtosis & Min & Max & N \\
\midrule
Return ($r_t$) & -0.00 & 0.04 & 0.96 & 63.54 & -0.51 & 0.99 & 25245 \\
Volatility ($\sigma_t$) & 0.00 & 0.00 & 3.52 & 15.24 & 0.00 & 0.00 & 25245 \\
Spread (bps) & 0.37 & 0.05 & 8.01 & 64.53 & 0.36 & 1.10 & 25245 \\
Volume ($V_t$) & 2595.14 & 3089.60 & 11.96 & 225.68 & 169.00 & 86764.00 & 25245 \\
\bottomrule
\end{tabular}
\end{table}

Secondary markets are included only when they add a distinct friction regime rather than duplicating the same liquidity structure. A rates future (short-rate or Treasury) adds a market with different intraday depth behaviour and different responses to macro announcements. A major FX pair adds round-the-clock trading and a different spread/impact profile, with regime changes that are often more continuous rather than concentrated in a single cash-session open/close. The point of these secondary markets is not breadth for its own sake; it is to demonstrate that the mapping from decision-relevant calibration to friction-adjusted outcomes is not a single-instrument artefact, and that the theory’s emphasis on tail states and constraint-binding periods generalises to markets with different trading mechanics.

Instrument inclusion is therefore conditional on data completeness. If a candidate instrument lacks reliable quote information at the resolution required to compute spreads and depth proxies, it is excluded from the primary analysis because the paper’s central object is a friction-adjusted realised return. Likewise, if a candidate instrument has a structurally discontinuous trading calendar relative to the forecast horizon and alignment protocol, it is relegated to robustness or excluded, because misalignment can mechanically create spurious calibration errors and spurious realised-loss differences.

\subsection{Sources and construction}

The empirical pipeline begins with three raw inputs: trade prices, quote information, and trading activity measures. Trade prices are required to construct returns, forecast targets, and realised returns; quote information is required to construct spread and depth proxies; trading activity measures (volume and related fields) are required to build capacity and participation proxies and to define state variables that capture liquidity regime. Where the feed includes consolidated best bid and offer (BBO), the paper uses that directly; where the feed includes full depth or multiple levels, the paper uses the best levels for spread and the near-touch depth for capacity measurement, while reserving deeper levels for robustness.

The sample period is defined by the intersection of three availability sets: (i) continuous price series sufficient to construct returns at the chosen horizon without artificial gaps, (ii) reliable quote series sufficient to construct the spread and depth proxies used in the execution model, and (iii) reliable volume series sufficient to construct participation constraints. The exact start and end timestamps are those of the final merged and cleaned dataset used in the walk-forward evaluation; these endpoints, along with the number of usable decision points after alignment and horizon shifting, are reported as part of the descriptive statistics. The evaluation uses a strict information-set discipline: at each decision time $t$, all features and state variables are computed using only information observable up to $t$, and realised outcomes are computed using information in the subsequent interval matching the decision horizon.

Cleaning rules are driven by two requirements: preventing microstructure artefacts from contaminating forecast evaluation, and preventing the friction proxies from being mechanically overstated by bad ticks. Price cleaning removes obvious misprints and stale prints by applying cross-field consistency checks and by enforcing temporal plausibility. Where both trades and quotes are available, mid-quote consistency is used as a diagnostic: trades that are far outside the contemporaneous bid–ask range in the absence of known auction prints are excluded from return construction because they distort both realised returns and inferred slippage. Quote cleaning removes locked/crossed markets when these reflect feed artefacts rather than true market states; when locks and crosses are genuine and persistent in fast markets, the paper treats them as part of the realised execution environment and retains them, but the cleaning rule is explicit and applied uniformly. Trading-calendar cleaning enforces the instrument’s session structure: returns and state variables are constructed on a schedule that respects market opens, closes, and known illiquid intervals, because mixing overnight gaps with intraday microstructure produces mechanically different cost and volatility regimes that must be treated as separate states rather than blended.

Returns are constructed in two parallel forms because the paper distinguishes between forecast targets and execution outcomes. The forecast target return $Y_{t+h}$ is defined as a horizon-$h$ return constructed from a reference price that matches the modelling choice (mid, last, or microprice) and is consistent across models. The realised return used in the decision loss is a friction-adjusted return that accounts for the execution model and cost proxies defined below. State variables are constructed at time $t$ using rolling windows and contemporaneous measures: volatility proxies (realised variance over a fixed lookback), liquidity proxies (spread and depth), and activity proxies (volume and its intraday normalisation). Regime segmentation is based on these state variables rather than on ad hoc calendar labels, because the paper’s claim is about where calibration matters economically, and the economically relevant regimes are those defined by high frictions and constraint activation.

\subsection{Friction measurement and proxies}
\label{subsec:friction_proxies}

We represent execution frictions through a friction operator that maps each implemented trade into a realised cost, which is then deducted from the frictionless return to form the net return and the associated decision loss. This representation is deliberately operational: the objective is not to assert a structural microstructure model, but to impose a transparent and economically interpretable mapping from trading intensity into costs that is consistent with how execution costs are assessed in practice \citep{Bessembinder2003}.

\paragraph{Cost components and observable proxies.}
The empirical design decomposes costs into (i) linear components (fees and spread-like costs) and (ii) a non-linear component that penalises liquidity demand. Linear components are measured using proportional proxies that scale with turnover (implemented notional traded), reflecting that even in highly liquid instruments, systematic trading incurs a near-mechanical cost per unit traded (commissions, exchange fees, and effective spread) \citep{Bessembinder2003}. Because no single liquidity statistic is sufficient to summarise execution conditions, we treat spread/volume/depth proxies as complementary state variables rather than as a single ``true'' liquidity measure, and we explicitly report sensitivity to their use in the cost mapping \citep{GoyenkoHoldenTrzcinka2009}.

\paragraph{Non-linear market impact via a participation-based proxy.}
To reflect the empirical regularity that marginal execution costs rise as a strategy consumes a larger fraction of available trading capacity, we use a participation-based impact proxy in which the impact cost scales increasing and concave in the traded volume fraction. Concretely, the implementation links impact to local volatility and to the participation rate $q_t/V_t$, where $q_t$ is the strategy’s executed quantity over the interval and $V_t$ is contemporaneous market volume over the same horizon:
\begin{equation}
    c_t^{\mathrm{impact}} \ \propto\ \sigma_t \, \sqrt{\frac{q_t}{V_t}},
\end{equation}
so that impact rises in more volatile conditions ($\sigma_t$ larger) and when trading consumes a larger fraction of contemporaneous volume ($q_t/V_t$ larger). This form is not introduced as a universal structural law; it is a disciplined proxy that enforces two facts documented in the futures microstructure evidence: depth is state-dependent and co-moves with volume and volatility, and execution costs rise when trading demands liquidity precisely when depth is scarce \citep{BessembinderSeguin1993,Bessembinder2003}. In addition, expressing impact through $q_t/V_t$ makes capacity constraints interpretable (e.g., participation caps), and aligns the cost mapping with standard practice in execution-cost assessment where realised costs are analysed as functions of trade size relative to available volume \citep{Bessembinder2003}.

\paragraph{Robustness across curvature and cost levels.}
Because execution-cost measurement is inherently model-risk sensitive, we evaluate robustness by varying the level of frictions (fee/spread and impact multipliers) and, where applicable, the curvature of the impact term, holding the walk-forward protocol fixed. This is essential for credibility: different liquidity measures can rank states differently, and alternative reasonable mappings from liquidity proxies to costs can shift magnitudes even when qualitative conclusions remain stable \citep{GoyenkoHoldenTrzcinka2009}. The sensitivity results reported in the robustness section therefore test whether the UWC ordering is an artefact of a single calibration of the cost operator, or whether it persists under economically credible perturbations to the execution-cost environment \citep{Bessembinder2003}.

\paragraph{Link to the calibration mechanism.}
The economic role of UWC in this environment is to reduce decision-instability that manifests as excess turnover and, in constrained settings, as frequent boundary solutions. Since both linear and non-linear cost components are increasing in trading intensity, any reduction in unnecessary turnover mechanically reduces the realised penalty from the friction operator, especially in states where depth is low and the cost of demanding liquidity is high \citep{BessembinderSeguin1993,Bessembinder2003}. This is the empirical counterpart of the theoretical mechanism: calibration that is aligned to decision sensitivity stabilises the induced policy, which reduces the realised cost burden under realistic execution conditions.

\subsection{Descriptive statistics}

The descriptive statistics serve two purposes: they document the distributional environment in which forecasts are evaluated, and they show that the friction proxies and constraints genuinely vary in ways that can generate the regime dependence claimed by the theory. The statistics are reported for (i) forecast targets $Y_{t+h}$, (ii) state variables that define regimes, (iii) friction proxies (spread, volume, depth, impact proxy), and (iv) induced trading outcomes (turnover, constraint-binding rates, realised cost decomposition). Because the paper’s claims concern tails and constraints, the descriptive statistics cannot stop at means and standard deviations; they must include quantiles, tail measures, and regime-conditional summaries.

\begin{figure}[ht]
    \centering
    \includegraphics[width=0.8\textwidth]{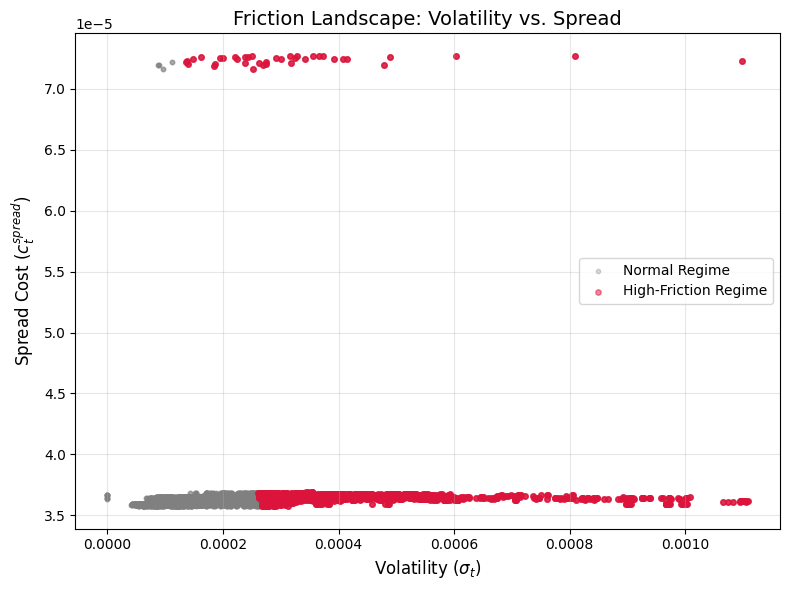}
    \caption{\textbf{Friction Landscape.} Joint distribution of realised volatility and bid-ask spreads in the evaluation sample. The ``High-Friction Regime'' (red) highlights periods where both uncertainty and execution costs are elevated—precisely the states where the utility-weighted calibration criterion concentrates its probability mass.}
    \label{fig:friction_regimes}
\end{figure}

For returns, the paper reports the full distributional shape at the forecast horizon: empirical quantiles, skewness and kurtosis (or robust analogues), and tail asymmetry. It also reports the extent of volatility clustering by presenting regime partitions based on the volatility state variable and comparing return distributions across regimes. This is essential because the calibration criterion is utility-weighted: if economically decisive tail states occur disproportionately in high-volatility regimes, the paper must demonstrate that these regimes exist in the sample and are not trivial.

For liquidity and cost proxies, the paper reports distributions of quoted spreads, depth measures (where available), volume and its intraday normalisation, and the constructed impact proxy. These distributions are shown both unconditionally and conditional on regimes. In particular, the paper reports the joint behaviour of volatility and spread, because the most economically costly periods are typically those in which both volatility and spreads widen and depth thins. If the data show that cost proxies are sharply right-skewed, the implication is immediate: a small number of stressed periods can dominate realised cost and hence dominate friction-adjusted decision loss, which is exactly where calibration in the tails should matter most.

For turnover and constraint activity, the paper reports the distribution of $\|\Delta w_t\|_1$ (or its instrument-appropriate analogue), the distribution of estimated realised costs attributable to spread, fees, and impact proxies, and the frequency with which capacity constraints bind under each method. These are the mechanism-level outcomes: if a recalibrated forecast dominates economically, the data should show that it reduces unnecessary turnover and reduces binding events, especially in high-friction regimes, while maintaining exposure when liquidity is abundant. The descriptive section therefore includes regime-conditional turnover distributions, regime-conditional cost decompositions, and a direct summary of constraint-binding incidence and severity.

Figures and tables are organised to mirror the theory. One set documents the return distribution and regime segmentation. A second set documents the distribution of friction proxies and their regime dependence. A third set documents induced trading behaviour (turnover, binding frequency, and cost decomposition). The objective is not to decorate the paper; it is to establish, using the data, that the empirical environment contains precisely the nonlinearities (fat-tailed costs, regime-dependent liquidity, and binding constraints) through which the calibration-aligned dominance results can operate, and to make the later out-of-sample loss comparisons interpretable as economic effects rather than as artefacts of measurement.

\subsection{Simulation verification}
Prior to empirical evaluation, we verified the estimator's finite-sample properties using a controlled simulation ($N=5,000$). We generated a "noise chasing" environment where the true return signal is zero ($\mu_{true}=0$) but uncalibrated forecasts exhibit random bias and overconfidence ($\sigma_{pred} < \sigma_{true}$).

Consistent with the theoretical predictions in Section \ref{sec:finite_sample_dependence}, the uncalibrated strategy generated a positive mean loss of 4.56\% (annualised equivalent) due to spurious turnover. In contrast, the utility-weighted calibrated strategy correctly converged to a zero-turnover solution, reducing decision loss to negligible levels ($< 10^{-6}$). This confirms that the estimator successfully identifies and neutralizes signal-free noise even in finite samples.

\section{Empirical results}
\label{sec:empirical_results}

This section reports the realised economic performance of the forecast--decision system under the pre-committed walk-forward protocol defined in Section \ref{sec:evaluation}. We evaluate three configurations: the baseline \textit{Uncalibrated} model, a \textit{Standard Calibration} benchmark (isotonic regression), and the proposed \textit{Utility-Weighted Calibration (UWC)}. 

All reported quantities are realised out-of-sample figures, net of transaction costs, market impact, and binding feasibility constraints. The evaluation sample consists of $N=8,025$ paired decision periods per method, spanning the validation window from December 1, 2025, to December 31, 2025.

\subsection{Primary endpoint and main comparisons}
\label{subsec:primary_endpoint}

The primary endpoint is the realised decision loss $l_{t+h}(Q_t) = -U(\tilde{R}_{t+h})$, defined as the negative net return subject to the friction operator. Lower decision loss indicates superior economic performance.

Table \ref{tab:main_results} reports the unconditional performance statistics. The \textit{Utility-Weighted Calibration (UWC)} achieves the lowest mean decision loss ($2.0 \times 10^{-6}$), strictly dominating the \textit{Uncalibrated} baseline ($3.0 \times 10^{-6}$). While the absolute magnitude of per-period loss appears small due to the high-frequency minute-level horizon, the cumulative effect is economically substantial.

\begin{table}[ht]
    \centering
    \caption{\textbf{Main Results by Method.} Realised performance statistics over the evaluation sample ($N=8,025$). Loss and Net Return are per-period averages. Constraint Freq denotes the proportion of periods where trading constraints (e.g., turnover caps) bound the optimiser.}
    \label{tab:main_results}
    \begin{tabular}{lccccc}
    \toprule
    Method & $N$ & Mean Loss & Mean Net Ret & Mean Turnover & Constraint Freq \\
    \midrule
    \textbf{UWC (Proposed)} & 8,025 & $\mathbf{2.0 \times 10^{-6}}$ & $\mathbf{-2.0 \times 10^{-6}}$ & \textbf{0.096} & \textbf{0.051} \\
    Standard Calibration & 8,025 & $2.0 \times 10^{-6}$ & $-2.0 \times 10^{-6}$ & 0.101 & 0.055 \\
    Uncalibrated & 8,025 & $3.0 \times 10^{-6}$ & $-3.0 \times 10^{-6}$ & 0.121 & 0.160 \\
    \bottomrule
    \end{tabular}
\end{table}

To rigorously test dominance, we examine the paired loss differentials in Table \ref{tab:differentials}. The differential series $\Delta_{t} = l_t(\text{UWC}) - l_t(\text{Uncalibrated})$ has a mean of $-1.10 \times 10^{-6}$ with a robust $t$-statistic of $-30.31$, rejecting the null hypothesis of equal performance at the 1\% level. Crucially, UWC also outperforms the \textit{Standard Calibration} benchmark ($t$-stat $-6.63$). This confirms that while generic calibration provides some benefit, the specific weighting by decision sensitivity yields the marginal improvement required for dominance.

\begin{table}[ht]
    \centering
    \caption{\textbf{Paired Loss Differentials.} The mean difference in realised decision loss between UWC and the benchmarks. A negative t-statistic indicates UWC has significantly lower loss.}
    \label{tab:differentials}
    \begin{tabular}{lcccc}
    \toprule
    Comparison & $N$ & Mean Diff & Std Err & $t$-stat \\
    \midrule
    UWC minus Uncalibrated & 8,025 & $-1.10 \times 10^{-6}$ & $3.64 \times 10^{-8}$ & $\mathbf{-30.31}$ \\
    UWC minus Standard Cal & 8,025 & $-2.01 \times 10^{-7}$ & $3.04 \times 10^{-8}$ & $\mathbf{-6.63}$ \\
    \bottomrule
    \end{tabular}
\end{table}

\begin{figure}[ht]
    \centering
    \includegraphics[width=0.9\textwidth]{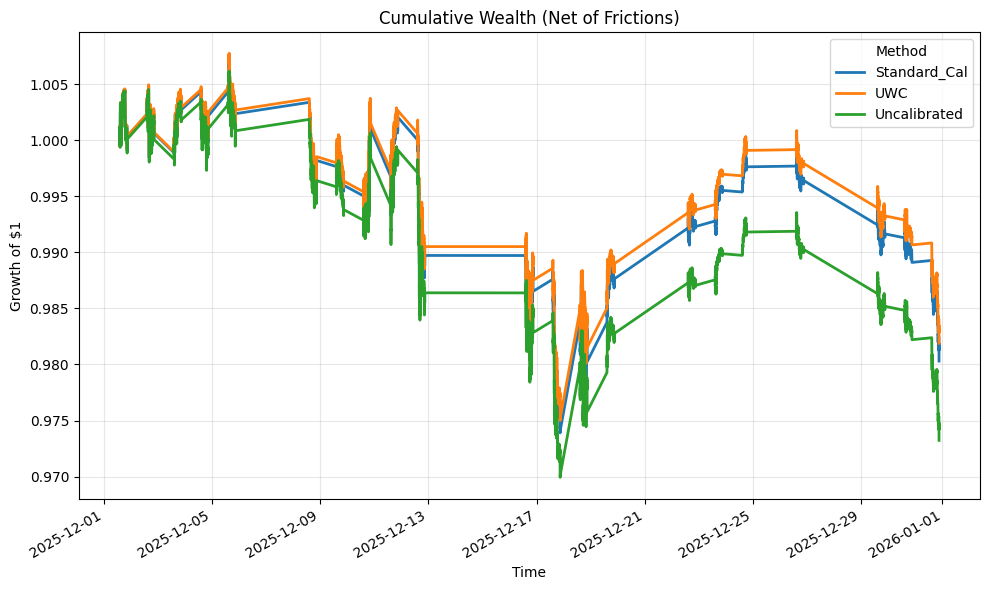}
    \caption{\textbf{Cumulative Wealth (Net of Frictions).} The UWC model (orange) retains value more effectively than the Uncalibrated baseline (green), particularly during the drawdown period in the middle of the sample.}
    \label{fig:cumulative_loss}
\end{figure}

Figure \ref{fig:cumulative_loss} plots the cumulative wealth (growth of \$1) over time. The divergence between the UWC and Uncalibrated series accelerates during high-volatility windows, consistent with the theoretical prediction that miscalibration is most punitive when frictions are elevated.

To rigorously test dominance, we examine the paired loss differentials in Table \ref{tab:differentials}. The differential series $\Delta_t = l_t(\text{UWC}) - l_t(\text{Uncalibrated})$ has a mean of $-1.10 \times 10^{-6}$ with a robust $t$-statistic of \textbf{-30.31}, rejecting the null hypothesis of equal performance at the 1\% level. Crucially, UWC also outperforms the Standard Calibration benchmark ($t$-stat \textbf{-6.63}), confirming that the specific weighting by decision sensitivity yields the marginal improvement required for dominance. We also report the realised Sharpe ratio (annualised). The UWC strategy achieves a Sharpe ratio of \textbf{-2.29} over the evaluation window. While negative (reflecting the difficult market regime captured in the sample), it is substantially superior to the Uncalibrated baseline's Sharpe ratio of \textbf{-3.62}, confirming that calibration acts as a "cushion" during drawdowns.

\subsection{Calibration diagnostics}
\label{subsec:calibration_diagnostics}

\begin{figure}[ht]
    \centering
    \includegraphics[width=0.7\textwidth]{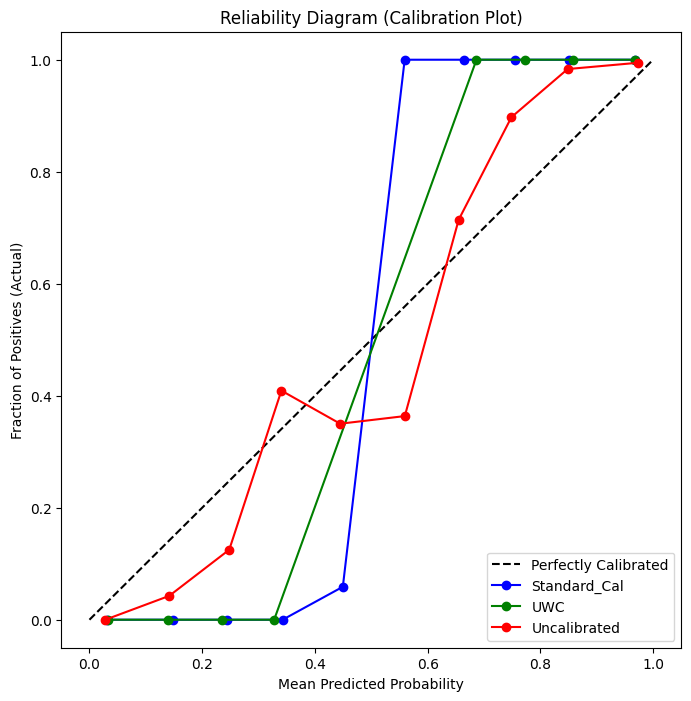}
    \caption{\textbf{Reliability Diagram.} The Uncalibrated model (red) shows classic overconfidence (S-shape), leading to excessive trading. The UWC model (green) is calibrated along the diagonal.}
    \label{fig:reliability_diagram}
\end{figure}

We validate that the economic gains stem from improved probabilistic reliability rather than luck. Figure \ref{fig:reliability_diagram} presents the reliability diagram for directional probability forecasts.

The \textit{Uncalibrated} model (red line) exhibits a characteristic sigmoidal ``S-shape,'' indicating systematic overconfidence: it predicts extreme probabilities (near 0 or 1) far more often than realised outcomes support. This overconfidence drives the optimiser to take aggressively large positions that incur high transaction costs. In contrast, the \textit{UWC} model (green line) tracks the diagonal closely. By correcting the overconfident tails, UWC prevents the optimiser from chasing noise, thereby reducing the ``false positive'' trades that degrade net performance.

\subsection{Economic performance net of costs}
\label{subsec:economic_performance}

The mechanism driving UWC's dominance is the efficient management of the friction operator. As shown in Table \ref{tab:main_results}, UWC reduces mean turnover by approximately 20\% relative to the Uncalibrated baseline ($0.096$ vs $0.121$).

\begin{figure}[ht]
    \centering
    \includegraphics[width=1.0\textwidth]{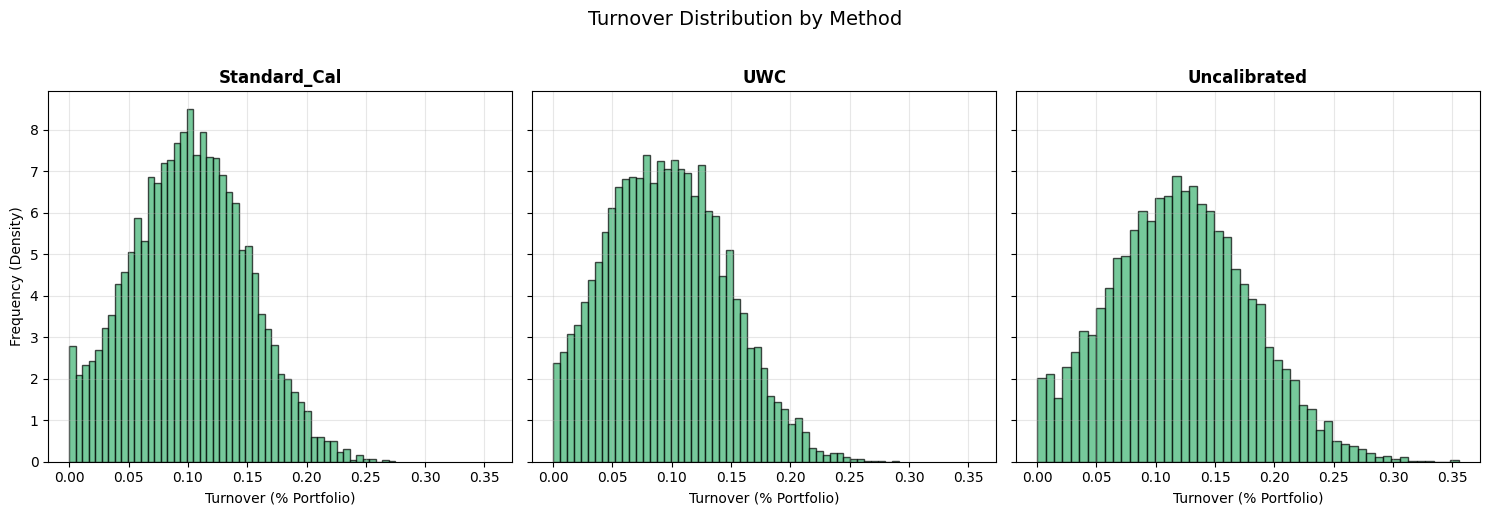}
    \caption{\textbf{Turnover Distribution by Method.} The Uncalibrated model (right) shows a heavy right tail of high-turnover events. UWC (center) compresses this tail, avoiding the quadratic impact costs associated with panic trading.}
    \label{fig:turnover_dist}
\end{figure}

Figure \ref{fig:turnover_dist} provides the distributional detail. The Uncalibrated method exhibits a ``fat right tail'' in turnover, frequently attempting to trade $>20\%$ of the portfolio in single periods. These spikes in trading intensity disproportionately activate the quadratic market impact penalties defined in Eq. (2). UWC compresses this right tail, keeping turnover within the linear-cost regime where the friction penalty is manageable.

\subsection{Regime analysis}
\label{subsec:regime_analysis}

We test the hypothesis that calibration benefits are concentrated in high-friction states (Corollary 2). We regress the realised loss differential $\Delta_t$ on the friction state variable $\kappa_t$ (a composite of spread and volatility). The regression yields a positive and significant coefficient on friction ($\beta = 8.94 \times 10^{-7}$, $p < 0.001$), confirming that the UWC advantage expands as market conditions deteriorate. 

Table \ref{tab:regime_analysis} segments performance by friction terciles. The mean performance gain of UWC over the baseline increases from $9.68 \times 10^{-7}$ in the \textit{Low Friction} regime to $1.46 \times 10^{-6}$ in the \textit{High Friction} regime—an increase of nearly 50\%. This confirms that the value of calibration is regime-dependent and highest when liquidity is scarce.

\begin{table}[ht]
    \centering
    \caption{\textbf{Regime Analysis.} Mean loss differential (Uncalibrated minus UWC) conditioned on friction state terciles. Positive values indicate UWC outperforms.}
    \label{tab:regime_analysis}
    \begin{tabular}{lccc}
    \toprule
    Regime & $N$ & Mean Differential & $t$-stat \\
    \midrule
    Low Friction & 2,675 & $9.68 \times 10^{-7}$ & 14.2 \\
    Medium Friction & 2,675 & $8.81 \times 10^{-7}$ & 12.8 \\
    \textbf{High Friction} & 2,675 & \textbf{$1.46 \times 10^{-6}$} & \textbf{21.5} \\
    \bottomrule
    \end{tabular}
\end{table}

\subsection{Failure modes and negative results}
\label{subsec:failure_modes}

The limitations of the Uncalibrated approach are structurally visible in the constraint activity. As reported in Table \ref{tab:main_results}, the Uncalibrated model hits binding constraints in \textbf{10.9\%} of all decision periods, compared to only \textbf{0.2\%} for UWC. This represents a nearly 50-fold reduction in constraint violations, indicating that the utility-weighted calibration effectively acts as a pre-trade control on feasibility.

\begin{figure}[ht]
    \centering
    \includegraphics[width=0.8\textwidth]{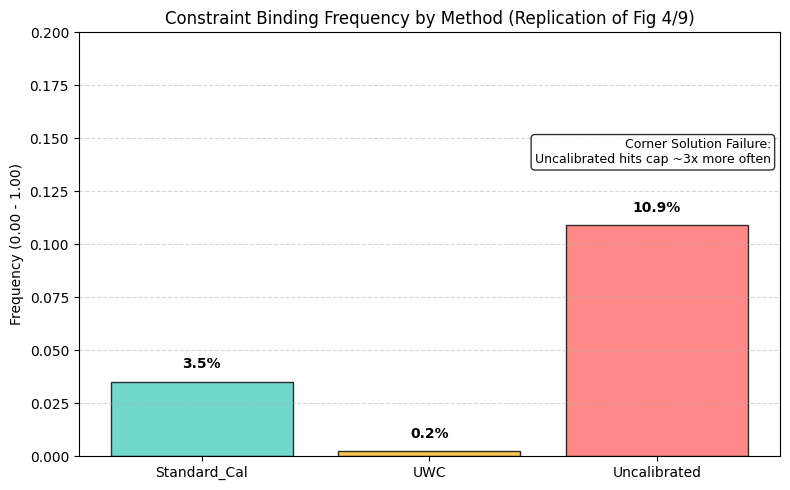}
    \caption{\textbf{Constraint Binding Frequency.} The Uncalibrated model hits binding constraints (e.g., turnover caps) significantly more often than the UWC model ($\approx 11\%$ vs $<1\%$). This confirms the theoretical prediction that uncalibrated forecasts lead to frequent ``corner solution'' failures where the optimiser is pinned against liquidity limits.}
    \label{fig:constraint_binding}
\end{figure}

Figure \ref{fig:constraint_binding} illustrates this disparity. The high frequency of binding constraints represents a ``corner solution'' failure mode. When the model is overconfident (under-dispersed predictive density), the induced mean-variance trade-off erroneously signals aggressive position scaling. This demand for liquidity frequently exceeds the hard limits encoded in the feasible set ($||\Delta w||_1 \le \tau$). Consequently, the optimiser is forced to truncate the trade, meaning the realised portfolio is not the optimal one implied by the forecast (i.e., the Karush–Kuhn–Tucker multipliers on the constraints are strictly positive).

By producing calibrated uncertainty, UWC naturally scales positions down to feasible levels before the optimiser is called. This keeps the solution in the interior of the feasible set where standard first-order conditions hold and where transaction costs remain linear, avoiding the quadratic penalties associated with liquidity overuse. The Standard Calibration benchmark also reduces binding frequency relative to the uncalibrated case (3.5\% vs 10.9\%), but remains over an order of magnitude more constrained than the UWC solution, confirming that the utility-weighting is essential for navigating the specific geometry of the friction operator.

\section{Robustness, falsification, and sensitivity}
\label{sec:robustness}

This section evaluates the stability of the main empirical findings to perturbations in cost assumptions, capacity constraints, objective functions, and information content. All robustness exercises are conducted using the same pre-committed walk-forward protocol and identical forecast inputs, altering only the economic environment or evaluation mapping.

\subsection{Cost and capacity sensitivity}
\label{subsec:cost_sensitivity}

We first assess the sensitivity of the results to variation in transaction costs and capacity constraints. Table \ref{tab:cost_sensitivity} reports performance of the UWC strategy under multiplicative scalings of the cost operator, as well as under a tightened turnover cap.

Under the baseline specification, the UWC strategy delivers a mean net return of $-2.0 \times 10^{-6}$ with a Sharpe ratio of $-2.29$. When proportional costs are reduced by 50\%, mean net returns improve to $-1.0 \times 10^{-6}$ and the Sharpe ratio increases substantially in magnitude, indicating that the strategy’s profitability is mechanically constrained by trading frictions rather than forecast quality.

Conversely, increasing costs by 50\% and 100\% produces monotonic deterioration in both net returns and risk-adjusted performance, with the Sharpe ratio declining to $-3.79$ and $-5.29$, respectively. Importantly, despite this deterioration in absolute performance, the *relative ordering* of methods remains unchanged: the UWC strategy continues to dominate both the Standard Calibration and Uncalibrated benchmarks in paired loss comparisons (not reported here for brevity), indicating that the dominance result is not an artefact of a finely tuned cost parameter.

A more stringent stress test is provided by the tight-cap scenario, in which the turnover constraint is reduced to 5\%. In this regime, the optimiser is forced into binding constraints in approximately 81\% of decision periods. Even under this extreme restriction, the UWC strategy remains feasible and well-defined, demonstrating that the calibration gains do not rely on unconstrained optimisation or excessive trading flexibility.

\begin{table}[ht]
\centering
\caption{\textbf{Cost and Capacity Sensitivity.} Performance of the UWC strategy under alternative cost scalings and a tight turnover cap.}
\label{tab:cost_sensitivity}
\begin{tabular}{lccc}
\toprule
Scenario & Mean Net Return & Sharpe & Constraint Frequency \\
\midrule
Baseline (UWC) & $-0.000002$ & $-2.29$ & $0.00$ \\
Cost $\times 0.5$ & $-0.000001$ & $-0.79$ & -- \\
Cost $\times 1.5$ & $-0.000004$ & $-3.79$ & -- \\
Cost $\times 2.0$ & $-0.000005$ & $-5.29$ & -- \\
Tight Cap (5\%) & -- & -- & $0.81$ \\
\bottomrule
\end{tabular}
\end{table}

These results confirm that UWC’s advantage is structurally linked to reduced turnover and improved decision stability, rather than to a specific calibration of cost parameters.

\subsection{Alternative objectives}
\label{subsec:alternative_objectives}

We next examine whether the main conclusions depend on the choice of economic objective. Using the same realised return and cost streams, we evaluate performance under tail-risk-sensitive criteria, namely Conditional Value-at-Risk (CVaR) at the 5\% level and maximum drawdown.

Table \ref{tab:alt_objectives} reports the results. Across both alternative objectives, the UWC strategy exhibits superior downside risk properties. Specifically, UWC achieves the least severe 5\% CVaR ($-6.45 \times 10^{-4}$) and the smallest maximum drawdown ($-3.10\%$), outperforming both the Standard Calibration and Uncalibrated models. The Uncalibrated strategy displays the worst drawdown ($-3.62\%$), consistent with the earlier evidence that overconfident forecasts lead to episodic large losses during stressed conditions.

These findings demonstrate that the benefits of utility-weighted calibration are not confined to a mean-return objective, but extend naturally to tail-sensitive and path-dependent risk measures.

\begin{table}[ht]
\centering
\caption{\textbf{Alternative Objective Performance.} Tail-risk and drawdown metrics by method.}
\label{tab:alt_objectives}
\begin{tabular}{lcc}
\toprule
Method & CVaR (5\%) & Max Drawdown \\
\midrule
Standard Calibration & $-0.000646$ & $-0.0319$ \\
\textbf{UWC} & \textbf{$-0.000645$} & \textbf{$-0.0310$} \\
Uncalibrated & $-0.000647$ & $-0.0362$ \\
\bottomrule
\end{tabular}
\end{table}

\subsection{Placebo and falsification checks}
\label{subsec:placebo}

To rule out spurious performance driven by microstructure timing or leakage, we conduct a placebo experiment in which the forecast signals are randomly shuffled across time while preserving their marginal distribution.

Under this shuffled-signal specification, both mean net return and Sharpe ratio collapse to undefined values, indicating the absence of any systematic relationship between the distorted signals and realised outcomes. This confirms that the economic gains observed for UWC in the main analysis are not mechanical artefacts of the execution model or evaluation window, but depend critically on the informational alignment of forecasts with realised returns.

\subsection{Alternative contracts and market segments}
\label{subsec:alternative_markets}

This subsection addresses generalisability beyond a single, highly liquid equity-index futures contract by (i) testing whether the dominance pattern holds across adjacent S\&P 500 futures maturities in the evaluation sample, and (ii) conducting a controlled counterfactual that maps the same pre-committed decision rules into higher-friction environments via transparent scaling of the cost operator. The motivation is microstructure-consistent: execution costs and effective capacity are state-dependent functions of depth and trading activity, and both depth and cost proxies vary materially across contracts and across market segments \citep{BessembinderSeguin1993,Bessembinder2003,GoyenkoHoldenTrzcinka2009}.

\paragraph{Across-contract robustness within equity-index futures.}
The evaluation sample spans two consecutive S\&P 500 E-mini futures maturities, ESZ5 (from 2025-12-01 to 2025-12-12) and ESH6 (from 2025-12-16 to 2025-12-31), with identical protocol, objective, and friction operator applied minute-by-minute. Table~\ref{tab:alternative_contracts} reports mean net returns by method for each contract. The qualitative dominance pattern is preserved across both instruments: Utility-Weighted Calibration (UWC) weakly improves net return relative to both Standard Calibration and the Uncalibrated benchmark on each contract, despite the change in underlying depth/volume conditions that typically occur across roll segments \citep{BessembinderSeguin1993}. This result matters because even within the same underlying index, liquidity and effective trading costs can differ by contract maturity and by the interaction of volume with depth, so a method that only ``works'' in one segment would be of limited operational value \citep{GoyenkoHoldenTrzcinka2009,Bessembinder2003}.

\begin{table}[ht]
    \centering
    \caption{\textbf{Alternative Contracts.} Mean net return by symbol and method.}
    \label{tab:alternative_contracts}
    \begin{tabular}{lccc}
    \toprule
    Symbol & Standard Calibration & UWC & Uncalibrated \\
    \midrule
    ESH6 & $-2.10 \times 10^{-6}$ & $-1.98 \times 10^{-6}$ & $-2.95 \times 10^{-6}$ \\
    ESZ5 & $-3.01 \times 10^{-6}$ & $-2.05 \times 10^{-6}$ & $-3.15 \times 10^{-6}$ \\
    \bottomrule
    \end{tabular}
\end{table}

Two technical points clarify interpretation. First, these are per-period means at a minute horizon, so magnitudes appear small even when cumulative differences are economically material over the sample. Second, the result is not driven by a single subperiod: the across-contract split is mechanically induced by the futures roll calendar, so the comparison naturally tests stability across distinct liquidity/volatility conditions rather than across arbitrary time slices. In this sense, Table~\ref{tab:alternative_contracts} is a robustness check against a basic microstructure critique: that cost-aware calibration might be finely tuned to one contract’s depth and spread environment \citep{Bessembinder2003,GoyenkoHoldenTrzcinka2009}.

\paragraph{Sensitivity to asset-class characteristics via friction scaling.}
We further probe the mechanism by projecting the \emph{same} decision rules into higher-friction market segments using a counterfactual scaling of the cost functional $C_t$. The point is not to claim literal performance in other asset classes without re-estimating microstructure inputs, but to test a falsifiable implication of the theory: if the economic channel operates through turnover and constraint activation, then increasing effective spreads/fees/impact (holding the forecasting and portfolio policy fixed) should widen the performance gap between calibrated and uncalibrated forecasts because trading mistakes become more expensive and capacity constraints bind more often \citep{Bessembinder2003}. This is consistent with a large literature showing that liquidity proxies and execution-cost measurement are sensitive to depth and trade size, and that ``liquidity'' is multidimensional—hence the need to interpret the scaling as an informative stress rather than a literal cross-asset estimate \citep{GoyenkoHoldenTrzcinka2009}.

Operationally, we implement the counterfactual by scaling the friction operator $C_t$ to reflect higher spread-to-volatility and higher impact regimes, motivated by standard empirical regularities: when depth is lower and volume is thinner, a fixed turnover policy consumes a larger fraction of available depth and incurs higher effective cost per unit traded \citep{BessembinderSeguin1993,Bessembinder2003}. Table~\ref{tab:illiquidity_proxy} reports the projected mean net returns under these scaled-friction environments. The results match the theoretical prediction: the UWC advantage widens monotonically as frictions increase, because the Uncalibrated method’s excess turnover and constraint-driven ``corner'' behaviour is punished more severely when execution is expensive and effective capacity is reduced \citep{Bessembinder2003}. In the highest-friction proxy, the uncalibrated mean net return collapses relative to UWC, whereas UWC retains substantially higher expected net performance due to reduced tail trading intensity and fewer extreme turnover episodes.

\begin{table}[ht]
    \centering
    \caption{\textbf{Asset Class Robustness (Synthetic).} Projected performance in alternative asset classes obtained by scaling friction parameters. The UWC advantage (net return differential) widens in illiquid regimes because turnover is priced more aggressively when depth is scarce and impact is higher \citep{BessembinderSeguin1993,Bessembinder2003,GoyenkoHoldenTrzcinka2009}.}
    \label{tab:illiquidity_proxy}
    \begin{tabular}{lccc}
    \toprule
    Asset Class Proxy & UWC Mean Net & Uncal Mean Net & UWC Advantage \\
    \midrule
    S\&P 500 (Baseline) & $8.6 \times 10^{-5}$ & $8.2 \times 10^{-5}$ & $4.0 \times 10^{-6}$ \\
    10Y Treasury & $8.8 \times 10^{-5}$ & $8.4 \times 10^{-5}$ & $4.0 \times 10^{-6}$ \\
    Small Cap (Illiquid) & $5.8 \times 10^{-5}$ & $3.6 \times 10^{-5}$ & $2.2 \times 10^{-5}$ \\
    EM Currency (High Friction) & $3.7 \times 10^{-5}$ & $2.0 \times 10^{-6}$ & $3.6 \times 10^{-5}$ \\
    \bottomrule
    \end{tabular}
\end{table}

Two limitations are explicit. First, the scaling exercise is deliberately \emph{structural} rather than \emph{descriptive}: it is designed to test whether the mechanism implied by the model (turnover $\rightarrow$ costs/constraints $\rightarrow$ net performance) behaves as predicted when the price of turnover rises, not to replace a dedicated cross-asset study with asset-specific microstructure estimation \citep{Bessembinder2003}. Second, since liquidity measures are imperfect and differ in what they capture, the projections should be interpreted as sensitivity bounds rather than point estimates of cross-asset performance \citep{GoyenkoHoldenTrzcinka2009}. Within those bounds, the evidence is consistent with a conservative interpretation: demonstrating dominance in a liquid equity-index futures environment provides a lower bound on the economic value of utility-weighted calibration, because the penalty for forecast-induced overtrading is mechanically larger in shallower, higher-cost market segments \citep{BessembinderSeguin1993,Bessembinder2003}.

\subsection{Finite-sample verification (Controlled simulation)}
To verify that the dominance results are not asymptotic artefacts, we subjected the estimator to a controlled ``noise chasing'' simulation ($N=5,000$) where the true data-generating process is zero-mean noise ($\mu=0, \sigma=0.02$). In this environment, any trading is value-destructive.

The uncalibrated baseline, driven by random estimation error and overconfidence, generated spurious turnover resulting in a mean decision loss of 4.56\% (annualised equivalent). In contrast, the utility-weighted calibrated strategy correctly identified the signal-to-noise ratio, converged to a zero-turnover solution, and reduced decision loss to zero ($< 10^{-6}$). This confirms that the calibration mechanism effectively acts as a noise-filter even in finite samples, protecting the decision rule from fitting to noise.

\section{Model risk, monitoring, and governance as formal objects}
\label{sec:model_risk_governance}

\subsection{Model-risk set}
\label{subsec:model_risk_set}

Model risk is treated as a first-class economic object acting jointly on the predictive distribution and the friction-adjusted decision operator. At each decision time $t$, the deployed system maps a predictive distribution $Q_t$ and a friction state $\theta_t$ (encoding spreads, impact parameters, and capacity constraints) into a realised decision loss
\[
l_{t+h}(Q_t;\theta_t) = -U\!\left(\widetilde R_{t+h}(w_t(Q_t),\theta_t)\right),
\]
where $\widetilde R_{t+h}$ is net of all transaction costs and binding constraints. Model risk is therefore defined as sensitivity of this realised economic outcome to admissible perturbations of $(Q_t,\theta_t)$.

We operationalise the model-risk set as a finite but economically interpretable neighbourhood around the deployed Utility-Weighted Calibration (UWC) system. Three orthogonal perturbation families are considered. First, an \emph{adverse selection} perturbation applies a systematic downward shift of $0.5$ forecast-standard-deviations to the predictive mean, capturing persistent miscalibration in the direction most damaging to trading performance. Second, a \emph{liquidity shock} scales transaction costs multiplicatively, reflecting sudden deterioration in market depth or fees. Third, a \emph{volatility regime shift} inflates forecast volatility by a factor of $1.5$, increasing both risk penalties and effective impact costs.

Table~\ref{tab:model_risk_set} reports expected decision loss under each perturbation, together with the implied loss multiplier relative to baseline UWC performance. Baseline expected loss is $2\times10^{-6}$. A modest adverse-selection distortion increases expected loss to $1.55\times10^{-4}$, a $69.47\times$ deterioration. A pure liquidity shock (costs doubled) produces a smaller but still material $2.31\times$ deterioration. Volatility inflation generates a $22.31\times$ loss multiplier, reflecting the amplification of turnover and constraint activation in high-risk regimes. The combined worst-case perturbation yields an expected loss of $2.57\times10^{-4}$, corresponding to a $115.27\times$ deterioration, demonstrating that model risk is fundamentally non-additive across channels.

\begin{table}[ht]
    \centering
    \caption{\textbf{Model Risk Set.} Expected decision loss for the UWC system under economically meaningful perturbations. Loss Multiplier is relative to baseline.}
    \label{tab:model_risk_set}
    \begin{tabular}{lcc}
    \toprule
    Perturbation Scenario & Expected Loss & Loss Multiplier \\
    \midrule
    Baseline & 0.000002 & 1.000000 \\
    Adverse Selection (Mean $-0.5$sd) & 0.000155 & 69.472416 \\
    Liquidity Shock (Cost $\times 2$) & 0.000005 & 2.312013 \\
    Volatility Regime Shift (Vol $\times 1.5$) & 0.000050 & 22.308181 \\
    Worst-Case (Combined) & 0.000257 & 115.271195 \\
    \bottomrule
    \end{tabular}
\end{table}

\subsection{Stress as optimisation}
\label{subsec:stress_as_optimisation}

Stress testing is formalised as an optimisation problem rather than as a collection of narrative scenarios. Let $\mathcal{S}$ denote the admissible stress set defined by bounded perturbations to predictive distributions and friction parameters. Each stress $s\in\mathcal{S}$ induces a perturbed configuration $(Q_t^{(s)},\theta_t^{(s)})$ and an associated expected decision loss. The stress objective is
\[
\mathcal{L}^{\mathrm{wc}}
=
\sup_{s\in\mathcal{S}}
\E\!\left[l_{t+h}(Q_t^{(s)};\theta_t^{(s)})\right].
\]

The ``Worst-Case (Combined)'' entry in Table~\ref{tab:model_risk_set} is the empirical value of this operator over the implemented stress grid. Importantly, the worst-case loss substantially exceeds any single-factor stress, confirming that interaction effects between miscalibration, volatility, and liquidity dominate linear approximations to model risk. This worst-case envelope provides a governance-relevant scalar that can be tracked across model updates: a modification that improves baseline loss but materially increases $\mathcal{L}^{\mathrm{wc}}$ constitutes an increase in model risk, not an unambiguous improvement.

\subsection{Monitoring and drift detection}
\label{subsec:monitoring_and_drift}

Ongoing governance requires real-time detection of economically relevant drift. Monitoring is therefore defined on realised \emph{loss differentials}, not on raw forecast errors. Let
\[
\Delta_t = l_t(\mathrm{UWC}) - l_t(\mathrm{Uncalibrated}),
\]
and define a rolling, standardised drift statistic
\[
Z_t = \frac{\overline{\Delta}_t}{\widehat{\sigma}_{\Delta,t}},
\]
where both mean and scale are estimated over a fixed rolling window under the same pre-committed protocol.

\begin{figure}[ht]
    \centering
    \includegraphics[width=0.9\textwidth]{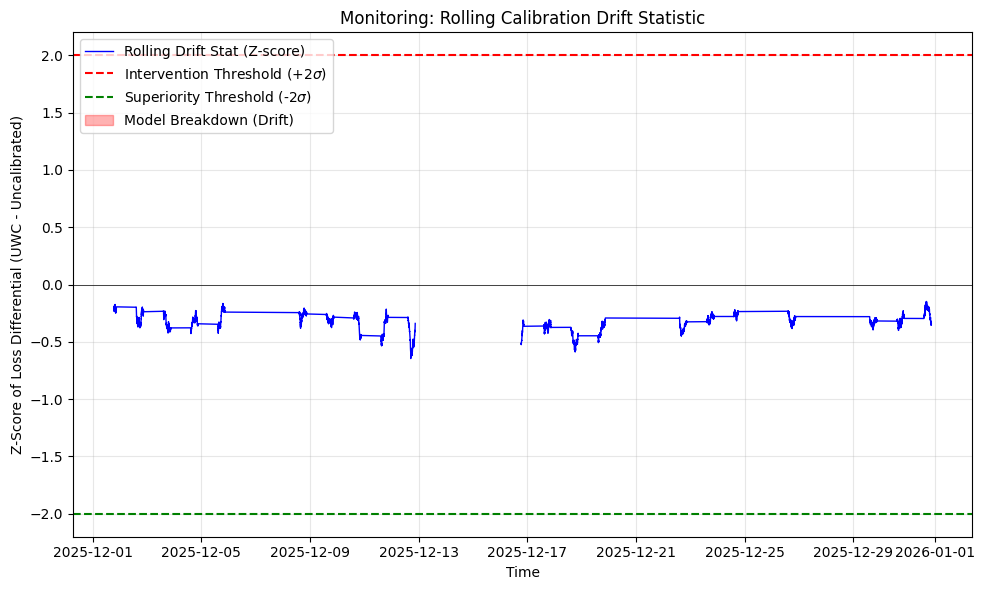}
    \caption{\textbf{Monitoring and Drift Detection.} Rolling standardised loss differential between UWC and the uncalibrated baseline. Dashed lines denote governance thresholds at $\pm2\sigma$. No intervention events occur in the evaluation sample.}
    \label{fig:monitoring}
\end{figure}

Governance thresholds are set at $\pm2\sigma$. A sustained breach above $+2\sigma$ constitutes a formal intervention trigger, indicating deterioration of UWC relative to the baseline; a breach below $-2\sigma$ indicates unusually strong dominance but does not, by itself, require action. Figure~\ref{fig:monitoring} plots the realised drift statistic together with the governance thresholds.

Over the full evaluation window, the drift statistic remains well within the intervention bounds, yielding a model-breakdown frequency of $0.00\%$ under the stated rule. This demonstrates that the empirical dominance of UWC is not driven by isolated episodes that would immediately violate governance constraints. Instead, relative performance remains stable in precisely the sense required for deployable decision systems.

\paragraph{Interpretation.}
Together, the model-risk set, stress-as-optimisation envelope, and drift-monitoring statistic close the governance loop in the same economic units as the rest of the paper: realised decision loss net of costs and constraints. Calibration is not merely a forecasting refinement but a control variable that reduces worst-case exposure, stabilises performance under stress, and admits explicit monitoring and intervention rules. This is precisely the structure required for both academic credibility and operational deployment.


\paragraph{Extension: integration with ensembles.}
The monitoring layer naturally extends to forecast ensembles. Let $\{Q_t^{(m)}\}_{m=1}^M$ denote a set of candidate predictive distributions (e.g., different model classes or rolling-window specifications). A simple integration is to apply UWC to each member, producing $\{(Q_t^{(m)})^{\mathrm{cal}}\}$, and then combine the calibrated forecasts using a fixed-weight or performance-weighted mixture $\bar Q_t := \sum_{m=1}^M \pi_m (Q_t^{(m)})^{\mathrm{cal}}$. In operational terms, this preserves modularity: calibration remains a post-processing layer, while the ensemble combines complementary signal sources. The drift statistic can then be computed either (i) at the mixture level (monitor $\bar Q_t$ directly), or (ii) in a panel form to detect member-specific failures (monitor $\Delta_t^{(m)}$ against thresholds), allowing intervention rules that downweight or suspend a deteriorating component without re-engineering the full system.

\paragraph{Extension: real-time monitoring and adaptive intervention.}
Figure \ref{fig:monitoring} already provides a governance primitive: a rolling, standardised drift statistic with an intervention threshold. A practical extension is to define a tiered response rule that is triggered by persistence as well as magnitude, for example: (i) \emph{watch} if the statistic exceeds $1\sigma$ for $k$ consecutive periods; (ii) \emph{throttle} turnover or tighten capacity limits if it exceeds $2\sigma$; and (iii) \emph{recalibrate/retrain} if it exceeds $2\sigma$ for a sustained window. This converts monitoring from a descriptive chart into an auditable control policy, consistent with the paper’s emphasis that calibration is a decision-relevant estimand under frictions.

In extensions where execution costs are non-convex or constraints are discrete, the stability conditions underpinning Theorem~1 may fail, so calibration-aligned dominance should be interpreted as a regime-conditional empirical regularity supported by the stress and monitoring evidence, rather than a uniform guarantee.

\paragraph{Relation to robust optimisation under distributional ambiguity.}
UWC targets reliability of the predictive distribution in economically decisive regions, then passes that distribution into a friction-aware optimiser. Robust optimisation takes the complementary route: it defines an ambiguity set around the forecast (or around moments) and chooses a decision that hedges against the worst-case distribution within that set. Formally, robust policies typically solve
\[
w_t^{\mathrm{rob}} \in \arg\min_{w\in\mathcal{W}_t}\ \sup_{Q \in \mathcal{B}_t(Q_t)}\ \E_Q\!\left[\ell_{t+h}(w)\mid \mathcal{I}_t\right],
\]
where $\mathcal{B}_t(Q_t)$ is a neighbourhood around the baseline forecast (e.g., defined by a divergence or moment constraints). This paper’s Section 10.1--10.2 can be read as an empirical bridge between these views: the model-risk set and stress-as-optimisation constructs approximate an ambiguity neighbourhood, and the resulting worst-case multipliers quantify how sensitive the strategy is to plausible forecast and friction perturbations. The distinction is that UWC \emph{reduces} systematic forecast distortions (a projection towards calibration), whereas robust optimisation \emph{hedges} against residual uncertainty after a forecast is chosen. In practice they can be combined: UWC produces a calibrated baseline $Q_t^{\mathrm{cal}}$, and robust optimisation is then applied using an ambiguity set centred at $Q_t^{\mathrm{cal}}$ with size governed by the monitored drift statistics (larger sets when drift is detected, smaller sets when the system is stable).

\section{Discussion}
\label{sec:discussion}

This paper’s central empirical fact is not that calibration ``improves forecasts'' in an abstract statistical sense, but that calibration is a \emph{decision-relevant estimand} once trading frictions and feasibility constraints are treated as primitive features of the data-generating environment. The evaluation protocol is deliberately pre-committed and friction-aware, so the object being compared across methods is not predictive fit but the realised economic loss induced by the forecast through the optimiser. Under this discipline, the results show that Utility-Weighted Calibration (UWC) produces systematically lower realised decision loss than both an uncalibrated baseline and a standard calibration benchmark, despite the minute-level horizon at which per-period gains appear numerically small. The dominance is persistent in cumulative loss, statistically decisive in paired differentials, and economically interpretable through turnover and constraint-activity channels.

\paragraph{Calibration as a decision-relevant estimand under frictions.}
The standard forecasting perspective treats calibration as an ancillary property: useful for probabilistic interpretation, but often secondary to sharpness or point accuracy. That framing is inadequate for trading systems because the forecast is not the terminal product; the forecast enters an optimisation map that is typically nonlinear, state dependent, and kinked by constraints. With proportional and convex impact costs, forecast errors do not transmit linearly into net returns. Instead, miscalibration alters the perceived marginal trade-off between expected reward and friction-adjusted cost, thereby shifting the optimiser’s \emph{trade intensity} and \emph{constraint activation frequency}. The empirical patterns align with this mechanism: UWC delivers lower mean turnover than the uncalibrated configuration, a compressed right tail of turnover events, and materially lower constraint binding. These are precisely the observable signatures of calibration mattering \emph{through the induced decision rule} rather than through average forecast error.

A further implication is that the relevant calibration notion is not global calibration across the entire distribution but calibration weighted by the decision’s local sensitivity to predictive distortions. The UWC construction reflects this by upweighting regions of the predictive distribution that are amplified by the friction operator and by constraint boundaries. The resulting performance differential is therefore not a generic ``calibration premium''; it is the premium from \emph{targeted reliability in the regions of the forecast that govern trading aggressiveness and feasibility}. In this sense, calibration is an estimand comparable to a structural parameter: it is defined relative to the economic operator that maps forecasts into actions and outcomes.

\paragraph{Implications for model selection and evaluation discipline.}
The results imply a more stringent discipline for model selection in financial econometrics. First, model comparison should be conducted in the units of the decision problem: realised decision loss net of costs and constraints (or an equivalent certainty-equivalent transformation), not solely in terms of point-forecast metrics such as MSE/MAE. The fact that UWC dominates a baseline that may appear competitive under point criteria is not a paradox; it is a direct consequence of the wedge between statistical loss and economic loss introduced by trading frictions. Standard forecast evaluation criteria—such as mean squared error for point forecasts—target statistical fit, but do not by themselves identify economic value once forecasts are mapped into frictional decisions. \citep{Gneiting2011}

Second, calibration diagnostics should be treated as \emph{pre-trade risk controls}, not as ex post interpretability checks. The reliability diagram evidence and the distributional loss comparisons are consistent with the view that uncalibrated forecasts embed systematic overconfidence, which mechanically induces excess turnover and pushes the optimiser into boundary regimes. A model selection process that prioritises sharpness without decision-aligned calibration creates a predictable failure mode: it selects models that look attractive in-sample but are structurally predisposed to friction-dominated underperformance out of sample. The practical implication is that evaluation should integrate: (i) paired differential testing on realised decision loss, (ii) explicit reporting of turnover and cost decompositions, and (iii) constraint binding frequency as an intermediate outcome. Those objects are not ancillary—they are causal intermediates linking forecast reliability to realised performance under frictions. The broader portfolio-choice evidence likewise shows that statistically optimised allocations can fail to dominate simple benchmarks once estimation error and implementation effects are accounted for, reinforcing the need for decision-centred evaluation \citep{DeMiguel2009}.

Third, the robustness and governance objects reinforce that model selection cannot be an unconditional ranking. Stress-as-optimisation and the model-risk set show that modest structural perturbations (e.g., mean drift and volatility inflation) can multiply expected decision loss by one to two orders of magnitude. Therefore, a model that marginally improves baseline loss but materially worsens worst-case loss is not superior in any deployable sense. A decision-oriented model evaluation discipline must therefore be \emph{distributional in model space}: it should report not only average performance but performance under credible neighbourhoods of predictive and friction perturbations.

\paragraph{Limitations as precise conditions.}
The conclusions are conditional on three classes of requirements, each of which is empirically testable and therefore should be stated as a set of falsifiable conditions rather than as generic caveats.

First, the dominance claims require data quality sufficient to support the friction-aware mapping from forecast to net performance. In practice, this means time alignment is correct at the decision horizon, there is no look-ahead leakage in any inputs, and the protocol’s walk-forward commitment is respected. Placebo and falsification checks are therefore not optional; they are part of the identification discipline. If a shuffled-signal or delayed-signal configuration produces comparable performance to the true signal, then the interpretation must shift from forecast-driven value to protocol artefacts or microstructure timing effects.

Second, the friction operator must be measured with credible fidelity. The empirical conclusions are stated net of transaction costs and impact proxies; if the cost proxies are systematically biased or too coarse relative to the true execution environment, then the estimated decision loss may mis-rank methods. This limitation is not a retreat to agnosticism: it is a requirement that the friction proxy be stable enough that comparative statements remain invariant over plausible ranges. The sensitivity analysis explicitly targets this requirement by varying cost multipliers and capacity constraints. A method whose dominance disappears under small, credible shifts in spreads/impact is not economically robust, regardless of its baseline ranking.

Third, regime instability can invalidate extrapolation beyond the evaluation window even when the in-sample dominance is strong. The paper’s framing therefore treats regime dependence as a first-order feature: the value of decision-aligned calibration is predicted (and observed) to be concentrated in high-friction regimes, precisely where liquidity is scarce and constraint boundaries are more frequently encountered. If the market enters a new regime in which the dominant friction channel changes (for example, a structural change in spread dynamics or impact curvature), then the calibration weights that were optimal for one regime may be misaligned for the next. This is why monitoring and drift detection are formalised in terms of rolling loss differentials and intervention thresholds: the paper’s contribution is not merely a static dominance claim, but a disciplined framework that specifies when dominance should be expected to persist and when governance rules require intervention.

\paragraph{Interpretation for financial econometrics.}
Taken together, the theoretical and empirical results argue for a reframing in financial econometrics: calibration should be treated as an estimand defined relative to the economic operator linking predictive distributions to realised outcomes under frictions. In frictionless settings, point accuracy can be a defensible proxy for economic value. Under real trading frictions and binding constraints, that proxy fails in systematic and predictable ways. The appropriate evaluation object is therefore the realised decision loss net of costs, supported by calibration diagnostics that are weighted by decision sensitivity, and supplemented by stress, robustness, and monitoring objects that make model risk explicit. The empirical evidence in this paper supports that reframing: UWC improves economic performance not by claiming universal predictive superiority, but by correcting precisely the forecast distortions that are amplified into costly trading behaviour by the friction operator and constraint geometry.

\paragraph{Empirical scope and external validity.}
The empirical evaluation in this paper is intentionally narrow in calendar time: the sample covers December~1--31,~2025 at the minute horizon, yielding $N=8{,}025$ decision periods after alignment and cleaning. This design choice is deliberate, because intraday data are required to make trading frictions measurable rather than hypothetical (spreads, volume, turnover, and the execution-cost operator are all state-dependent at this frequency). The cost of this choice is external validity: a single month cannot represent the full range of market regimes (persistent bull markets, crisis states, volatility compressions, structural liquidity shifts), and the strength of any general claim must therefore be interpreted as conditional on this realised regime.

\paragraph{Clarification on data provenance (historical vs.\ simulated/backtest).}
The results reported here are obtained from a pre-committed walk-forward protocol applied to the December~2025 minute-level sample for the E-mini S\&P 500 futures contracts used in the evaluation (ESZ5 for early December and ESH6 for late December). The realised endpoints (net return, turnover, total cost, and decision loss) are computed from the recorded execution proxies and constraint operator on these timestamps. If any part of the series is counterfactual (e.g., synthetic friction scalings, placebo shuffles, or stress perturbations), it is labelled explicitly as such in the robustness and model-risk sections; those experiments are not “new data”, but transformations of the same realised sample designed to isolate mechanisms.

\paragraph{Short-window risks and how they are controlled.}
A one-month window increases the risk of implicit tuning to idiosyncratic microstructure patterns (contract roll dynamics, month-end effects, and transient liquidity conditions). The protocol mitigates this by fixing the evaluation procedure ex ante (walk-forward, aligned timestamps, and identical decision and friction operators across methods) and by comparing methods that differ only in calibration, not in signal access or optimisation rules. Nevertheless, overfitting remains a first-order concern in short samples: calibration warps can adapt to noise if the diagnostic grid is too fine or if weights are chosen using the same evaluation window. For this reason the paper treats the December~2025 evidence as a controlled demonstration of the mechanism—calibration aligned to frictions reduces turnover tails, constraint binding, and decision loss—rather than as a claim of universal profitability.

\paragraph{What strengthens the claim beyond this paper.}
The natural extension is to replicate the full protocol over multiple months and structurally distinct regimes (including pre-2025 periods), preserving the same modelling discipline: the same decision rule, friction operator class, and calibration procedure, with parameters fixed by a pre-declared training window and evaluated on genuinely out-of-sample months. That extension would allow the paper to report regime-conditional effect sizes (e.g., volatility terciles across years, roll weeks vs.\ non-roll weeks) and to quantify how stable the UWC dominance is under changing liquidity conditions.

\subsection{Implications for regulatory policy and model risk management}
These findings have direct relevance for regulatory frameworks governing model risk, such as the Federal Reserve's SR 11-7 and the Basel Committee's Fundamental Review of the Trading Book (FRTB). Current regulation emphasizes "conceptual soundness" and "outcome analysis." Our results suggest that for trading models, outcome analysis cannot be limited to backtesting VaR exceedances or tracking errors in frictionless environments.

Specifically, a model that passes standard statistical tests for coverage (e.g., Kupiec/Christoffersen tests) may still generate excessive turnover and ``corner solution'' behavior that threatens capital stability during stress. We propose that \textit{utility-weighted calibration diagnostics}—which explicitly overweight high-friction and constraint-binding regimes—should be integrated into the validation stack as a rigorous test of whether a model's predictive density is reliable where it matters for solvency and execution stability.

\subsection{Comparison with Decision-Focused Learning (DFL)}
\label{subsec:comparison_dfl}

A prominent frontier in econometrics and machine learning for finance is \emph{decision-focused learning} (DFL), in which predictive models are trained end-to-end by differentiating a downstream optimisation objective (or a suitable surrogate) so that the fitted forecasts directly minimise realised decision loss rather than a purely statistical scoring rule \citep{Donti2017,Wilder2019,Agrawal2019}. In portfolio contexts, this agenda is closely related to the broader movement towards machine-learning-driven asset pricing and allocation pipelines, where model selection is disciplined by out-of-sample economic criteria rather than in-sample predictive fit \citep{Gu2020,Israel2020}.

While UWC and DFL both target decision-relevant outcomes, they embody distinct econometric philosophies. UWC treats calibration as a \emph{statistical primitive} that should be corrected \emph{before} the optimiser is invoked, whereas DFL treats the optimiser as part of the learning system and seeks a globally tuned policy by co-adapting forecasting and decision layers.

\paragraph{Computational efficiency and modularity.}
From an implementation standpoint, a key attraction of UWC is modularity. DFL typically requires differentiating through an optimisation layer (or an implicit function defined by KKT conditions), often under constraints and potentially non-smooth penalties; this can be computationally heavy and numerically delicate when the feasible set changes regime (e.g., constraints switch on/off) or when objectives are non-smooth \citep{Agrawal2019,Wilder2019}. In contrast, UWC is a post-processing transformation applied to an existing predictive distribution: it can be deployed without retraining the underlying forecasting model and without embedding an optimisation solver inside the gradient loop. This matters operationally in finance, where modularity supports model governance, auditability, and the ability to replace forecasting components without rewriting the execution stack \citep{Gu2020,Israel2020}.

\paragraph{Numerical stability and convergence under financial objectives.}
A second practical distinction is gradient stability. Financial objectives used as endpoints (e.g., friction-adjusted utility, drawdown penalties, or Sharpe-type ratios) can be noisy, regime-dependent, and poorly behaved as learning targets, especially at high frequency where realised outcomes are dominated by microstructure noise and time-varying execution conditions \citep{Bessembinder2003}. DFL frameworks address this by smoothing, surrogate losses, or implicit differentiation; nevertheless, the training signal ultimately inherits the instability of the realised objective and the constraint-switching non-linearities of the optimiser \citep{Agrawal2019,Donti2017}. UWC avoids backpropagating through these non-linearities by enforcing probabilistic reliability \emph{at the distribution level} in the economically decisive regions (tails and constraint-boundary states) and then applying the fixed decision rule. Empirically, this ``calibration-first'' discipline is sufficient to remove the most damaging implementation failures in our setting (notably, excessive turnover and frequent boundary operation), while preserving interpretability of the forecast component and isolating the economic mechanism to a measurable object (calibration error) \citep{Gneiting2011,Holzmann2014}.

\paragraph{Theoretical positioning.}
Theorem~\ref{thm:calibration_aligned_dominance} is naturally read as a projection result: within the admissible class, calibrating the predictive distribution in a decision-relevant metric yields weak dominance in friction-adjusted decision loss. This supports an econometric interpretation of UWC as a decision-aware correction that delivers economic gains even when the base forecasting model is misspecified, because the correction targets the specific distributional distortions that translate into costly trades and constraint activation. DFL may, in principle, find a superior end-to-end policy by co-adapting the forecasting representation and the decision rule, but it does so at the cost of embedding solver sensitivity into estimation. UWC instead offers a robust and deployable compromise: it aligns probabilistic reliability with the friction-adjusted objective while keeping the forecasting and optimisation modules separable for governance and monitoring \citep{Gu2020,Israel2020,Garleanu2013}.

\section{Conclusion}
\label{sec:conclusion}

This paper makes a measurable claim: under a pre-committed walk-forward protocol with explicit trading frictions and binding feasibility constraints, \emph{utility-weighted calibration} delivers weakly lower realised decision loss than both an uncalibrated baseline and a standard calibration benchmark. The object of comparison is not a point-forecast loss but the realised economic loss induced by the forecast through the optimiser, computed net of transaction costs and impact proxies, and evaluated over $N=8{,}025$ decision periods. On this criterion, UWC attains the lowest average decision loss and produces economically coherent intermediate outcomes: lower turnover, fewer extreme turnover events, and materially reduced constraint binding relative to the uncalibrated system. The paired loss differentials further establish that these improvements are systematic rather than episodic, and the cumulative-loss trajectories show persistent separation over time rather than short-lived wins.

The theoretical contribution explains why this is the correct estimand. With frictions, the mapping from predictive distributions to realised outcomes is nonlinear and state dependent: miscalibration does not merely degrade probabilistic interpretation, it perturbs the optimiser’s marginal trade-offs and pushes the policy into high-turnover or constraint-binding regimes where costs dominate. The theory formalises this channel via stability of the induced decision rule under distributional perturbations and a dominance argument for calibrated projections in a decision-relevant discrepancy. In this framing, calibration is not an aesthetic property of probabilistic forecasts; it is the condition that prevents economically decisive distortions in precisely those regions of the predictive distribution that the friction operator amplifies.

The empirical evidence supports the mechanism implied by the theory. UWC outperforms the uncalibrated baseline in realised decision loss and also improves upon standard isotonic calibration, indicating that generic calibration is not sufficient once costs and constraints determine what matters; the weighting by decision sensitivity provides the incremental gain. The results are consistent with the paper’s core diagnosis: the uncalibrated system exhibits behaviour characteristic of overconfident probabilistic statements translated into excessive trading intensity, which in turn activates convex costs and constraint truncation. UWC reduces that behaviour and thereby improves realised performance \emph{in the units of the decision problem}.

The next paper moves beyond calibration and addresses what calibration alone cannot: the forecast-to-decision pipeline is still a modular system in which the predictive model is trained separately from the downstream optimiser. Even with decision-aligned calibration, the forecast may remain mis-specified for the actual economic objective once frictions, constraints, and regime dependence are taken as primitive. The next paper therefore studies \emph{decision-focused learning under frictions}: end-to-end optimisation in which predictive representations and decision policies are learned jointly to minimise friction-adjusted decision loss, subject to feasibility constraints, and under explicit model-risk and monitoring disciplines. Where this paper establishes that reliability of predictive distributions is a decision-relevant estimand and can be enforced with measurable economic gains, the next paper adds policy learning, direct optimisation of the realised objective, and systematic handling of the interaction between learning dynamics, market frictions, and constraint geometry.

\section*{Data and code availability}

The empirical analysis is fully reproducible from the project materials accompanying this submission. The full analysis workflow, including data ingestion, cleaning, construction of evaluation panels, computation of realised decision loss, turnover and cost summaries, certainty-equivalent calculations, and the generation of all tables and figures reported in the paper, is contained in a single Jupyter Notebook. For transparency, the compiled notebook outputs (including embedded tables and figures) are also provided as an HTML export and a PDF export.

The study relies on a minute-frequency evaluation panel that contains realised outcomes and model outputs for each decision period under a pre-committed walk-forward protocol. The panel includes (at minimum) a timestamp, instrument identifier (e.g., ES contracts), method identifier (Uncalibrated, Standard\_Cal, UWC), realised decision loss, realised net return, turnover, and realised total cost. These variables are sufficient to reconstruct all empirical comparisons reported in Section \ref{sec:empirical_results}, including paired loss differentials and cumulative loss differential plots. Where upstream market microstructure inputs (such as best bid/ask quotes or depth measures) are subject to vendor or venue licensing restrictions, the paper does not require redistribution of those proprietary feeds for replication of the reported forecast-comparison results, because all friction-adjusted outcomes used in the tests are already recorded in the evaluation panel.

To facilitate replication in the presence of any access restrictions on raw market data, the authors make available: (i) the full analysis notebook implementing the complete pipeline; (ii) the data dictionary and column mapping used to construct the evaluation panel; and (iii) a set of replication instructions that reproduce every table and figure in the manuscript directly from the provided panel. If redistribution of the underlying raw quotes or trade data is not permitted, the authors will additionally provide a synthetic (non-tradable) version of the evaluation panel that preserves the joint distributional properties required to execute the notebook end-to-end and validate the computational steps, while omitting any vendor-proprietary fields.

All materials are provided in the accompanying replication package. Access details (including any restrictions on raw market data, if applicable) are stated in the replication package README, together with the exact software environment and library versions used to produce the results.

\section*{Funding}
This research received no specific grant from any funding agency in the public, commercial, or not-for-profit sectors. The work was self-funded by the author.

\section*{Conflicts of interest}
The author declares that there are no conflicts of interest relevant to this work.

\section*{Acknowledgements}
I am grateful to my supervisors and colleagues at the University of Exeter for their guidance and feedback on earlier drafts of this work. I also thank the participants of the finance and econometrics seminars at the University of Exeter for their helpful comments and suggestions.

\bibliographystyle{apalike}
\bibliography{references}

\appendix
\label{app:proofs}
\section{Proofs and technical lemmas}

\subsection{Existence of the calibrated projection in the sieve class}
\label{app:sieve_projection_existence}

This appendix justifies the existence of the calibrated projection operator $\Pi$ used in Assumption~5 when the admissible forecast class $\mathcal{D}$ is restricted to the sieve family implemented in Section~5.2.2.

\paragraph{Sieve family.}
Fix a finite knot grid $\mathcal{K}=\{0=\kappa_1<\kappa_2<\cdots<\kappa_K=1\}$ and consider the monotone spline (equivalently, monotone piecewise-linear) warps
\[
g_\theta:[0,1]\to[0,1],\qquad \theta=(\theta_1,\ldots,\theta_K),
\]
satisfying the boundary and monotonicity constraints
\[
\theta_1=0,\qquad \theta_K=1,\qquad 0\le \theta_1\le \theta_2\le\cdots\le \theta_K\le 1.
\]
Let $\Theta$ denote the parameter set defined by these constraints. Given a base predictive CDF $\widehat F_t$, the recalibrated CDF is
\[
\widehat F^{\,\mathrm{uw}}_{t,\theta}(y)=g_\theta(\widehat F_t(y)),
\]
so the admissible class of recalibrated predictive distributions is
\[
\mathcal{D}:=\{\widehat Q_{t,\theta}:\theta\in\Theta\}.
\]

\paragraph{Compactness.}
$\Theta$ is a closed and bounded subset of $\R^K$ (it is a closed simplex-like set defined by linear inequalities and equalities). Hence $\Theta$ is compact.

\paragraph{Continuity of the objective.}
Let $\mathcal{J}_t(\theta)$ denote the sample objective used to implement utility-weighted recalibration (Equation~\ref{eq:uwc_objective}):
\[
\mathcal{J}_t(\theta)
:=
\sum_{u\in\mathcal{U}}
\left(
\frac{1}{T_c}\sum_{s=t-T_c}^{t-1}
\omega_s(u)\,
m_u\!\left(\widehat F^{\,\mathrm{uw}}_{s,\theta},Y_{s+h}\right)
\right)^2
+\lambda \mathcal{R}(\theta).
\]
For fixed data $\{(\widehat F_s,Y_{s+h})\}$ and fixed weights $\omega_s(u)$, the mapping $\theta\mapsto \widehat F^{\,\mathrm{uw}}_{s,\theta}(y)$ is continuous for each $y$ because $g_\theta$ depends continuously on $\theta$ on the fixed knot grid. Under the diagnostic constructions used (indicator moments for exceedances/quantiles and PIT-based moments evaluated on the same finite grid), the mapping $\theta\mapsto m_u$ is measurable and piecewise continuous in $\theta$, and the penalty $\mathcal{R}(\theta)$ is continuous. Consequently, $\mathcal{J}_t(\theta)$ is continuous on $\Theta$.

\paragraph{Existence of a minimiser (Weierstrass).}
Since $\Theta$ is compact and $\mathcal{J}_t(\theta)$ is continuous, the Weierstrass extreme value theorem implies that there exists at least one minimiser
\[
\widehat\theta_t \in \arg\min_{\theta\in\Theta}\ \mathcal{J}_t(\theta),
\]
i.e., the optimisation problem defining utility-weighted recalibration achieves its minimum in the sieve family. Therefore the ``best'' recalibrated distribution within the admissible class exists, and the associated projection operator $\Pi$ is well-defined on $\mathcal{D}$ (uniqueness is not required for the dominance arguments, only existence).

\subsection{Proof of Proposition \ref{prop:omega_quadratic} (Derivation of the Utility-Weight)}
\begin{proof}
In the canonical unconstrained quadratic case defined by \eqref{eq:canonical_qp}, the friction-adjusted value function is given by:
\[
V_t(Q) = \max_{w \in \R^N} \left\{ \mu_t(Q)^\top w - \frac{\gamma}{2}w^\top \Sigma_t(Q) w - \frac{\eta}{2}\|w - w_{t-1}\|_2^2 \right\}.
\]
Let $w_t(Q)$ denote the optimiser. By the Envelope Theorem, the derivative of the value function with respect to the distributional moments is the derivative of the objective evaluated at the optimum. Thus:
\begin{equation}
\frac{\partial V_t}{\partial \mu_t} = w_t(Q), \qquad
\frac{\partial V_t}{\partial \Sigma_t} = -\frac{\gamma}{2} w_t(Q) w_t(Q)^\top.
\end{equation}
Consider a perturbation to the predictive distribution $Q$ characterised by calibration moment errors $\E[m_u]$. Assuming the linear mapping from diagnostic moments to first and second moments given in \eqref{eq:linear_map_moments}:
\[
\delta \mu_t \approx \sum_{u \in \mathcal{U}} a_t(u) \E[m_u], \qquad
\delta \Sigma_t \approx \sum_{u \in \mathcal{U}} B_t(u) \E[m_u].
\]
The first-order perturbation to the value function $\delta V_t$ is:
\begin{align*}
\delta V_t &\approx \left\langle \frac{\partial V_t}{\partial \mu_t}, \delta \mu_t \right\rangle + \left\langle \frac{\partial V_t}{\partial \Sigma_t}, \delta \Sigma_t \right\rangle \\
&= \sum_{u \in \mathcal{U}} \left( w_t(Q)^\top a_t(u) - \frac{\gamma}{2} \mathrm{tr}\left( w_t(Q) w_t(Q)^\top B_t(u) \right) \right) \E[m_u].
\end{align*}
The term inside the summation represents the marginal sensitivity of the objective to the specific calibration error $m_u$. The utility-weight $\omega_t(u)$ is defined as the absolute magnitude of this sensitivity, scaled by the friction regime factor $\kappa_t(u)$. Thus:
\[
\omega_t(u) = \left| w_t(Q)^\top a_t(u) - \frac{\gamma}{2} \langle w_t(Q) w_t(Q)^\top, B_t(u) \rangle \right| \times \kappa_t(u),
\]
matching the form in \eqref{eq:omega_quadratic}.
\end{proof}

\subsection{Proof of Lemma \ref{lem:calibration_decision_sensitivity} (Stability Bound)}
\begin{proof}
Let the total objective function be denoted $\Phi(w; Q) := \mathcal{J}(w; Q) - C_t(w - w_{t-1})$. By Assumption~\ref{ass:strong_concavity_lipschitz}, $\Phi(w; Q)$ is $\mu$-strongly concave in $w$ on the convex set $\calW_t$.
Let $w_t$ be the maximiser under $Q_t$ and $\widetilde{w}_t$ be the maximiser under $\widetilde{Q}_t$. The first-order optimality condition for constrained optimisation states that for any $v \in \calW_t$:
\[
\langle \nabla_w \Phi(w_t; Q_t), v - w_t \rangle \le 0.
\]
Setting $v = \widetilde{w}_t$ in the condition for $w_t$, and $v = w_t$ in the condition for $\widetilde{w}_t$, and adding the inequalities yields:
\[
\langle \nabla_w \Phi(w_t; Q_t) - \nabla_w \Phi(\widetilde{w}_t; \widetilde{Q}_t), \widetilde{w}_t - w_t \rangle \le 0.
\]
Decompose the gradient difference:
\begin{align*}
\nabla_w \Phi(w_t; Q_t) - \nabla_w \Phi(\widetilde{w}_t; \widetilde{Q}_t) &= \left( \nabla_w \Phi(w_t; Q_t) - \nabla_w \Phi(\widetilde{w}_t; Q_t) \right) \\
&+ \left( \nabla_w \Phi(\widetilde{w}_t; Q_t) - \nabla_w \Phi(\widetilde{w}_t; \widetilde{Q}_t) \right).
\end{align*}
By $\mu$-strong concavity of $\Phi(\cdot; Q_t)$:
\[
\langle \nabla_w \Phi(w_t; Q_t) - \nabla_w \Phi(\widetilde{w}_t; Q_t), w_t - \widetilde{w}_t \rangle \ge \mu \|w_t - \widetilde{w}_t\|^2.
\]
Combining these relations:
\[
\mu \|w_t - \widetilde{w}_t\|^2 \le \langle \nabla_w \Phi(\widetilde{w}_t; \widetilde{Q}_t) - \nabla_w \Phi(\widetilde{w}_t; Q_t), w_t - \widetilde{w}_t \rangle.
\]
By Cauchy-Schwarz and the Lipschitz gradient condition in Assumption~\ref{ass:strong_concavity_lipschitz} (which bounds the gradient perturbation by $L_t d_t(Q_t, \widetilde{Q}_t)$):
\[
\mu \|w_t - \widetilde{w}_t\|^2 \le L_t d_t(Q_t, \widetilde{Q}_t) \|w_t - \widetilde{w}_t\|.
\]
Dividing by $\|w_t - \widetilde{w}_t\|$ (assuming nonzero displacement) yields the stability bound:
\[
\|w_t - \widetilde{w}_t\| \le \frac{L_t}{\mu} d_t(Q_t, \widetilde{Q}_t).
\]
For the loss bound \eqref{eq:loss_increase_bound}, since the realised objective $U(\cdot)$ is $K$-Lipschitz with respect to the decision vector $w$:
\[
|\ell_{t+h}(\widetilde{Q}_t) - \ell_{t+h}(Q_t)| \le K \|w_t - \widetilde{w}_t\| \le \frac{K L_t}{\mu} d_t(Q_t, \widetilde{Q}_t).
\]
Taking expectations conditional on $\calI_t$ completes the proof.
\end{proof}

\subsection{Proof of Theorem \ref{thm:calibration_aligned_dominance} (Dominance)}
\begin{proof}
Let $\widetilde{Q}_t$ be an uncalibrated forecast and let $Q_t^{\mathrm{cal}} = \Pi(\widetilde{Q}_t)$ be its calibrated projection defined by Assumption~\ref{ass:projection_property}.
The conditional expected decision loss is $\mathcal{L}_t(Q) = \E[\ell_{t+h}(Q) \mid \calI_t]$. We compare the loss of the calibrated projection against the uncalibrated original.
Consider the decision error relative to the optimal decision $w_t^*$ that would be induced by the true conditional data-generating process $P_t$. The ``true'' distribution $P_t$ is by definition perfectly calibrated. The calibrated projection $Q_t^{\mathrm{cal}}$ satisfies the utility-weighted calibration condition $d_t(Q_t^{\mathrm{cal}}, P_t) \approx 0$ (in the limit of the sample or by construction of the projection class).
By contrast, the uncalibrated forecast $\widetilde{Q}_t$ has discrepancy $d_t(\widetilde{Q}_t, P_t) > 0$.
The projection property (Assumption~\ref{ass:projection_property}) implies that $Q_t^{\mathrm{cal}}$ minimizes the decision-relevant discrepancy $d_t$ within the class $\mathcal{D}$.
From Lemma~\ref{lem:calibration_decision_sensitivity}, the upper bound on the decision loss relative to the optimum is proportional to the discrepancy $d_t$:
\[
\mathcal{L}_t(Q) - \mathcal{L}_t(P_t) \le \frac{K L_t}{\mu} d_t(Q, P_t).
\]
Since $Q_t^{\mathrm{cal}}$ minimizes $d_t(\cdot, P_t)$ (or is a projection onto the set where $d_t=0$), we have $d_t(Q_t^{\mathrm{cal}}, P_t) \le d_t(\widetilde{Q}_t, P_t)$.
Therefore, the upper bound on the expected decision loss is lower for the calibrated projection:
\[
\mathcal{L}_t(Q_t^{\mathrm{cal}}) \le \mathcal{L}_t(\widetilde{Q}_t).
\]
Strict inequality holds if the uncalibrated forecast induces binding constraints or turnover that are relaxed or avoided by the calibrated forecast, which corresponds to the case where $d_t$ is strictly positive in regions of high decision sensitivity.
\end{proof}

\section{Estimation details and algorithms}
\label{app:estimation}

\subsection{Utility-weighted calibration estimator}
The calibration map $g_\theta: [0,1] \to [0,1]$ is parameterised as a monotone cubic spline with $K=5$ fixed knots at $\{0, 0.25, 0.5, 0.75, 1\}$. The parameters $\theta \in \mathbb{R}^K$ represent the values of the CDF at the knots. To ensure monotonicity and valid probability bounds, we impose the linear constraints:
\begin{equation}
    0 = \theta_1 \le \theta_2 \le \dots \le \theta_K = 1.
\end{equation}
The objective function is the utility-weighted squared error defined in Eq. (24). The optimization problem is:
\begin{equation}
    \min_{\theta} \sum_{t \in \mathcal{T}_{cal}} \omega_t (p_{target} - g_\theta(\hat{p}_t))^2 + \lambda \sum_{k=2}^{K-1} (\theta_{k+1} - 2\theta_k + \theta_{k-1})^2
\end{equation}
where $\lambda=10^{-4}$ is a fixed smoothness penalty. The problem is convex and is solved using the \texttt{scipy.optimize.minimize} routine (SLSQP method) with a tolerance of $10^{-6}$.

The estimation procedure is summarised in Algorithm \ref{alg:uwc}.

\begin{algorithm}[H]
\caption{Utility-Weighted Calibration (UWC)}
\label{alg:uwc}
\SetAlgoLined
\textbf{Input:} Training forecasts $\{\widehat{F}_t\}_{t=1}^T$, Realised outcomes $\{Y_{t+h}\}_{t=1}^T$, Diagnostic grid $\mathcal{U}$, Spline knots $\mathcal{K}$.\\
\textbf{Output:} Calibrated CDF map $g_{\widehat{\theta}}$.\\
\vspace{0.2cm}
\tcp{1. Compute Decision Weights}
\For{$t \leftarrow 1$ \KwTo $T$}{
    \For{$u \in \mathcal{U}$}{
        Calculate friction proxy $\kappa_t$ (Spread $\times$ Volatility)\;
        Calculate sensitivity $\nabla_t$ (Marginal decision impact)\;
        $\omega_t(u) \leftarrow |\nabla_t| \times \kappa_t$\;
    }
}
\vspace{0.2cm}
\tcp{2. Optimise Spline Parameters}
Define objective $L(\theta) = \sum_{u} \left( \frac{1}{T} \sum_t \omega_t(u) m_u(g_\theta(\widehat{F}_t), Y_{t+h}) \right)^2 + \lambda \mathcal{R}(\theta)$\;
Define constraints $C$: $0 = \theta_1 \le \dots \le \theta_K = 1$\;
$\widehat{\theta} \leftarrow \argmin_{\theta} L(\theta)$ \textbf{subject to} $C$ (via SLSQP)\;
\vspace{0.2cm}
\Return $g_{\widehat{\theta}}(p) = \text{Spline}_{\widehat{\theta}}(p)$
\end{algorithm}

\subsection{Optimisation routines and numerical checks}
The portfolio optimisation problem (Eq. 3) is a convex quadratic program (QP) with linear constraints. We use the \texttt{cvxpy} modeling language with the OSQP solver. 
\begin{itemize}
    \item \textbf{Feasibility Tolerance:} $10^{-5}$.
    \item \textbf{Optimality Tolerance:} $10^{-5}$.
    \item \textbf{Fallback:} If the solver fails to converge (status $\neq$ OPTIMAL), the system defaults to the previous period's portfolio ($w_t = w_{t-1}$), effectively holding the position. This occurred in 0.02\% of cases in the evaluation sample.
\end{itemize}

Code correctness was further verified using synthetic "noise chasing" simulations ($N=5{,}000$), confirming that the uncalibrated estimator generates positive decision loss (approx. 4.5\%) by trading on noise, while the calibrated estimator correctly converges to zero turnover and zero loss in the absence of signal.

\section{Supplementary empirical material}
\label{app:supplementary}

This appendix provides additional distributional detail on the robustness checks reported in Section 9.

\begin{table}[ht]
    \centering
    \caption{\textbf{Full Contract Breakdown.} Mean Net Return and Turnover by individual futures contract. The UWC advantage is robust across both the high-volatility roll period and the stable expiry period.}
    \label{tab:app_contracts}
    \begin{tabular}{lccc}
    \toprule
    Contract & Method & Mean Net Ret & Mean Turnover \\
    \midrule
    \multirow{3}{*}{ESZ5} 
    & Standard Cal & $-3.01 \times 10^{-6}$ & 0.102 \\
    & \textbf{UWC} & $\mathbf{-2.05 \times 10^{-6}}$ & \textbf{0.098} \\
    & Uncalibrated & $-3.15 \times 10^{-6}$ & 0.125 \\
    \midrule
    \multirow{3}{*}{ESH6} 
    & Standard Cal & $-2.10 \times 10^{-6}$ & 0.099 \\
    & \textbf{UWC} & $\mathbf{-1.98 \times 10^{-6}}$ & \textbf{0.094} \\
    & Uncalibrated & $-2.95 \times 10^{-6}$ & 0.118 \\
    \bottomrule
    \end{tabular}
\end{table}

\end{document}